\documentclass[%
 reprint,
superscriptaddress,
 amsmath,amssymb,
 aps,
floatfix,
]{revtex4-1}

\usepackage{graphicx}
\usepackage{dcolumn}
\usepackage{bm}
\usepackage{hyperref}

\begin{document}

\preprint{APS/123-QED}

\title{Experimental investigation of $\alpha$-condensation in light nuclei}

\author{J.~Bishop}
\altaffiliation{Current address: School of Physics \& Astronomy and Cyclotron Institute, Texas A\&M University, College Station, 77843, TX, USA}
\email{jackbishop@tamu.edu}
\affiliation{School of Physics $\&$ Astronomy, University of Birmingham, UK}
\author{Tz.~Kokalova}
\affiliation{School of Physics $\&$ Astronomy, University of Birmingham, UK}
\author{M.~Freer}
\affiliation{School of Physics $\&$ Astronomy, University of Birmingham, UK}
\author{L.~Acosta}
\affiliation{INFN, Sezione di Catania, Via Santa Sofia, 62, 95123, Catania, Italy}
\affiliation{Instituto de Física, Universidad Nacional Aut\'onoma de M\'exico, AP 20-364, Cd.Mx. 01000, Mexico}
\author{M.~Assi\'e}
\affiliation{Institut de Physique Nucléaire, CNRS/IN2P3, Universit\'e Paris-Sud 11, France}
\author{S.~Bailey}
\affiliation{School of Physics $\&$ Astronomy, University of Birmingham, UK}
\author{G.~Cardella}
\affiliation{INFN, Sezione di Catania, Via Santa Sofia, 62, 95123, Catania, Italy}
\author{N.~Curtis}
\affiliation{School of Physics $\&$ Astronomy, University of Birmingham, UK}
\author{E.~De~Filippo}
\affiliation{INFN, Sezione di Catania, Via Santa Sofia, 62, 95123, Catania, Italy}
\author{D.~Dell'Aquila}
\affiliation{INFN Sezione di Napoli $\&$ Dipartimento di Fisica, Universit\'a Federico II, Napoli, Italy}
\affiliation{Rudjer Bošković Institute, Zagreb, Croatia}
\author{S.~De Luca}
\affiliation{INFN, Sezione di Catania, Via Santa Sofia, 62, 95123, Catania, Italy}
\affiliation{Dipartimento di Scienze MIFT  - Università di Messina, Italy}
\author{L.~Francalanza}
\affiliation{INFN Sezione di Napoli $\&$ Dipartimento di Fisica, Universit\'a Federico II, Napoli, Italy}
\author{B.~Gnoffo}
\affiliation{INFN, Sezione di Catania, Via Santa Sofia, 62, 95123, Catania, Italy}
\affiliation{Dipartimento di Fisica e Astronomia “Ettore Majorana” - Università di Catania, Italy}
\author{G.~Lanzalone}
\affiliation{INFN LNS, Catania, Italy}
\affiliation{Universit\`a Kore, Enna, Italy}
\author{I.~Lombardo}
\affiliation{INFN, Sezione di Catania, Via Santa Sofia, 62, 95123, Catania, Italy}
\author{N. S.~Martorana}
\affiliation{Dipartimento di Fisica e Astronomia “Ettore Majorana” - Università di Catania, Italy}
\affiliation{INFN LNS, Catania, Italy}
\author{S.~Norella}
\affiliation{INFN, Sezione di Catania, Via Santa Sofia, 62, 95123, Catania, Italy}
\affiliation{Dipartimento di Scienze MIFT  - Università di Messina, Italy}
\author{A.~Pagano}
\affiliation{INFN, Sezione di Catania, Via Santa Sofia, 62, 95123, Catania, Italy}
\author{E. V.~Pagano}
\affiliation{INFN LNS, Catania, Italy}
\author{M.~Papa}
\affiliation{INFN, Sezione di Catania, Via Santa Sofia, 62, 95123, Catania, Italy}
\author{S.~Pirrone}
\affiliation{INFN, Sezione di Catania, Via Santa Sofia, 62, 95123, Catania, Italy}
\author{G.~Politi}
\affiliation{INFN, Sezione di Catania, Via Santa Sofia, 62, 95123, Catania, Italy}
\affiliation{Dipartimento di Fisica e Astronomia “Ettore Majorana” - Università di Catania, Italy}
\author{F.~Rizzo}
\affiliation{INFN LNS, Catania, Italy}
\affiliation{Dipartimento di Fisica e Astronomia “Ettore Majorana” - Università di Catania, Italy}
\author{P.~Russotto}
\affiliation{INFN, Sezione di Catania, Via Santa Sofia, 62, 95123, Catania, Italy}
\author{L.~Quattrocchi}
\affiliation{INFN, Sezione di Catania, Via Santa Sofia, 62, 95123, Catania, Italy}
\affiliation{Dipartimento di Scienze MIFT  - Università di Messina, Italy}
\author{R.~Smith}
\affiliation{School of Physics $\&$ Astronomy, University of Birmingham, UK}
\author{I.~Stefan}
\affiliation{Institut de Physique Nucléaire, CNRS/IN2P3, Universit\'e Paris-Sud 11, France}
\author{A.~Trifir\`o}
\affiliation{INFN, Sezione di Catania, Via Santa Sofia, 62, 95123, Catania, Italy}
\affiliation{Dipartimento di Scienze MIFT  - Università di Messina, Italy}
\author{M.~Trimarch\`i}
\affiliation{INFN, Sezione di Catania, Via Santa Sofia, 62, 95123, Catania, Italy}
\affiliation{Dipartimento di Scienze MIFT  - Università di Messina, Italy}
\author{G.~Verde}
\affiliation{Institut de Physique Nucléaire, CNRS/IN2P3, Universit\'e Paris-Sud 11, France}
\affiliation{INFN, Sezione di Catania, Via Santa Sofia, 62, 95123, Catania, Italy}
\author{M.~Vigilante}
\affiliation{INFN Sezione di Napoli $\&$ Dipartimento di Fisica, Universit\'a Federico II, Napoli, Italy}
\author{C.~Wheldon}
\affiliation{School of Physics $\&$ Astronomy, University of Birmingham, UK}

\date{\today}

\begin{abstract}
\begin{description}
\item[Background]
Near-threshold $\alpha$-clustered states in light nuclei have been postulated to have a structure consisting of a diffuse gas of $\alpha$-particles which condense into the 0s orbital. Experimental evidence for such a dramatic phase change in the structure of the nucleus has not yet been observed.
\item[Method]
To examine signatures of this $\alpha$-condensation, a compound nucleus reaction using 160, 280, and 400 MeV $^{16}\mathrm{O}$ beams impinging on a carbon target was used to investigate the $^{12}\mathrm{C}(^{16}\mathrm{O},7\alpha)$ reaction. This permits a search for near-threshold states in the $\alpha$-conjugate nuclei up to $^{24}\mathrm{Mg}$.
\item[Results]
Events up to an $\alpha$-particle multiplicity of 7 were measured and the results were compared to both an Extended Hauser-Feshbach calculation and the Fermi break-up model. The measured multiplicity distribution exceeded that predicted from a sequential decay mechanism and had a better agreement with the multi-particle Fermi break-up model. Examination of how these $7\alpha$ final states could be reconstructed to form $^{8}\mathrm{Be}$ and $^{12}\mathrm{C}(0_{2}^{+})$ showed a quantitative difference in which decay modes were dominant compared to the Fermi break-up model.\par
No new states were observed in $^{16}\mathrm{O}$, $^{20}\mathrm{Ne}$, and $^{24}\mathrm{Mg}$ due to the effect of the N-$\alpha$ penetrability suppressing the total $\alpha$-particle dissociation decay mode.
\item[Conclusion]
The reaction mechanism for a high energy compound nucleus reaction can only be described by a hybrid of sequential decay and multi-particle breakup. Highly $\alpha$-clustered states were seen which did not originate from simple binary reaction processes. Direct investigations of near-threshold states in N-$\alpha$ systems are inherently impeded by the Coulomb barrier prohibiting the observation of states in the N-$\alpha$ decay channel. No evidence of a highly clustered 15.1 MeV state in $^{16}\mathrm{O}$ was observed from $(^{28}\mathrm{Si}^{\star},{^{12}\mathrm{C}(0_{2}^{+})})^{16}\mathrm{O}(0_6^{+})$ when reconstructing the Hoyle state from 3 $\alpha$-particles. Therefore, no experimental signatures for $\alpha$-condensation were observed.
\end{description}
\end{abstract}

\maketitle

\section{\label{sec:Introduction}Introduction}
The role of the $\alpha$-particle in the structure of light nuclei has been well explored for over 80 years. Observation of the binding energy per nucleon shows that $\alpha$-conjugate nuclei are much more strongly bound and can be described by a model of a tightly packed geometry of $\alpha$-particles \cite{HafstadTeller}. Since this initial investigation, the importance of the $\alpha$-particle in clustering, by virtue of its inert nature and own high binding energy, has been demonstrated experimentally and theoretically. The lightest (non-trivial) $\alpha$-conjugate system, $^{8}\mathrm{Be}$, has been shown to have a structure comprising of a dumbbell configuration of $\alpha$-particles \cite{Sp2R} with a large $\alpha$-$\alpha$ separation distance of 4.4 fm \cite{Be8_cluster,FreerRMP}. This dilute $\alpha$-particle behavior is then extended to the Hoyle state in carbon-12 ($^{12}\mathrm{C}(0_{2}^{+})$), a resonance above the 3$\alpha$ threshold of  astrophysical importance, which has been the subject of extensive studies \cite{Hoyle1, Hoyle2, Hoyle3, Hoyle4, Hoyle5,Raduta}. Recent experimental efforts have provided evidence for the structure of $^{12}\mathrm{C}$ as an equilateral triangle formation of 3 $\alpha$-particles \cite{Lambarri} but a $^{8}\mathrm{Be}{+}\alpha$ configuration is favored by some theoretical works \cite{Epelbaum,Kanada}. \par
This idea of describing the structure of $\alpha$-conjugate nuclei in terms of a dilute arrangement of $\alpha$-particles has been accompanied by a study of the nuclear equation of state in infinite systems. It was demonstrated \cite{infinite} that a nuclear liquid can undergo a phase change associated with the scattering length of the $\alpha$-particle when the density of a nuclear system $\rho$, drops below a specific value relative to the nuclear saturation density $\rho_{0}$, the current estimation is $\frac{\rho_{0}}{3}$ to $\frac{\rho_{0}}{5}$  \cite{infinite,Critical2,Critical3,Critical4}. Below this value, the system is no longer described by the fermionic interactions of the nucleons but the dominant degrees of freedom are those of the $\alpha$-particles. Their bosonic nature allows for a macroscopic occupation of the ground-state. This system is analogous to a Bose-Einstein Condensate which has been well studied in atomic systems \cite{BECLondon, BECExpt}. This work has been furthered in the nuclear realm by several experiments involving high-multiplicity particle decays \cite{Borderie,Morelli,Akimune} and experiments to observe an $\alpha$-gas via ``Coulomb explosion'' of $^{40}\mathrm{Ca}$ appear possible from a theoretical perspective \cite{Explosion}. Such an exotic state of matter is therefore postulated to describe the well-studied light N-$\alpha$ clustered states ($^{8}\mathrm{Be}$(g.s) and $^{12}\mathrm{C}(0_{2}^{+})$) and is known as an $\alpha$-condensate.  This presents a unique probe into understanding the nuclear force and generates two questions. Firstly, \textbf{how can such an unusual state of matter be detected}? Secondly, \textbf{can these N-$\boldsymbol{\alpha}$ states behave as an $\boldsymbol{\alpha}$-condensate}?
\section{\label{sec:signatures}Signatures of $\alpha$-condensation}
To understand the signatures associated with condensation, it is important to first comprehend the differences in the properties of an $\alpha$-condensate and a clustered state. When discussing $\alpha$-condensation, the system being described is one whereby the bosonic degrees of freedom are dominant \cite{FirstTHSR}. An additional degree of selectivity, necessary to describe the phase change of the system as mentioned above, is the need for a more dilute structure. In this case, the individual $\alpha$-particles have a large average separation where the underlying fermionic structure of the $\alpha$-particles is not resolved and the system behaves like an $\alpha$-gas. When choosing to describe the observed N-$\alpha$ states, a large separation is required and therefore $\alpha$-gas is chosen as an apt description for the resonances of interest.\par
\begin{figure}[h]
\includegraphics[width=0.5\textwidth]{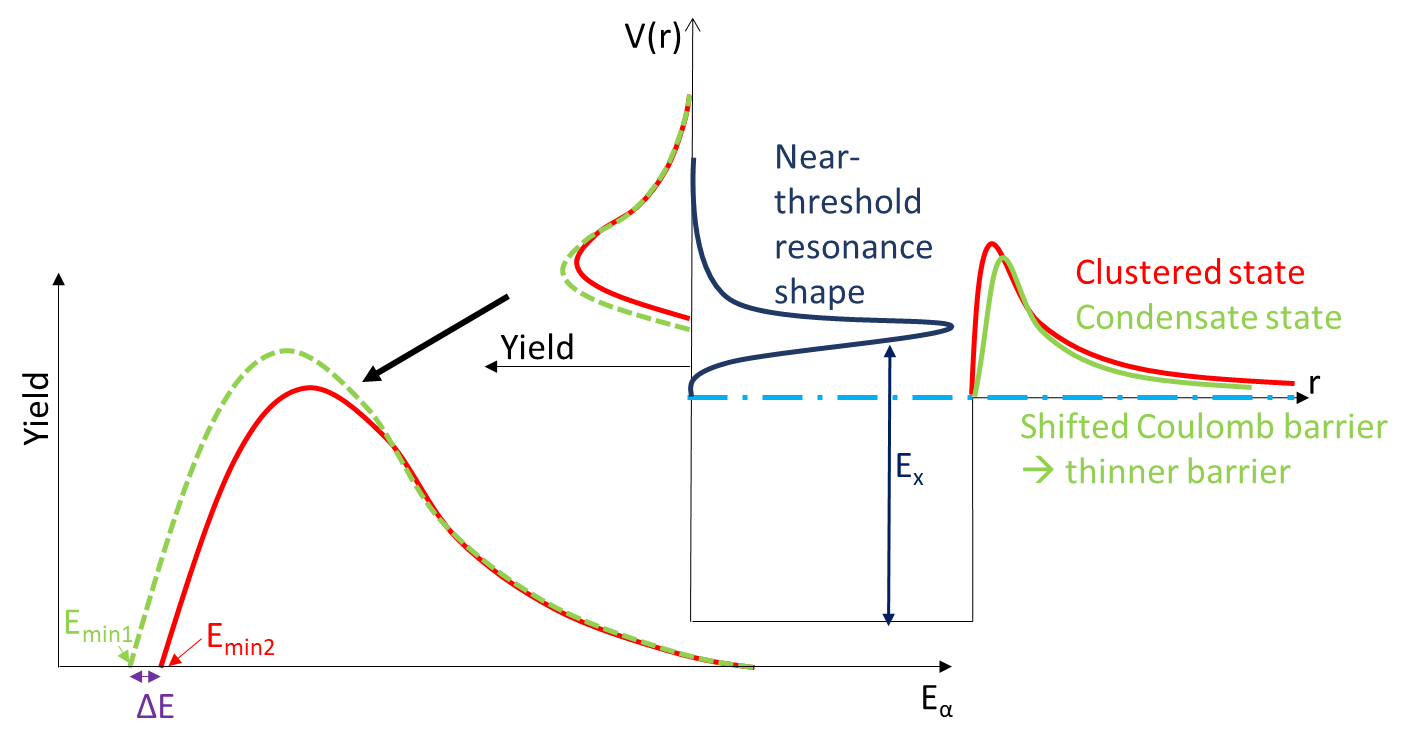}
\caption{\label{fig:reducedcoulomb}Schematic representation of the effect a condensate state has on modifying the Coulomb barrier for $\alpha$-gas emission and therefore the yield of the decay products at low kinetic energies compared to a clustered state. The near-threshold resonance has a width which is spread around the Coulomb barrier. The observed yield of such a decay through the Coulomb barrier is shown to the left for the two situations of a clustered and a condensate state. The shift in the energy of the decaying products between the two different types of state can be a signature of $\alpha$-condensation.}
\end{figure}
Such a change in the size of a nuclear system can have a profound effect on the decay properties. Previous investigations into this phenomena \cite{Tzany} have demonstrated $\alpha$-gas states will exhibit an increased preference for emission of $\alpha$-gas states ($\alpha$, $^{8}\mathrm{Be}$(g.s), and $^{12}\mathrm{C}(0_{2}^{+})$) as a consequence of the modification to the Coulomb barrier for these light clusters. This effect is demonstrated in Fig.~\ref{fig:reducedcoulomb} where a more dilute nuclear structure modifies the penetrability factor. Such an enhancement in the yield of these states is therefore indicative of a more dilute system congruent with an $\alpha$-gas structure. This can be compared to predictions from statistical decay models. The reduction of the barrier in proximity to the state also produces decay products with lower kinetic energies than those expected from a geometric cluster configuration \cite{Tzany}. Finally, the direct observation of states in the $\alpha$-conjugate nuclei in proximity to the N-$\alpha$ threshold and their decay modes can also be compared to theoretical predictions. Due to the nature of the $\alpha$-gas structure, the non $\alpha$-gas decay modes should be extremely inhibited and as such, their reduced widths should be extremely small due to the small overlap in the wavefunction.
\section{\label{sec:Exp}Experimental investigation}
To examine the signatures mentioned in Section~\ref{sec:signatures}, an experiment was performed at Laboratori Nazionali del Sud, Catania, Italy. An $^{16}\mathrm{O}$ beam at lab energies of 160, 280, and 400 MeV was provided by the K800 cyclotron with an average beam current of 100 pA. This beam was then impinged on a natural carbon target of thickness 58 $\mu \mathrm{g/cm}^2$ for the 280 and 400 MeV beam energies and 92 $\mu \mathrm{g/cm}^2$ for the 160 MeV beam energy. A gold target of thickness 174 $\mu \mathrm{g/cm}^2$ was also used for calibration purposes to provide elastic scattering events. The $^{12}\mathrm{C}(^{16}\mathrm{O}$,$^{28}\mathrm{Si}^{\star}$) reaction was then investigated. To measure the reaction products, two detector arrays were combined to provide a large solid angle coverage, high granularity system with good particle identification (PID) and energy range. These two arrays used were the CHIMERA \cite{CHIMERA} and FARCOS \cite{FARCOSpaper,FARCOSLuis} detector systems, which combined 1192 Si-CsI(Tl) telescopes with 4 DSSD-DSSD-CsI(Tl) telescopes placed at forward angles. The combined system provided an excellent combination of high angular resolution at small angles and a high detector coverage. An overall solid angle coverage of 28$\%$ (of the full 4$\pi$) was achieved.
\subsection{CHIMERA}
The CHIMERA (Charged Heavy Ion Mass and Energy Resolving Array) detector \cite{CHIMERA} is comprised of two sections, the first being the forward rings. This section consists of 18 rings (i.e. full annular azimuthal coverage) of detectors covering lab scattering angles from $1^{\circ} \rightarrow 30^{\circ}$ at varying distances with the smallest angles the furthest away. The second section is the backwards ball component which has rings covering from $30^{\circ} \rightarrow 176^{\circ}$ at a constant distance from the target of 40 cm. The rings consist of a series of individual telescopes which allow for high dynamic-energy range PID and momentum measurements. A single telescope is made from two detectors, the first is a 300 $\mu$m (nominal thickness) n-type planar Si detector (25 cm$^{2}$) behind which is a CsI(Tl) scintillator read out via a silicon photodiode.\par
\begin{figure}[ht]
\includegraphics[width=0.5\textwidth]{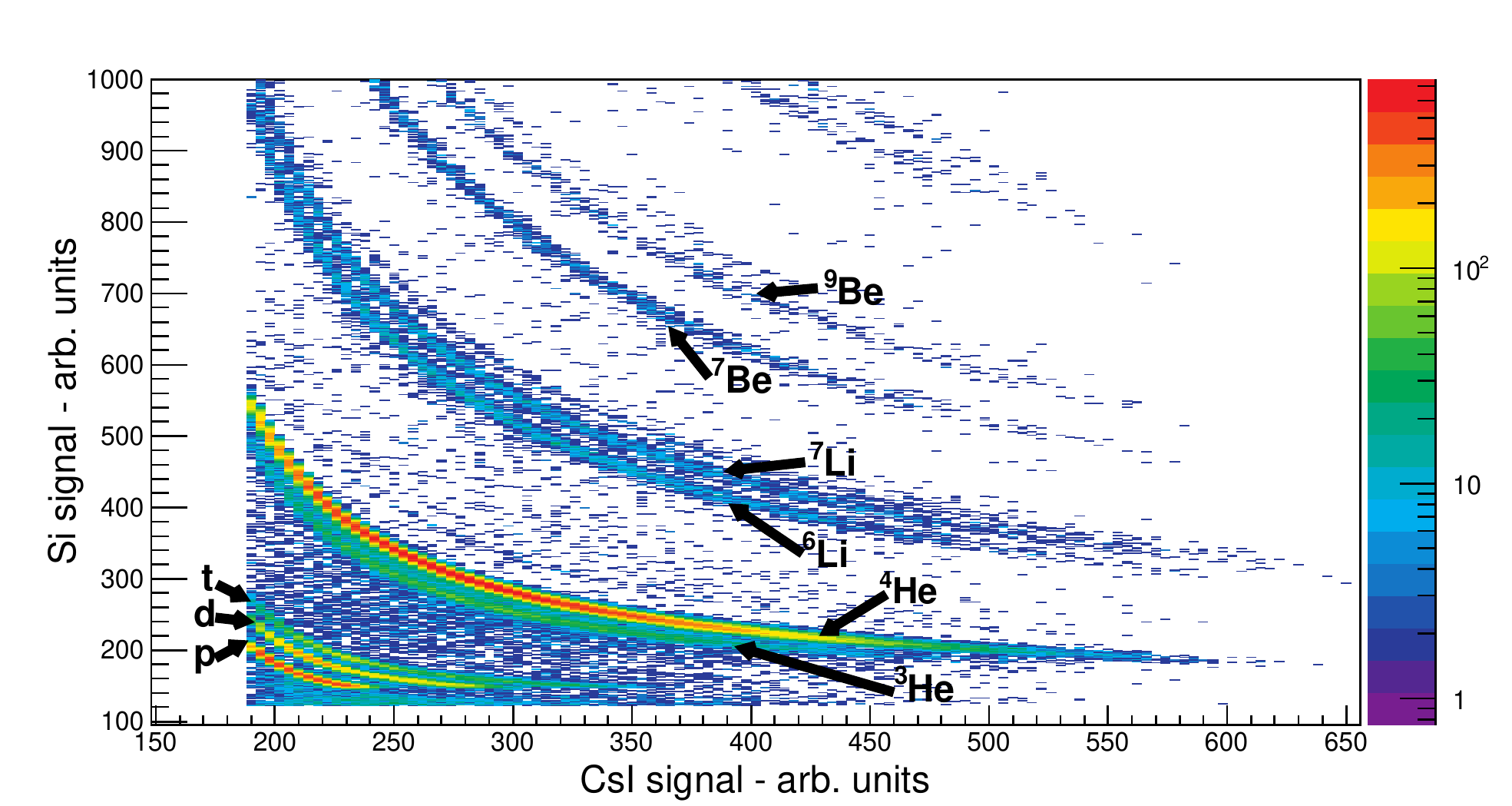}
\caption{Particle identification plot from CHIMERA showing the signal in the silicon stage against that in the cesium iodide stage for a single telescope. The different loci corresponding to nuclei of interest are labeled. The $^{4}\mathrm{He}$ can be separated from the $^{3}\mathrm{He}$ locus with 99.4$\%$ purity.\label{fig:deeCHIMERA}}
\end{figure}
\begin{figure}[h]
\includegraphics[width=0.5\textwidth]{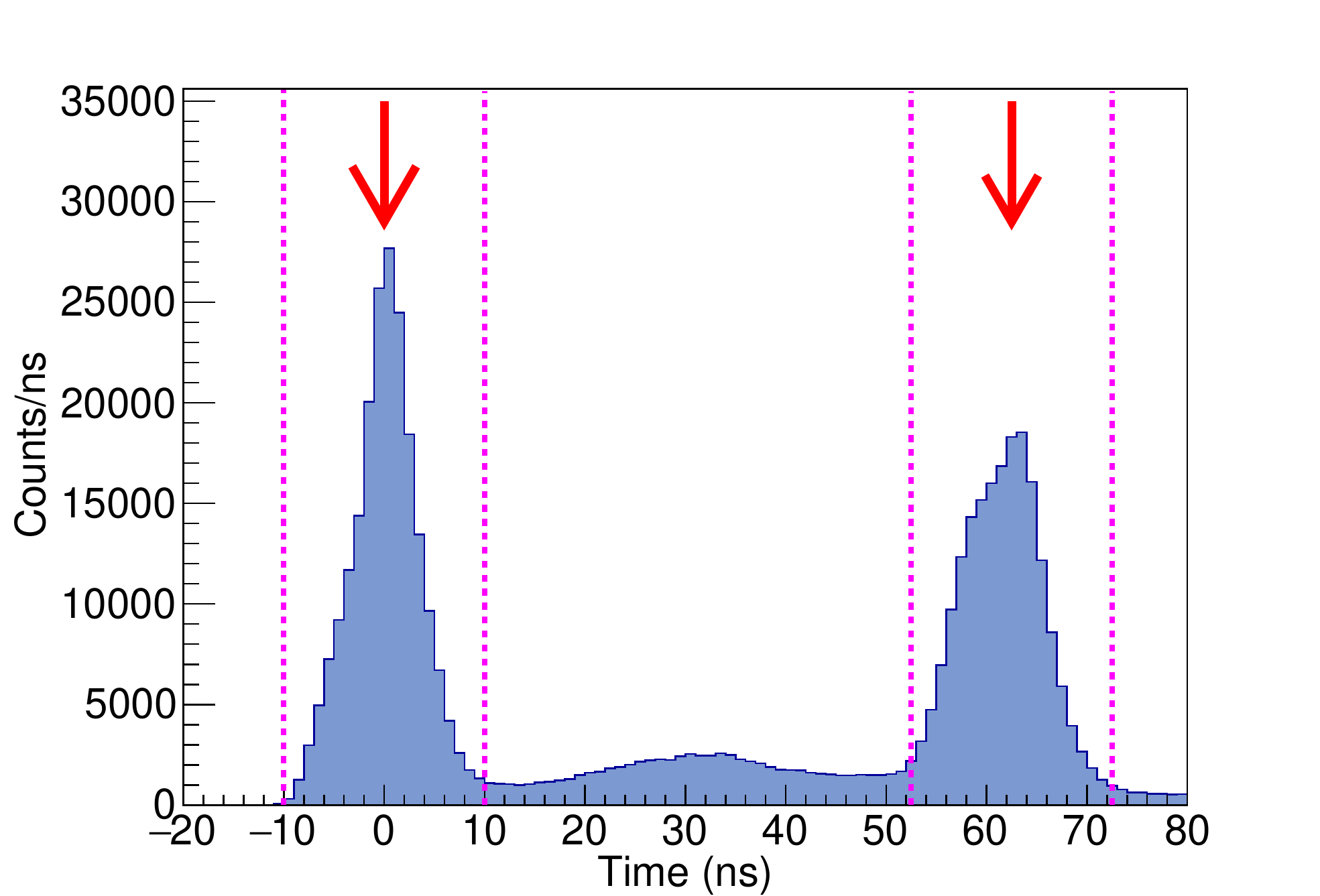}
\caption{Interaction time plot for an example CHIMERA ring 9I ($24^{\circ} < \theta < 27^{\circ}$). The two peaks (with the second peak arising from events where the original beam pulse was misidentified at $\Delta t$ = 62.5 ns) are labeled with a red arrow and correspond to correct identification of $\alpha$-particles whereas those outside the pair of dashed magenta lines are not classified as such and other masses are tested.\label{fig:TOF}}
\end{figure}

Both of these detectors in the telescope utilize a dual-gain setup which allows for a high energy-range while avoiding ADC (Analogue-to-Digital Converter) discretization effects.
The silicon detectors were calibrated using a combination of elastic scattering data from the beam impinging on a gold target, a triple-alpha calibration source ($^{239}\mathrm{Pu}$, $^{241}\mathrm{Am}$, and $^{244}\mathrm{Cm}$), a pulser calibration and time-of-flight (TOF) information. A timing signal was also taken from the silicon detectors in reference to the cyclotron radiofrequency signal. This energy and timing information allowed for particle identification across a wide energy range. For particles which have sufficient energy to penetrate the 300 $\mu$m nominal thickness of silicon, the energy deposited in the silicon ($\Delta$E) was plotted against that left in the scintillator ($E$). This then allowed for PID via the observation of different loci according to the expected energy loss $\delta E$, through a small distance $\delta x$, given by the Bethe-Bloch formula $\frac{\delta E}{\delta x} \propto mZ^2/E$ \cite{CHIMERAPID}. This is the $\Delta$E-E PID method. These loci are shown in Fig.~\ref{fig:deeCHIMERA}. For those particles where the energy was insufficient to be identified this way ($E \lesssim 24$ MeV for an $\alpha$-particle traversing 300 $\mu$m of silicon), the timing signal $t$, was used in addition to the energy deposited in the silicon $E$, to identify the mass of the particle using the time-of-flight particle identification - TOF PID. To do this, the interaction time $t_0$, was calculated assuming the hit corresponded to an $\alpha$-particle ($m=4$) by:
\begin{equation}
t_{0}=d\sqrt{\frac{m}{2E}}-t,
\end{equation}
for a detector of distance $d$ from the target. The result from this TOF PID method is shown in Fig.~\ref{fig:TOF} to identify $\alpha$-particles. Those events which lay outside of the gate shown undergo the same procedure assuming $m=$ 6, 7, 12, and 16. The timing and energy resolution were insufficient to differentiate other mass nuclei. These can however still be identified using the $\Delta$E-E method.\par
To determine the energy of the particle where $\Delta$E-E PID was possible, this value was calculated via the signal in the silicon stage due to the non-linear response of the CsI(Tl) crystal as well as its mass and charge signal dependence \cite{CsIresponse}. This gives an energy resolution of approximately 400 keV for a 30 MeV $\alpha$-particle increasingly approximately linearly to 1000 keV for a 60 MeV $\alpha$-particle.

\subsection{FARCOS}
The FARCOS (Femtoscope ARray for COrrelation and Spectroscopy) detector \cite{FARCOSpaper} is a high granularity, high resolution detector system mainly designed for heavy-ion collisions and studies of light-clustering. The FARCOS system comprises of three detector stages, the first being a nominal 300 $\mu$m DSSD (double-sided silicon strip detector) with 32 strips in both the vertical and horizontal direction. The second stage is a thicker nominal 1500 $\mu$m DSSD of the same strip number. Behind this is a set of four 6 cm-thick CsI(Tl) crystals with a Si PIN diode read-out arranged in a 2 $\times$ 2 grid. The overall system is 64 $\times$ 64 mm in extent giving an angular resolution of around $0.1^{\circ}$ for a single strip at a distance of $1$ m. Timing information was not available from these detectors therefore particle identification in the FARCOS detector relied solely on the $\Delta$E-E method from the energy deposited in the 300 and 1500 $\mu$m silicon detectors as discussed above for CHIMERA.\par
The presence of the FARCOS detectors provided excellent angular and energy resolution ($\delta \theta \approx 0.1^{\circ}$, $\delta E/E < 1\%$ at 5.5 MeV) at low angles ($2^{\circ} \rightarrow 8^{\circ}$) to study the different reaction mechanisms, which will be covered in Section~\ref{sec:FARCOSDEE}.


\section{\label{sec:theorymodels}Theoretical models of expected reaction mechanisms}
In order to demonstrate the observed yields in the experiment performed were congruent with an $\alpha$-gas state, theoretical input was needed to provide a benchmark. To understand the reaction, the sequential decay mechanism, multi-particle break-up and the 7-$\alpha$ decay through predicted $\alpha$-gas states were modeled to compare to the experimental data.
\subsection{\label{sec:EHF}Sequential decay}
To model the sequential decay mechanism, an Extended Hauser-Feshbach (EHF) code was used \cite{EHFcode} to model the p, n, $\alpha$, and $\gamma$ decay from the compound system following fission. The compound nucleus is populated with a range of angular momenta values as given by the input channel's masses and energy. In this code, the fission yields are calculated from the density of states at the scission point then sequential emission is modeled until all decay paths are exhausted.\par
A calculation was performed for each of the three beam energies and provided a number of observables. The first of these, the final yields for different nuclei are summarized in Table~\ref{tab:EHFyields} where the dominance of the $\alpha$-conjugate nuclei can clearly be seen for the lowest beam energy of 160 MeV. At higher beam energies, the non $\alpha$-conjugate nuclei become more readily populated as their larger Q-values are less inhibited by virtue of the increased excitation energy in the compound nucleus.\par
\begin{figure}[h]
\includegraphics[width=0.5\textwidth]{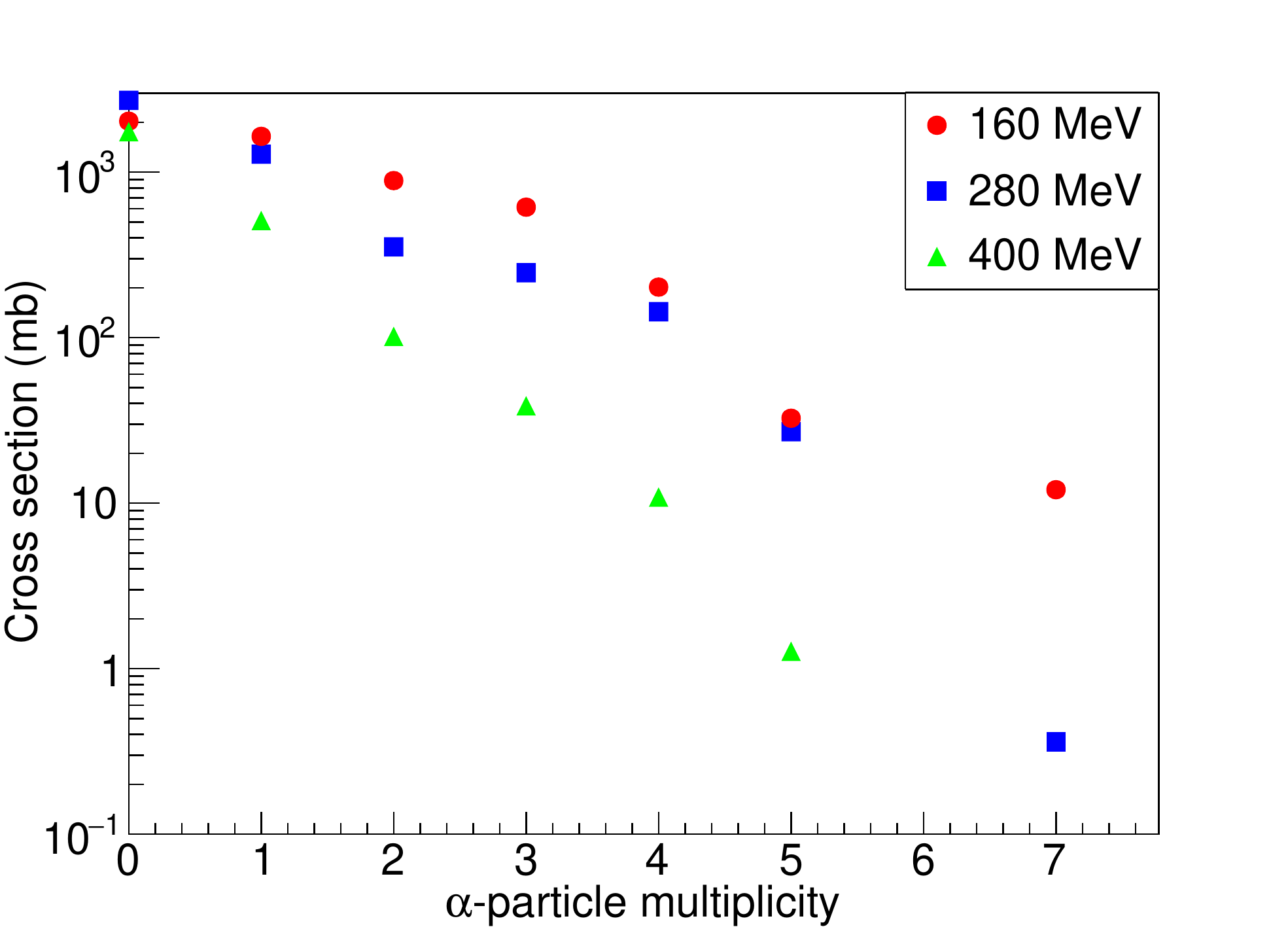}
\caption{Predicted $\alpha$-particle multiplicity from the EHF calculation for the three beam energies. The higher beam energies can be seen to suppressed alpha decay cross sections.\label{fig:EHFmult}}
\end{figure}

An important prediction to extract from the calculation is the expected $\alpha$-particle multiplicity. As discussed in Section~\ref{sec:signatures}, from an $\alpha$-gas structure, the modification of the Coulomb barrier would create a higher multiplicity of $\alpha$-particles in a decay than normally expected. The EHF code models sequential $\alpha$-decay by summing the total cross section for $n$ sequential $\alpha$-decays for different nuclei following fission, hence the expected multiplicity from the sequential decay model can be extracted. Decay chains where $n$ sequential decays were interspersed with a proton or neutron decay were also examined but shown to have a negligible contribution to the total cross section. These values can be seen in Fig.~\ref{fig:EHFmult} where the behavior for different beam energies can be observed to vary significantly. As the excitation energy in the compound nucleus increases because of a higher beam energy, the cross section for high multiplicity decays can be seen to dramatically decrease. Examining the dynamics of the sequential decay demonstrated this reduction is due to the decreased preference for $\alpha$-decay as the effect of the Q-value for each decay stage becomes less important. As such, the system becomes to more equally prefer proton, neutron and $\alpha$-decay as the beam energy increases. Consequently, the largest multiplicity one might expect at a beam energy of 400 MeV is 5. In comparison, at the lowest beam energy of 160 MeV, the smaller amount of energy in the system means that when a nucleus still has sufficient energy to decay, the preferred path is that of $\alpha$-decay. This results in a still sizeable cross section for multiplicity 7 decays. It should also be noted that the cross section for multiplicity 6 events is $<0.1$ mb. This is a consequence of the need to break apart an $\alpha$-particle to have a final state which is comprised of 6 $\alpha$-particles and d+d or $^{3}\mathrm{He}$+n etc.\par
\begin{figure}[h]
\includegraphics[width=0.425\textwidth]{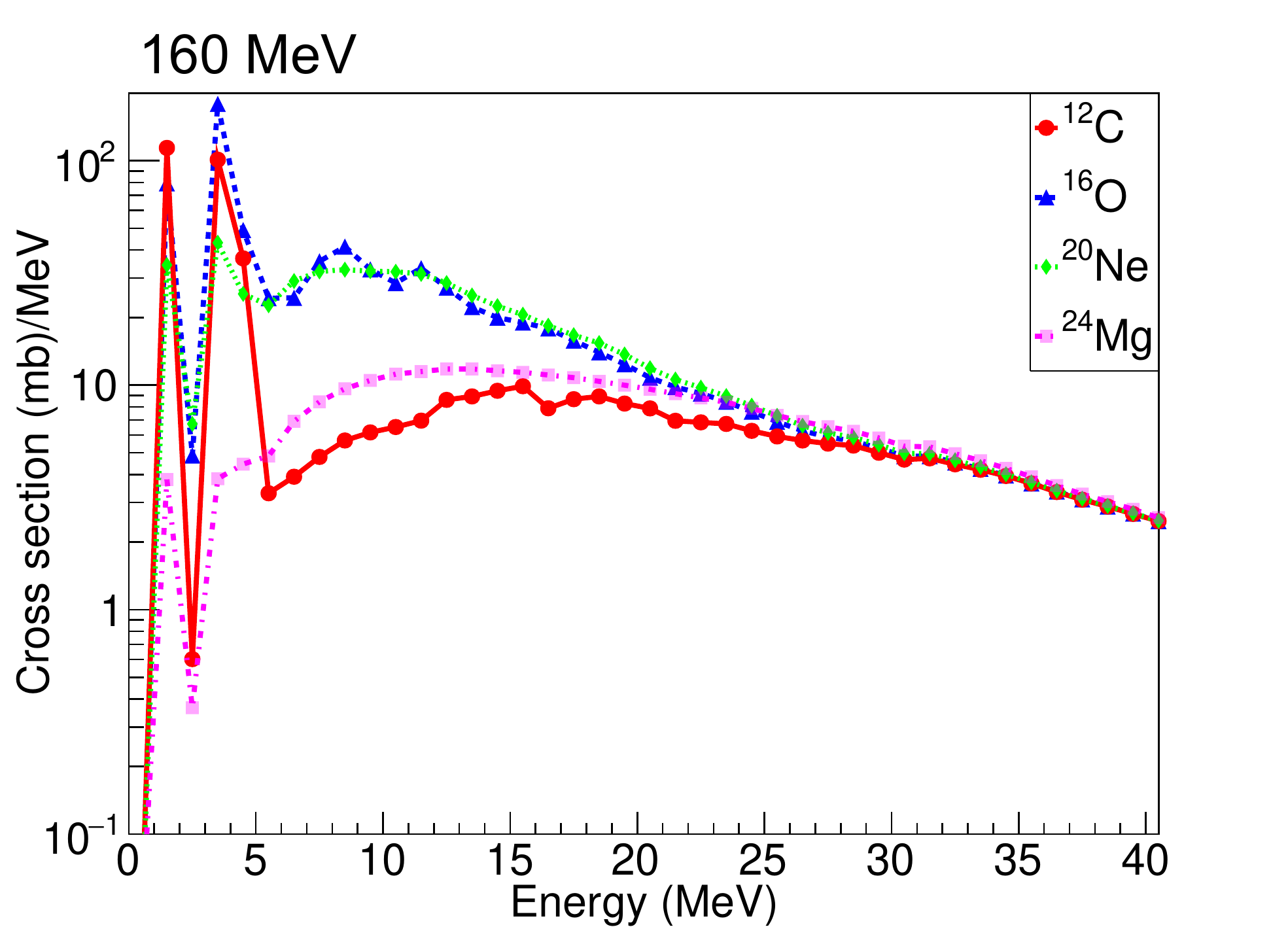}\\
\includegraphics[width=0.425\textwidth]{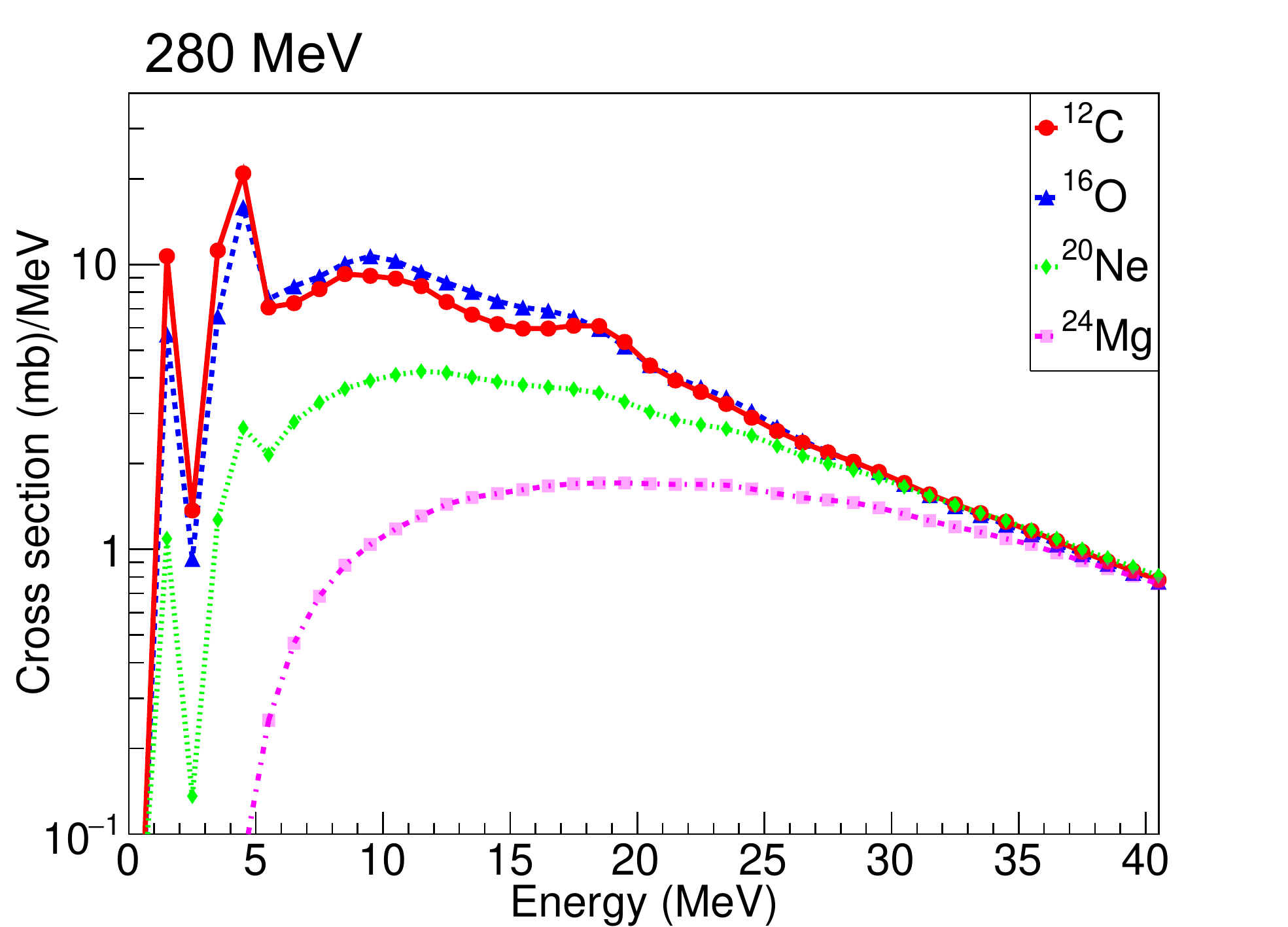}\\
\includegraphics[width=0.425\textwidth]{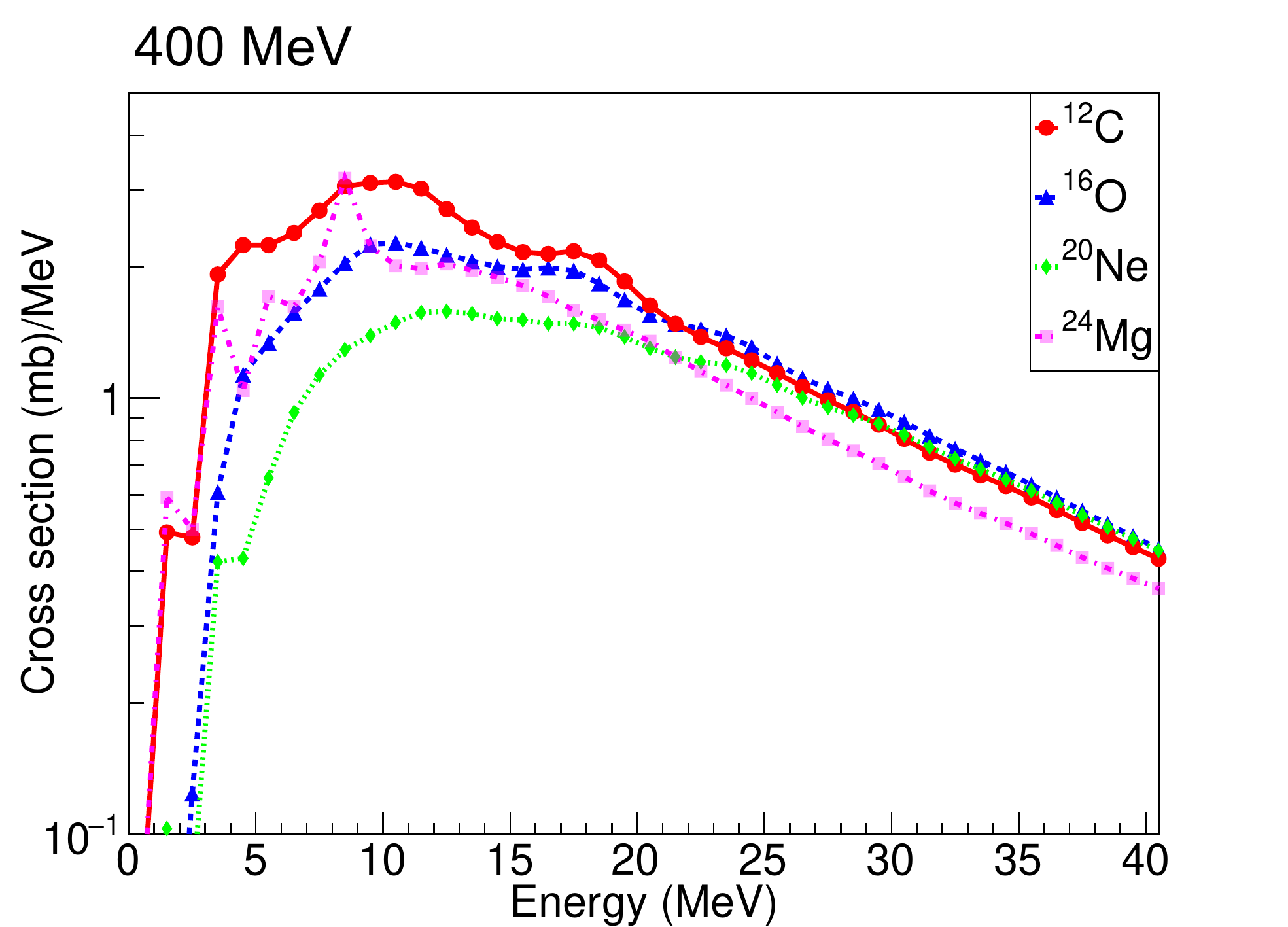}\\
\caption{Distribution of $\alpha$-particle energies in the center-of-mass system from the EHF for a beam energy of a) 160 MeV, b) 280 MeV and c) 400 MeV. The yields from the population of different initial nuclei are shown in different colors with $^{12}\mathrm{C}$ (red solid with circles), $^{16}\mathrm{O}$ (blue short-dashed with triangles), $^{20}\mathrm{Ne}$ (green dotted with diamonds) and $^{24}\mathrm{Mg}$ (magenta dot-dashed with squares) with the lines guiding the eye between the data points.\label{fig:EHFalphaKE}}
\end{figure}

\begin{table}[h]
\setlength\extrarowheight{1.5pt}
\caption{Final yields as predicted from the EHF calculation for the three different beam energies (160, 280, and 400 MeV). The $\alpha$-conjugate nuclei masses are shown in \textbf{bold}.\label{tab:EHFyields}}
\begin{ruledtabular}
\begin{tabular}{cccc}
\textbf{Decay} & \multicolumn{3}{c}{\textbf{Final cross section (mb)}} \\
\textbf{product mass} & \textbf{160 MeV} & \textbf{280 MeV} & \textbf{400 MeV}\\
\hline
1 & 1.92 $\times 10^{3}$ & 3.69 $\times 10^{3}$ & 4.78 $\times 10^{3}$\\
2 & 2.58 $\times 10^{2}$ & 6.56 $\times 10^{2}$ & 1.10 $\times 10^{3}$\\
3 & 1.67 $\times 10^{2}$ & 3.21 $\times 10^{2}$ & 5.30 $\times 10^{2}$\\
\textbf{4} & $\mathbf{3.79 \times 10^{3}}$ & $\mathbf{3.27 \times 10^{3}}$ & $\mathbf{2.74 \times 10^{3}}$\\
5 & 8.31 $\times 10^{1}$ & 1.52 $\times 10^{2}$ & 2.41 $\times 10^{2}$ \\
6 & 7.82 $\times 10^{1}$ & 2.07 $\times 10^{2}$ & 2.71 $\times 10^{2}$ \\
7 & 3.60 $\times 10^{1}$ & 1.23 $\times 10^{2}$ & 1.27 $\times 10^{2}$ \\
\textbf{8} & $\mathbf{3.69 \times 10^{1}}$ & $\mathbf{1.19 \times 10^{2}}$ & $\mathbf{7.36 \times 10^{1}}$ \\
9 & 7.64 $\times 10^{0}$ & 4.43 $\times 10^{1}$ & 6.29 $\times 10^{1}$ \\
10 & 7.81 $\times 10^{1}$ & 2.34 $\times 10^{2}$ & 1.31 $\times 10^{2}$ \\
11 & 2.47 $\times 10^{1}$ & 1.06 $\times 10^{2}$ & 4.76 $\times 10^{1}$ \\
\textbf{12} & $\mathbf{7.91 \times 10^{2}}$ & $\mathbf{7.01 \times 10^{2}}$ & $\mathbf{2.14 \times 10^{2}}$ \\
13 & 4.46 $\times 10^{1}$ & 1.93 $\times 10^{2}$ & 1.23 $\times 10^{3}$ \\
14 & 2.12 $\times 10^{2}$ & 3.35 $\times 10^{2}$ & 1.51 $\times 10^{2}$ \\
15 & 1.51 $\times 10^{2}$ & 1.32 $\times 10^{2}$ & 5.14 $\times 10^{1}$ \\
\textbf{16} & $\mathbf{8.75 \times 10^{2}}$ & $\mathbf{1.32 \times 10^{2}}$ & $\mathbf{5.27 \times 10^{1}}$ \\
17 & 6.87 $\times 10^{1}$ & 2.20 $\times 10^{2}$ & 1.00 $\times 10^{1}$ \\
18 & 1.57 $\times 10^{2}$ & 3.71 $\times 10^{1}$ & 9.05 $\times 10^{0}$ \\
19 & 2.59 $\times 10^{1}$ & 5.02 $\times 10^{1}$ & 1.42 $\times 10^{0}$ \\
\textbf{20} & $\mathbf{1.02 \times 10^{2}}$ & $\mathbf{9.91 \times 10^{0}}$ & $\mathbf{8.52 \times 10^{-2}}$ \\
\end{tabular}
\end{ruledtabular}
\end{table}
Finally, the distribution of $\alpha$-particle energies was also extracted according to the initial population of different nuclei. This can be observed in Fig.~\ref{fig:EHFalphaKE}. The figure shows that when populating $^{24}\mathrm{Mg}$, the kinetic energy distribution is a smooth continuum whereas when lighter $\alpha$-conjugate nuclei are populated (e.g. $^{16}\mathrm{O}$), peaks can be seen which correspond to the population of discrete resonances which are included automatically in the code. This demonstrates that discrete levels in nuclei heavier than $^{16}\mathrm{O}$ are not expected to be strongly populated above the strength of the continuum. The behavior with increasing beam energy also demonstrates that the continuum becomes increasingly dominant as the magnitudes of the discrete peaks become negligible in comparison to the smooth statistical decay contribution. In the experimental data, the larger beam energies should therefore correspondingly be almost entirely dominated by uncorrelated $\alpha$-particles.

\subsection{\label{sec:FBU}Multi-particle break-up}
To model the dissociation of the compound nucleus into a large number of particles, the Fermi break-up model \cite{FBU} was also employed. This calculation is driven by modeling the decay according to Fermi's golden rule \cite{FermiBU}:
\begin{equation}
R_{if}=\frac{2\pi}{\hbar}|A_{if}|^2 \omega_{f},
\end{equation}
with $|A_{if}|^2$ corresponding to the transition matrix element between the initial state $i$ and the final state $f$. The important factor in the Fermi break-up calculations originates from the $\omega_{f}$ term which describes the density of states. The phase space available for a decay of the system of mass number $A$ into $n$ fragments is given by \cite{FBU2}:
\begin{eqnarray}
\omega_{f} = && \prod_{b=1}^{n} (2s_b + 1) \prod_{j=1}^{k} \frac{1}{n_{j}!} \left (\frac{r^3_{0} A}{6 \pi^2 \hbar^3} \right)^{n-1} \nonumber \\
&& \left ( \frac{1}{\sum_{b=1}^{n} m_b} \prod_{b=1}^{n} m_{b} \right )^{\frac{3}{2}} \frac{(2 \pi)^{3(n-1)/2}}{\Gamma(3(n-1)/2)} E_{\mathrm{kin}}^{\frac{3n}{2}-\frac{5}{2}} \label{eq:FBU},
\end{eqnarray}
with $s_b$ being the spin of particle $b$ and $n_j$ is the number of particles of type $j$. There is a very strong dependence therefore on $n$, particularly in the term $ \left (\frac{r^3_{0} A}{6 \pi^2 \hbar^3} \right)^{n-1}$ which describes the partition of the phase space of the decay. Additionally, the energy available in the decay also provides a very large density of states for high $n$ decays. For a break-up of a system into 7-$\alpha$, one generates an 8$^{\mathrm{th}}$ order power dependence on the kinematically available energy. This is usually offset by a correspondingly smaller energy available for higher multiplicity decays. In our case, for a decay to provide 1 $\alpha$-particle, 10.0 MeV is needed to break-up the system whereas for 7-$\alpha$ particles one requires 38.5 MeV. Therefore, for a state with an excitation energy of $\sim$ 40 MeV, the small available energy is therefore prohibitive to the 7-$\alpha$ break-up and preferential towards a smaller $n$ break-up.\par

To investigate the dynamics of this break-up mode, the phase space for the dissociation into different $\alpha$-conjugate nuclei was investigated, both in their ground state and for the predicted $\alpha$-gas states above the N-$\alpha$ threshold. The $\alpha$-gas states in $^{8}\mathrm{Be}$ and $^{12}\mathrm{C}$ have been given energies of 0.0 and 7.65 respectively. These reaction modes are summarized in Table~\ref{tab:FBUmodes}. Due to the experiment being a high excitation energy compound nucleus reaction, there is a large amount of energy present which means the system prefers a higher $n$ body break-up. To demonstrate this, the sum of the binary fission modes to the ground state (i.e. p+$^{27}\mathrm{Al}$, $^{11}\mathrm{C}$+$^{17}\mathrm{O}$ etc.) was compared to the sum of those modes in Table~\ref{tab:FBUmodes}. These binary fission modes were demonstrated to only constitute 9.6$\%$, 0.2$\%$ and 0.02$\%$ for the 160, 280, and 400 MeV beam energies respectively. For the lowest beam energy, the binary fission can be observed to be important however for the higher beam energies, the majority of the phase space is associated with $n > 2$ decays, a fact which will be important later.

\begin{table}[h]
\setlength\extrarowheight{1.5pt}
\caption{Fermi break-up modes of $\alpha$-conjugate and the associated number of break-up particles. The final column details the number of ${\alpha}$-particles such a decay would yield in the final state (due to unbound resonances).\label{tab:FBUmodes}}
\begin{ruledtabular}
\begin{tabular}{ccc}
\textbf{Decay path} & \multicolumn{1}{p{2cm}}{\centering \textbf{Break-up} \\ \textbf{particles}} & $\pmb{\alpha}$-\textbf{particles} \\
\hline
$^{24}$Mg + $\alpha$  & 2 & 1 \\ 

$^{20}$Ne + $^{8}\mathrm{Be}$  & 2 & 2 \\ 

$^{16}\mathrm{O}$+$^{12}\mathrm{C}$  & 2 & 0 \\ 

$^{16}\mathrm{O}$+$^{12}\mathrm{C}(0_{2}^{+})$  & 2 & 3 \\ 

$^{20}$Ne+2$\alpha$ & 3 & 2 \\ 

$^{16}\mathrm{O}$+$^{8}\mathrm{Be}$+$\alpha$  & 3 & 3 \\ 

$^{12}\mathrm{C}$+$^{12}\mathrm{C}$+$\alpha$  & 3 & 1 \\ 

$^{12}\mathrm{C}(0_{2}^{+})$+$^{12}\mathrm{C}(0_{2}^{+})$+$\alpha$  & 3 & 7 \\ 

$^{12}\mathrm{C}$+$^{8}\mathrm{Be}$+$^{8}\mathrm{Be}$  & 3 & 4 \\ 

$^{12}\mathrm{C}(0_{2}^{+})$+$^{8}\mathrm{Be}$+$^{8}\mathrm{Be}$ & 3 & 7 \\ 

$^{12}\mathrm{C}$+$^{8}\mathrm{Be}$+2$\alpha$& 4 & 4 \\ 

$^{12}\mathrm{C}(0_{2}^{+})$+$^{8}\mathrm{Be}$+2$\alpha$ & 4 & 7 \\ 

$^{8}\mathrm{Be}$+$^{8}\mathrm{Be}$+$^{8}\mathrm{Be}$+$\alpha$ & 4 & 7 \\

$^{16}\mathrm{O}$+3$\alpha$ & 4 & 3\\

$^{12}\mathrm{C}$+4$\alpha$ & 5 & 4\\

$^{12}\mathrm{C}(0_{2}^{+})$+4$\alpha$ & 5 & 7\\

$^{8}\mathrm{Be}$+$^{8}\mathrm{Be}$+3$\alpha$ & 5 & 7\\

$^{8}\mathrm{Be}$+5$\alpha$& 6 & 7\\

7$\alpha$ & 7 & 7\\
\end{tabular}
\end{ruledtabular}
\end{table}
\begin{figure*}[ht]
\includegraphics[width=0.9\textwidth]{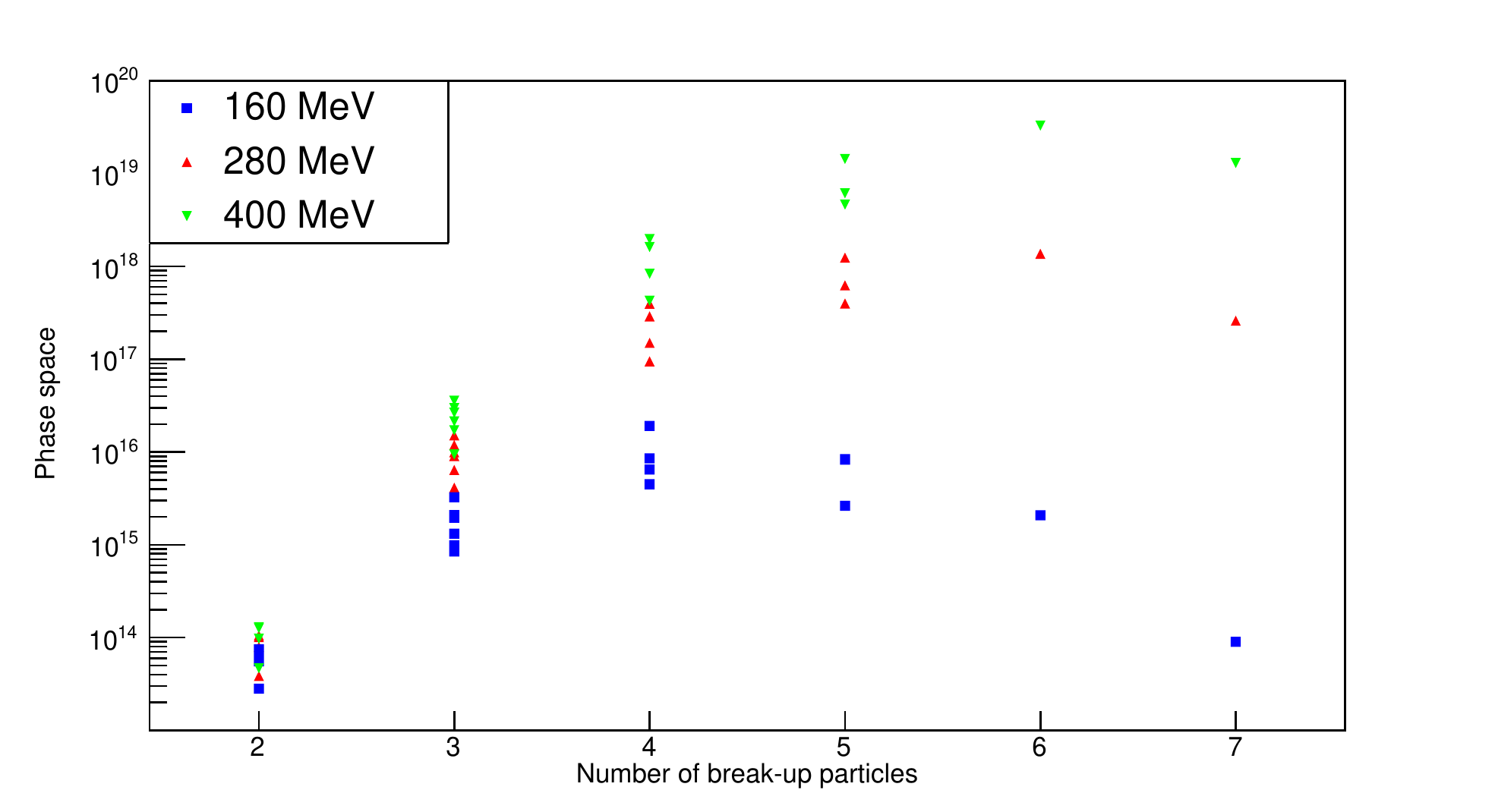}
\caption{Phase space for the different break-up partitions in Table~\ref{tab:FBUmodes} showing how modifying the beam energy (160 MeV - green inverted triangles, 280 MeV - red triangles, 400 MeV - blue squares) changes the preference for a different number of break-up particles.\label{fig:FBUparticles}}
\end{figure*}
This high preference for larger $n$ particle decay can also be seen in Fig.~\ref{fig:FBUparticles} where the phase space against the number of break-up particles is shown for the three beam energies. The largest phase space is associated with $n$~=~6 for the 280 and 400 MeV beam energy but a smaller $n$~=~4 is preferred at the lowest beam energy of 160 MeV.\par
An underlying assumption of the $\alpha$-condensate wave function is that the dissociation of an $\alpha$-gas into a set of sub-units which are also $\alpha$-gas states has the same transition matrix \cite{Tzany}. This can therefore be applied to the Fermi golden rule to compare the expected strength of the different break-up paths with a final state comprising of 7 $\alpha$-particles, i.e. those where the sub-units are expected to be $\alpha$-gas states. This prediction can then be compared to the experimental data as a probe of the $\alpha$-condensate. In order to accurately compare the predictions from the Fermi break-up model to those from the experimental data, detector effects and the efficiency of these different channels needed to be well understood to make a meaningful comparison on the efficiency-corrected data. For this purpose, a Monte Carlo simulation was performed to bridge the connection between the predictions of the EHF and the Fermi break-up models to the data.

\subsection{\label{sec:MC}Monte Carlo simulations}
As discussed above, the detector effects (resolution, detector coverage and particle pileup) play an important part in the comparison between experimental and theoretical data. To model the particle break-up, the \textsc{RESOLUTION8.1} code \cite{RES8a,RES8b} was used which models sequential break-up reactions. The output from the simulation allows for ``truth-level'' (i.e. the true values that were generated from the Monte Carlo) momentum vectors for the different reaction products to be generated. These truth level vectors were then passed through a wrapper which simulates the CHIMERA and FARCOS position and energy resolution which was then fed through the analysis code. As well as verifying the rigor of the analysis code, which requires care in this type of high-multiplicity environment, this also allows for an investigation into the reaction mechanisms seen in the data and the efficiency correction needed to compare to the 7-$\alpha$ decay channels calculated in Section~\ref{sec:FBU} using the Fermi break-up model. It is worth noting that the Monte Carlo simulation only models sequential decay, whereas the Fermi break-up is a single stage decay. The kinematics of the two systems are comparable as both produce an isotropic spray of particles in the center of mass system. There are 11 different decay paths through the $\alpha$-gas states which were used to compare to the data which are tabulated in Table~\ref{tab:MCpaths}.

\begin{table}
\begin{center}
\caption{\label{tab:MCpaths} Different decay paths through $\alpha$-gas type states until the system has decayed to $^8$Be or an $\alpha$-particle. Only the decay of the excited states marked with a star are shown at each subsequent stage.}
\begin{ruledtabular}
\setlength\extrarowheight{1.5pt}
\begin{tabular}{cc}

\textbf{Label} & \textbf{Path} \\ 
\hline
1&$^{24}\mathrm{Mg}^\star {+} \alpha {\rightarrow}^{20}\mathrm{Ne}^\star {+} \alpha {\rightarrow}$$ ^{16}\mathrm{O}^\star {+} \alpha {\rightarrow}$$ ^{12}\mathrm{C}^\star {+} \alpha {\rightarrow}$$ ^8\mathrm{Be} {+} \alpha$ \\

2&$^{24}\mathrm{Mg}^\star {+} \alpha {\rightarrow}$$ ^{20}\mathrm{Ne}^\star {+} \alpha {\rightarrow}$$ ^{16}\mathrm{O}^\star {+} \alpha {\rightarrow} $$^{8}\mathrm{Be} {+}$$^{8}\mathrm{Be}$ \\

3&$^{24}\mathrm{Mg}^\star {+} \alpha {\rightarrow}$$ ^{20}\mathrm{Ne}^\star {+} \alpha {\rightarrow}$$ ^{12}\mathrm{C}^\star {+} ^8\mathrm{Be}{\rightarrow}$$ ^8\mathrm{Be} {+} \alpha$ \\

4&$^{24}\mathrm{Mg}^\star {+}\alpha {\rightarrow}$$ ^{16}\mathrm{O}^\star {+} $$^8\mathrm{Be} {\rightarrow} $$^{12}$C$^\star {+} \alpha {\rightarrow}$$ ^8\mathrm{Be} {+} \alpha$ \\

5&$^{24}\mathrm{Mg}^\star {+} \alpha {\rightarrow}$$ ^{16}\mathrm{O}^\star {+} $$^8\mathrm{Be} {\rightarrow} $$ ^8\mathrm{Be} {+} ^8\mathrm{Be}$ \\

6&$^{24}\mathrm{Mg}^\star {+} \alpha {\rightarrow}$$ ^{12}\mathrm{C}^\star {+} $$^{12}\mathrm{C}^\star {\rightarrow}  $$ ^8\mathrm{Be} {+} ^8\mathrm{Be}$ \\

7&$^{20}\mathrm{Ne}^\star {+}^8\mathrm{Be} {\rightarrow} $$^{16}\mathrm{O}^\star {+} \alpha {\rightarrow}$$ ^{12}\mathrm{C}^\star {+} \alpha {\rightarrow} ^8\mathrm{Be} {+} \alpha$\\

8&$^{20}\mathrm{Ne}^\star {+} ^8\mathrm{Be} {\rightarrow} $$^{16}\mathrm{O}^\star {+} \alpha {\rightarrow}  $$ ^8\mathrm{Be} {+} $$^8\mathrm{Be}$ \\

9&$^{20}\mathrm{Ne}^\star {+} ^8\mathrm{Be} {\rightarrow} $$^{12}\mathrm{C}^\star {+} ^8\mathrm{Be} {\rightarrow} ^8\mathrm{Be} {+} \alpha$ \\

10&$^{16}\mathrm{O}^\star {+} ^{12}$C$^\star {\rightarrow} (^{12}\mathrm{C}^\star {+} \alpha) {+} (^8\mathrm{Be} {+} \alpha) {\rightarrow}  $$ ^8\mathrm{Be} {+} \alpha$ \\

11&$^{16}\mathrm{O}^\star{+} ^{12}\mathrm{C}^\star {\rightarrow} (^{8}\mathrm{Be} {+} $$^{8}\mathrm{Be}) {+} (^8\mathrm{Be} {+}\alpha$) \\

\end{tabular}
\end{ruledtabular}
\end{center}
\end{table}

\section{\label{sec:exp}Experiment}
By virtue of the high incident energy involved in this experiment, the probability of compound nucleus formation is reduced and other direct reaction mechanisms may become prevalent. To separate and quantify the effect of these reactions, the FARCOS detector, due to its low angle and high angular resolution, was used to investigate whether particles hitting this detector originate from a direct reaction or sequential decay.
\begin{figure}[h]
\includegraphics[width=0.5\textwidth]{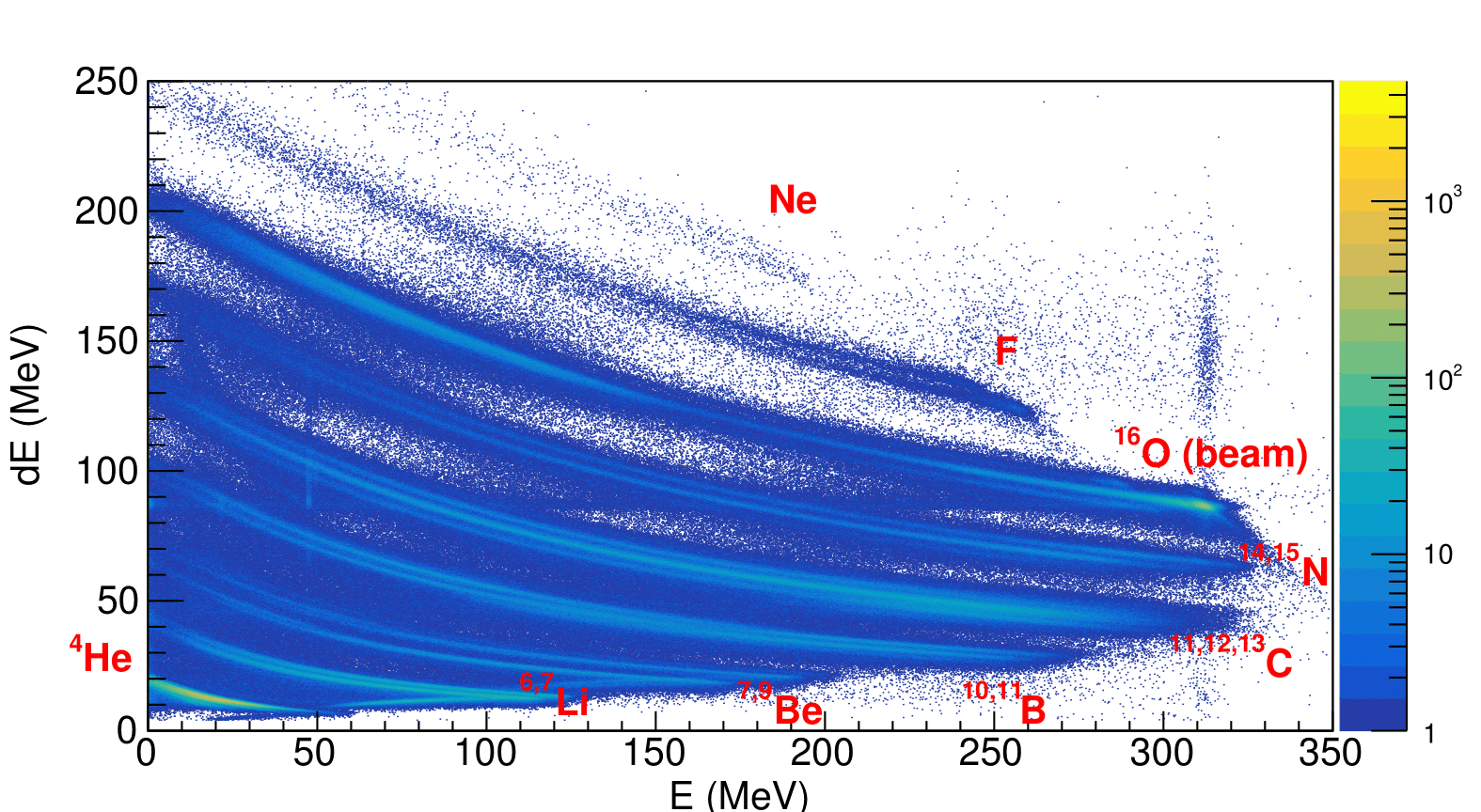}
\caption{Yields in the FARCOS detector for $E_b$ = 400 MeV with the different loci labeled.\label{fig:FARCOS400}}
\end{figure}
\begin{figure}[h]
\includegraphics[width=0.5\textwidth]{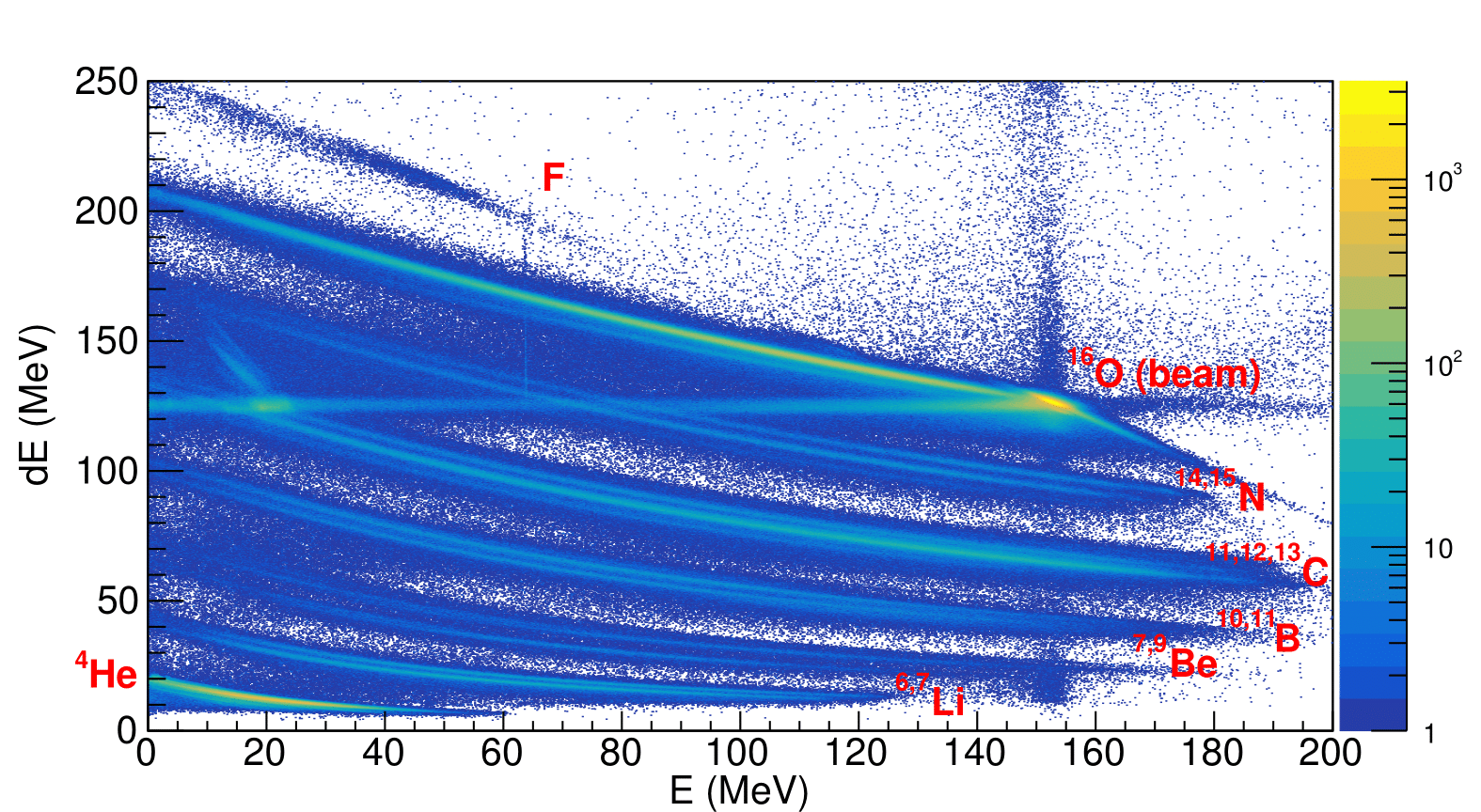}
\caption{Yields in the FARCOS detector for $E_b$ = 280 MeV with the different loci labeled.\label{fig:FARCOS280}}
\end{figure}
\begin{figure}[h]
\includegraphics[width=0.5\textwidth]{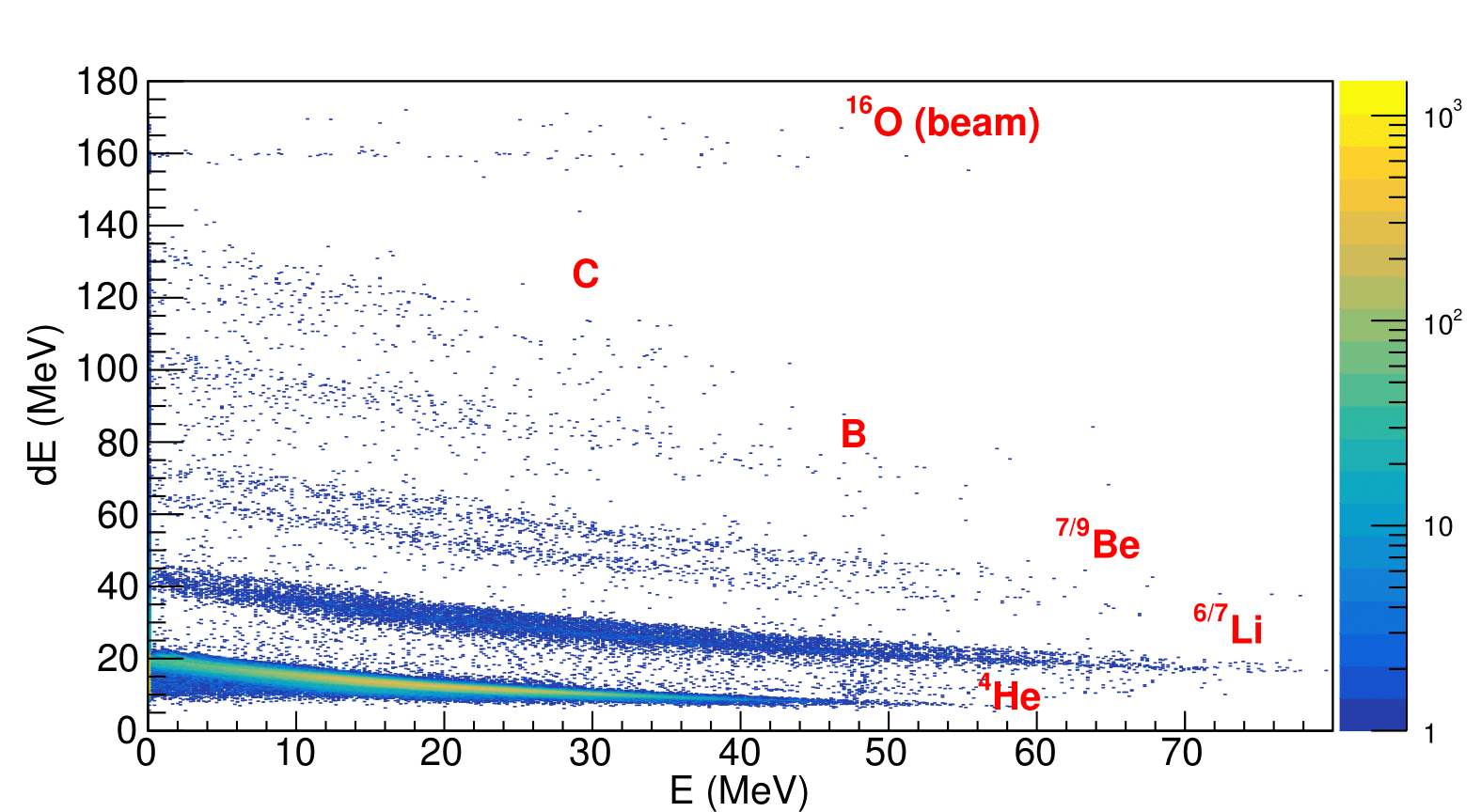}
\caption{Yields in the FARCOS detector for $E_b$ = 160 MeV with the different loci labeled.\label{fig:FARCOS160}}
\end{figure}
\subsection{\label{sec:FARCOSDEE}FARCOS particle yields}

The three beam energies show different particle yields (as seen in Figs.~\ref{fig:FARCOS400},~\ref{fig:FARCOS280} and~\ref{fig:FARCOS160}), mainly as a result of the differing excitation energy of the compound nucleus due to the change in beam energy. Two-body decays from a higher initial excitation energy are much more likely to produce daughter nuclei with an excitation energy above their particle decay thresholds therefore undergoing secondary decay processes. Additionally, as the energy of the system increases, the phase space for 3 body decays becomes significantly larger. The FARCOS detectors give a (polar) coverage between $2^{\circ}$ and $8^{\circ}$ which allows for the low angle behavior to be investigated where direct reactions are expected to be more dominant. To examine this, one can firstly examine the data taken with a beam energy ($E_b$) of 400 MeV (Fig.~\ref{fig:FARCOS400}).\par
These data show a large swathe of particles produced including nuclei heavier than the beam demonstrating these results are not solely due to beam break-up. In particular, the $^{4}\mathrm{He}$ locus is observed to be extremely strong with large contributions also arising from $^{12}\mathrm{C}$ and $^{16}\mathrm{O}$. By reconstructing the missing momentum, the possibility that these particles originate from a binary fission mode to the ground-state nuclei can be investigated by examining the Q-value for the reaction. As suggested by the Fermi break-up calculations, this mode is not seen in the data apart from a small component for $^{12}\mathrm{C}(^{16}\mathrm{O}{,}^{10}\mathrm{B}){^{18}}\mathrm{F}$ (with around $1/3$ of the measured $^{18}\mathrm{F}$ reproducing the expected Q-value of -17.7 MeV) and the elastic scattering channel which can clearly be observed where the $^{16}\mathrm{O}$ scattered beam is labeled in the figure.\par

The data from the 280 MeV beam energy (Fig.~\ref{fig:FARCOS280}) also show a very similar behavior with a strong yield in the $\alpha$-conjugate nuclei. The beam energy here is insufficient for the reaction products heavier than the beam to be detected by $\Delta$E-E (they are stopped in the silicon 300 $\mu$m stage). As with the 400 MeV beam energy data, there is also a small component for the $^{12}\mathrm{C}(^{16}\mathrm{O}{,}^{10}\mathrm{B}){^{18}}\mathrm{F}$ break-up measured from the $^{10}\mathrm{B}$ although this corresponds to a much smaller fraction of $< 5 \times 10^{-3} \%$ of the total measured $^{10}\mathrm{B}$. Other direct break-up components are also absent in these data.\par

In the 160 MeV beam energy data (Fig.~\ref{fig:FARCOS160}), the degree to which the $^{4}\mathrm{He}$ yield dominates is largely increased. Despite the lower beam energy meaning that isotopes heavier than carbon cannot be identified using this method, there is a very small contribution from $^{12}\mathrm{C}$ ($< 0.03 \%$) with the yield from $^{6/7}\mathrm{Li}$ becoming more dominant to constitute $\sim 3\%$ of the total counts.\par
These results mirror very strongly the predictions from the EHF calculations where the $\alpha$-decay mode was seen to be more important at the lowest beam energies where the proton and neutron decay modes are ``frozen-out'' because of the larger Q-value required. The Fermi break-up model also predicted a reasonable contribution for direct 2-body decay at a beam energy of 160 MeV but, as with the higher beam energies, no such decay channel is observed.\par
In addition to the decay of the compound nucleus, the possibility of these particles originating from the beam break-up was investigated by calculating the Q-value. These Q-value plots also demonstrate that the detected particles do not originate from the break-up of the beam. This is therefore indicative that the observed reaction mechanism is related to a higher multiplicity reaction such as that of sequential decay or direct multi-particle break-up.

\section{\label{sec:CHIMERAyield}CHIMERA particle yields}

\begin{figure}[h]
\includegraphics[width=0.5\textwidth]{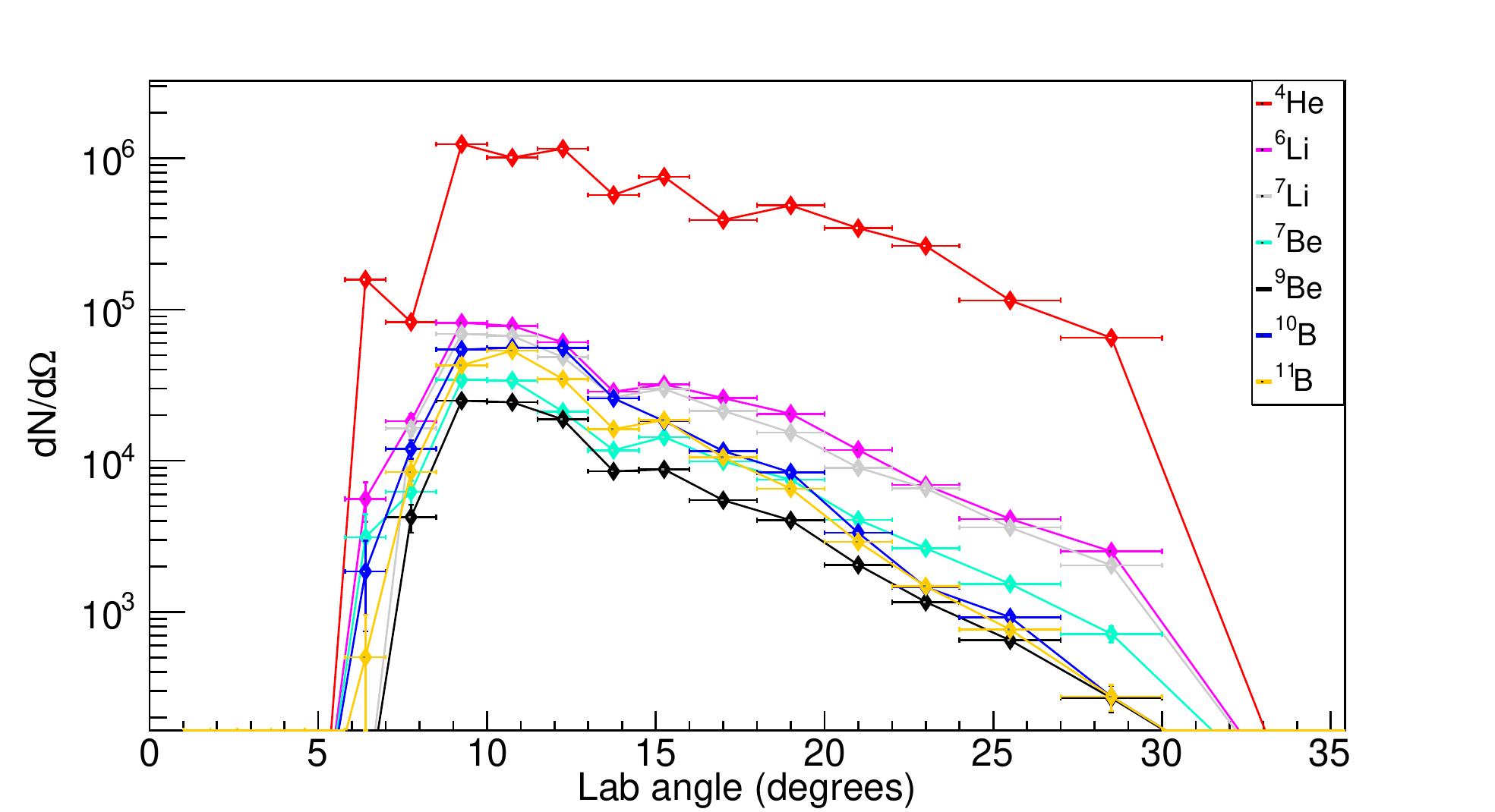}
\caption{Yields per solid angle in the CHIMERA detector for $E_b$ = 400 MeV as a function of angle for the non $\alpha$-conjugate nuclei in comparison to $^{4}\mathrm{He}$.\label{fig:CHIMERA_400}}
\end{figure}

\begin{figure}[h]
\includegraphics[width=0.5\textwidth]{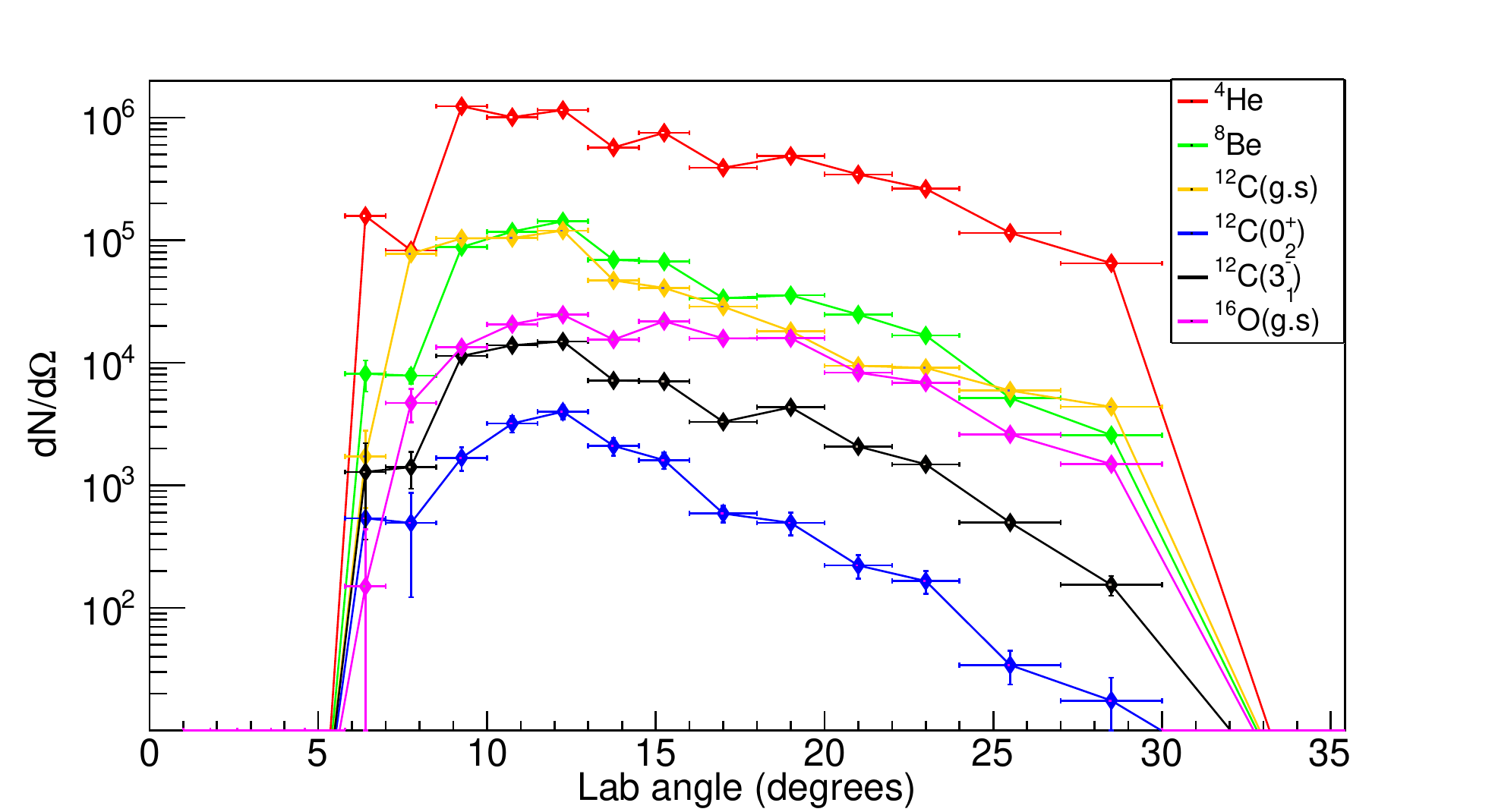}
\caption{Yields per solid angle in the CHIMERA detector for $E_b$ = 400 MeV as a function of angle for the $\alpha$-conjugate nuclei.\label{fig:CHIMERA_400ac}}
\end{figure}

As well as investigating the low angle particle yields, the high coverage of CHIMERA can be utilized to provide information about how the yield varies as a function of angle. This can give a much better signature for the dominance of compound nucleus formation which, as mentioned previously, will give an isotropic spray of particles in the center-of-mass. The CHIMERA detector covers from $6^{\circ}$ and the forward rings were used to study how the relative yields change up to $30^{\circ}$. The results for the three beam energies show the same features. Therefore, only the results for the 400 MeV data are shown in Figs.~\ref{fig:CHIMERA_400} and \ref{fig:CHIMERA_400ac} where missing telescopes in the CHIMERA rings were corrected for by considering the azimuthal solid angle coverage. For the lower beam energies there is a threshold effect observed. This occurs where the nuclei are created with a lower average energy in the compound nucleus decay and therefore may have insufficient energy to pass through the 300 $\mu$m silicon stage to be detected using $\Delta$E-E in CHIMERA. No correction is made for this effect.
When examining the angular dependence of the non $\alpha$-conjugate nuclei in Fig.~\ref{fig:CHIMERA_400}, it can clearly be seen that all the nuclei follow the same behavior with a small decrease in yield with increasing angle. The amount of $^{4}\mathrm{He}$ measured far exceeds that of lithium, beryllium, and boron which all have roughly the same yields. This similar situation is also mirrored when examining the contributions from the $\alpha$-conjugate nuclei and their associated excited states are discussed in Section~\ref{sec:reconstruction}. The $^{8}\mathrm{Be}$(g.s) is observed to be measured at a level consistent with the $^{12}\mathrm{C}$ when reconstructed from 2 $\alpha$-particles. When the efficiency of measuring both these $\alpha$-particles (compared to needing to detect only one particle for $^{12}\mathrm{C}$ rather than reconstructing $^{8}\mathrm{Be}$ from 2 $\alpha$-particles), the strength of $^{8}\mathrm{Be}$(g.s) exceeds that of  $^{12}\mathrm{C}$(g.s). Reconstruction of events where 3 $\alpha$-particles are detected, the strength of contributions from the Hoyle state ($^{12}\mathrm{C}(0_{2}^{+})$) and a broad contribution at 9.6 MeV ($^{12}\mathrm{C}(3_{1}^{-})$) also show the same fall-off with angle as the ground state nuclei detected showing the reaction mechanism for their population is common.\par
\begin{figure}[h]
\includegraphics[width=0.5\textwidth]{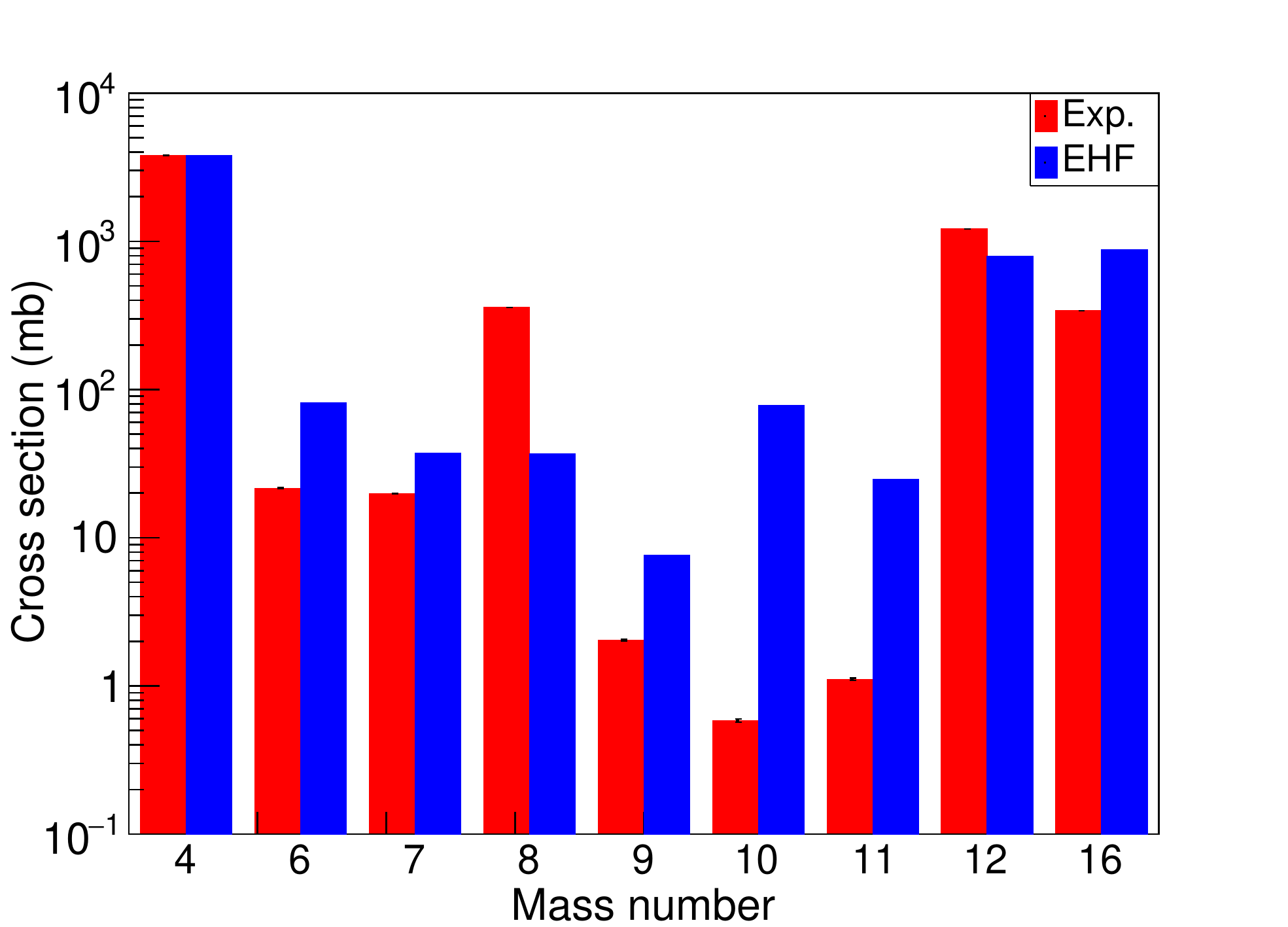}
\caption{Yields in the CHIMERA detector for $E_b$ = 160 MeV in comparison to the EHF calculation predictions. Errors appear in the center of the bars but may be too small to be visible.\label{fig:CHIMERA_160_yield}}
\end{figure}
\begin{figure}[h]
\includegraphics[width=0.5\textwidth]{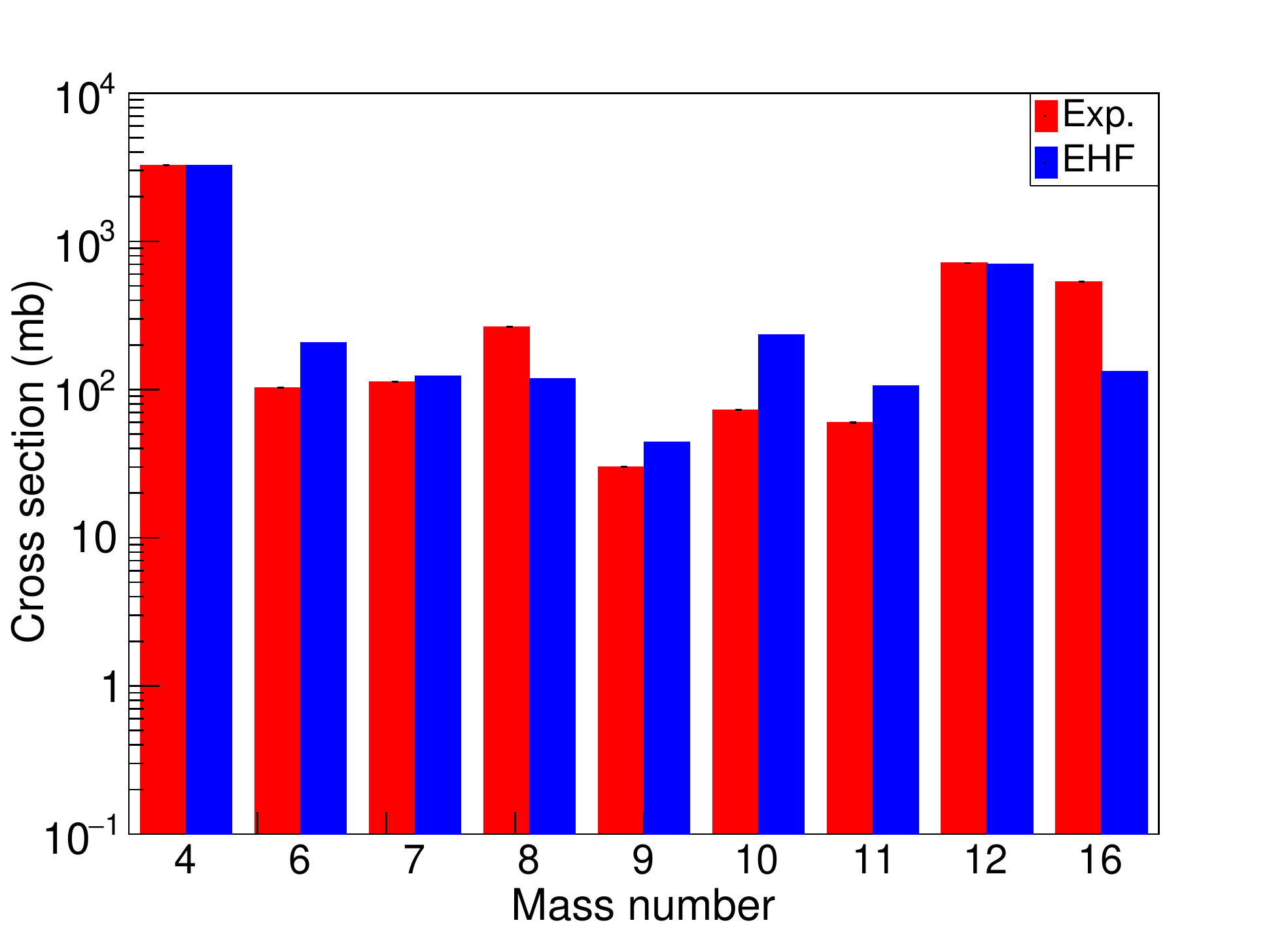}
\caption{Yields in the CHIMERA detector for $E_b$ = 280 MeV in comparison to the EHF calculation predictions. Errors appear in the center of the bars but may be too small to be visible.\label{fig:CHIMERA_280_yield}}
\end{figure}
\begin{figure}[h]
\includegraphics[width=0.5\textwidth]{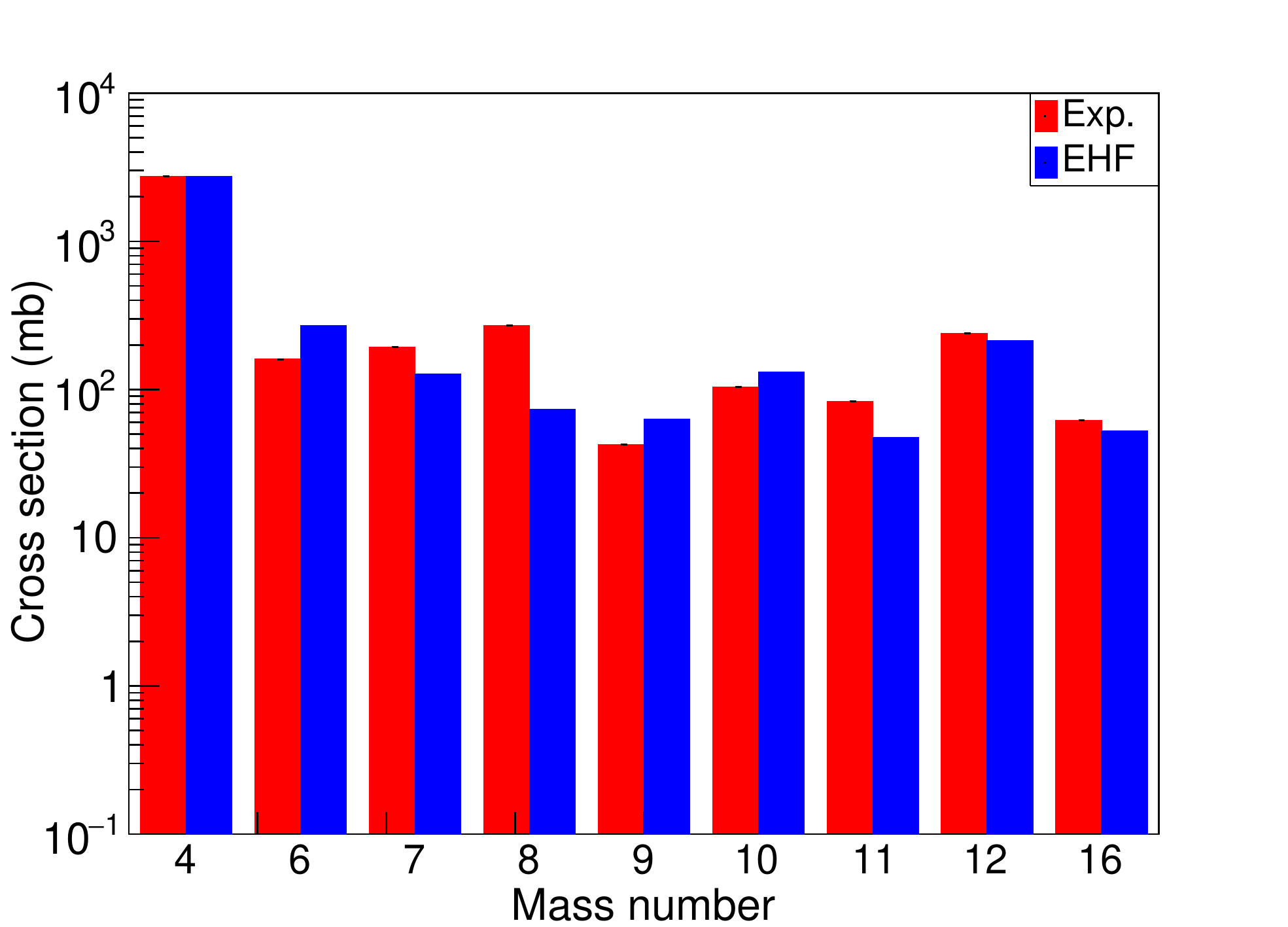}
\caption{Yields in the CHIMERA detector for $E_b$ = 400 MeV in comparison to the EHF calculation predictions. Errors appear in the center of the bars but may be too small to be visible.\label{fig:CHIMERA_400_yield}}

\end{figure}

The total strength of these yields in CHIMERA were then compared to the predictions from the EHF calculations by normalizing the experimental yield to the mass-4 EHF cross section. The results for this are seen in Figs.~\ref{fig:CHIMERA_160_yield}, \ref{fig:CHIMERA_280_yield} and \ref{fig:CHIMERA_400_yield}. The influence of the threshold effect can be observed here for the 160 MeV beam energy data (Fig.~\ref{fig:CHIMERA_160_yield}) where the experimental yield for the heavier mass nuclei is lower than expected for mass 9, 10, and 11 where the TOF PID was not possible. Aside from this discrepancy due to experimental effects, there is a good agreement between the predictions from the EHF and the experimental results for the three beam energies although the yield for $^{8}\mathrm{Be}$ can clearly be seen to exceed the predictions for all beam energies. When taking into account the reduced efficiency for this path by virtue of needing to reconstruct $^{8}\mathrm{Be}$ from 2 $\alpha$-particles, this increase over the sequential decay predictions is even larger showing an enhancement inconsistent with the EHF calculations. This effect has been observed previously in other experiments \cite{TzanyEPJ,Tzanythesis} in the $^{13}\mathrm{C}(^{18}\mathrm{O},{^{8}\mathrm{Be}})$ and $^{24}\mathrm{Mg}(^{28}\mathrm{Si},^{12}\mathrm{C}^{\star})$ reactions employing charged-particle and gamma-decay spectroscopy in tandem.  In the latter of these, the $\gamma$-rays associated with $^{36}\mathrm{Ar}$ were observed corresponding to the removal of 4 $\alpha$-particles. This 4 $\alpha$-particle removal was not predicted to be observed in the EHF model for the $^{52}\mathrm{Fe}$ compound nucleus decay at that energy.
\par
Overall, the majority of the reaction products can be seen to originate via the compound nucleus formation. For the $^{12}\mathrm{C}  + {^{16}\mathrm{O}}$ exit channel, there will undoubtedly be a component of inelastic scattering and other direct reaction mechanisms which cannot be easily disentangled from the compound nucleus formation. The angular coverage of CHIMERA + FARCOS to large lab angles means that the contribution of this mechanism to the data can be assigned as negligible to the first-order. However, care must be taken to ensure that the origins of any states observed in the $\alpha$-conjugate nuclei (i.e. did they arise from direct/compound nucleus formation) is taken into consideration to validate the assumptions made in Sec.~\ref{sec:theorymodels}. This will be examined later in Sec.~\ref{sec:Origins} where an attempt to understand the reaction mechanism used to generate the events of interest is made.
\section{\label{sec:mult}Multiplicity of alpha-particles}
\begin{figure}[h]
\includegraphics[width=0.5\textwidth]{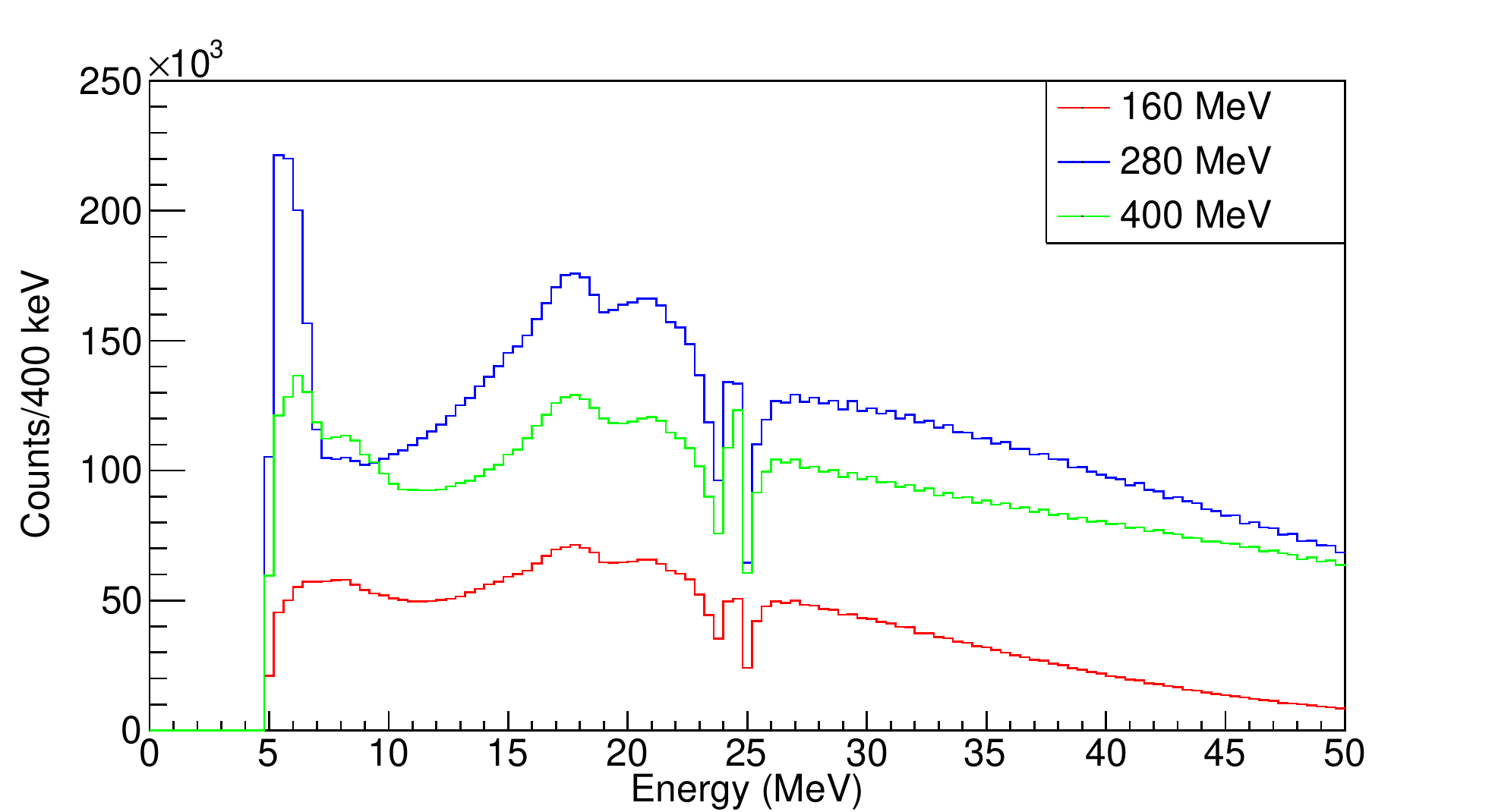}
\caption{Kinetic energy distribution of the $\alpha$-particles measured in CHIMERA and FARCOS. The discontinuities around 24 MeV are due to a gap in the TOF PID and the $\Delta$E-E PID for varying thicknesses of the silicon ($\Delta$E) detector.\label{fig:alpha_KE}}
\end{figure}

It was shown in Section~\ref{sec:signatures} that an enhancement in the production of $\alpha$-particles in comparison with that expected from theory is a potential signature of an $\alpha$-gas due to the modification to the Coulomb barrier. The calculations from the EHF code showed a large decrease is expected as the beam energy (and therefore excitation energy of the compound nucleus) is increased. Conversely, the Fermi break-up model showed that as the energy in the system increases, large $n$-body break-ups are preferred as a consequence of the increased phase space. To firstly demonstrate the compound nucleus description of the reaction was correct, the energy distribution of the $\alpha$-particles measured in the experiment was examined, the results of which are visible in Fig.~\ref{fig:alpha_KE}. It can clearly be seen that a smooth distribution of energies is seen, a key signature that the mechanism seen is indeed due to compound nucleus formation. Additionally, the discrete peaks predicted from the EHF code in the center of mass are not visible. This is either: conversion to the lab frame smoothing these discrete peaks or the statistical emission of $\alpha$-particles having a larger effect than that predicted within the sequential decay framework.
There are a few peaks which are visible on top of the broad continuum, most noticeably around 5 MeV, just above the threshold. By separately analyzing these events, it was demonstrated that (despite a larger impurity in the TOF PID around this energy due to the inclusion of particles from previous beam pulses) this increase corresponds to genuine events where particles are produced at backwards angles in the center of mass.\par
\begin{figure}[h]
\includegraphics[width=0.5\textwidth]{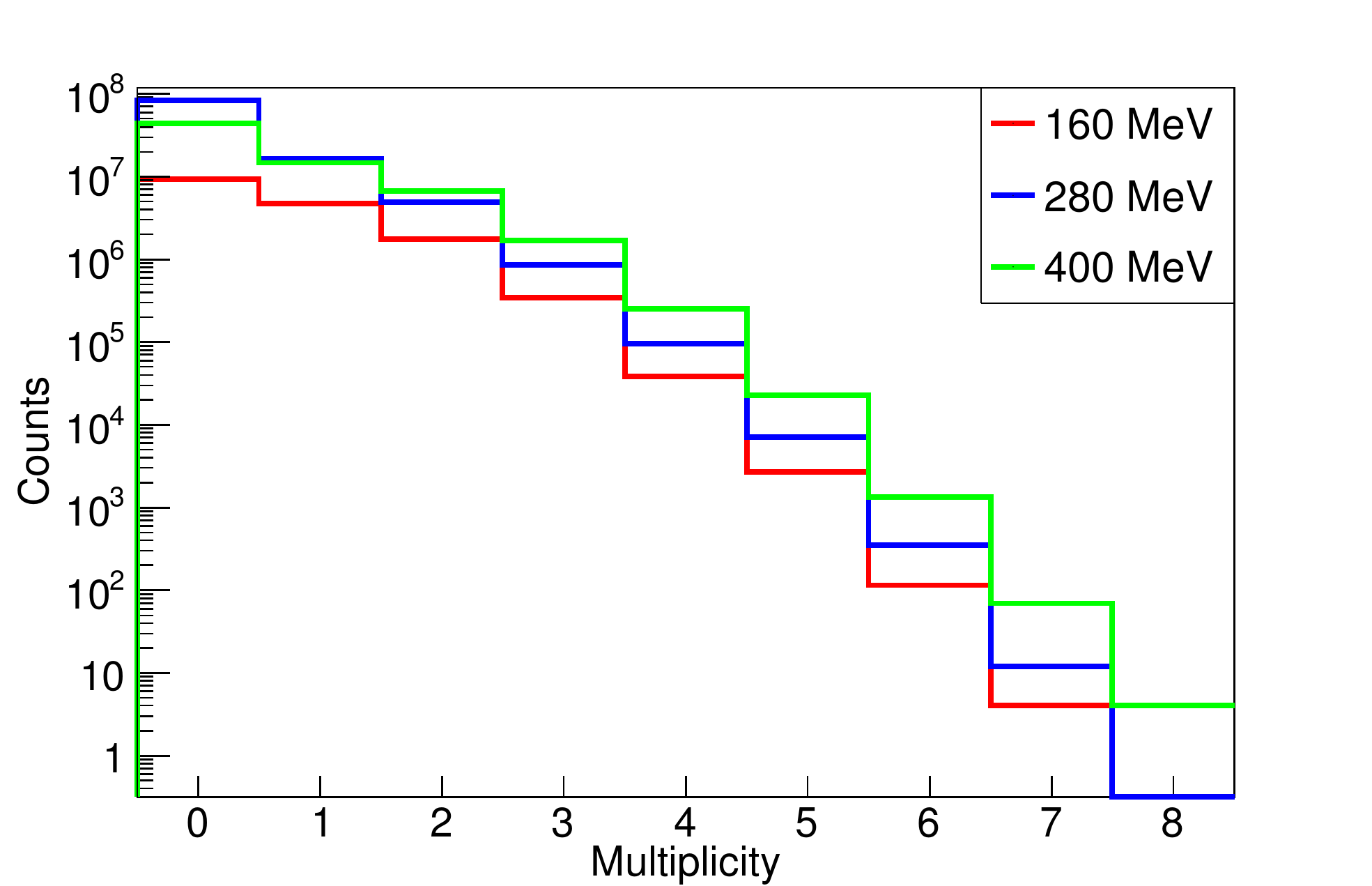}
\caption{Raw $\alpha$-particle multiplicity for the three beam energies of 160 (red), 280 (blue), and 400 (green) MeV.\label{fig:raw_mult}}
\end{figure}

As well as observing the distribution of the $\alpha$-particle energies, the more important probe is that of the $\alpha$-particle multiplicities. This involves the $\alpha$-particles observed in FARCOS (via $\Delta$E-E PID) and CHIMERA (via $\Delta$E-E and TOF PID), the results of which can be seen in Fig.~\ref{fig:raw_mult}. For all three beam energies, a very similar behavior is seen. As the multiplicity increases, the yield decreases by roughly a factor of 10 for each additional $\alpha$-particle. The full multiplicity events are seen for all beam energies and the data from the 400 MeV beam energy also demonstrate multiplicity 8 events. There are three possible reasons for this extraneous yield: firstly, it is possible that contaminants in the target are also causing a contribution from the $^{16}\mathrm{O}(^{16}\mathrm{O},$$^{32}\mathrm{S}^{\star})$ reaction. This is unlikely as any contaminant is expected to constitute a very small percentage of the target and is therefore insufficient to provide the observed level of 8 $\alpha$-particle events. A second explanation is the impurity of the PID classifying non $\alpha$-particle events incorrectly as $\alpha$-particles. At low energies, there is an observed impurity in the TOF PID (Fig.~\ref{fig:TOF}) although as the event multiplicity is increased, the effect of this impurity is decreased therefore suggesting this is not the dominant effect. The final possibility is that event pileup has a small effect sufficient to modify the normalized single event-multiplicity $\mathbf{M}$ to the observed raw event multiplicity $\mathbf{N}$ in Fig.~\ref{fig:raw_mult}. By calculating the probability of pileup for a single event, this effect can be removed.
The single event-multiplicity $\mathbf{M}$ can be related to the raw event multiplicity $\mathbf{N}$ by the pileup probability $\gamma_{\mathrm{pileup}}$ by:
\begin{eqnarray}
\mathbf{N} =&& \mathbf{N}_{\mathrm{no pileup}} + \mathbf{N}_{\mathrm{pileup}}, \\
=&& (1-\gamma_{\mathrm{pileup}})\mathbf{M} + \gamma_{\mathrm{pileup}}  \sum_{ij}^{7} \delta_{i+j,k} \left( \mathbf{M}_{i} \mathbf{M}_{j} \right). \label{eq:efficiencycorrect}
\end{eqnarray}
The Kronecker delta selects combinations of events of multiplicity $i$ and $j$ that combine to make an event of total multiplicity $k$. To obtain $\mathbf{M}$ from $\mathbf{N}$, a fit was performed with $\mathbf{N}$ being a free parameter which effectively inverts Eq.~\ref{eq:efficiencycorrect}. The pileup probabilities are extracted from the cross section given from the EHF code, the beam current used for each beam energy, the target thickness and the cyclotron time period. These probabilities are $4.0 \times 10^{-4}$, $8.0 \times 10^{-5}$ and $3.2 \times 10^{-4}$ for the beam energies of 160, 280, and 400 MeV respectively. These values are low, albeit sufficient, to reproduce the observed 8 $\alpha$-particle events for the 400 MeV beam energy. The modification of $\mathbf{M}$ from $\mathbf{N}$ is very small, particularly for the smaller multiplicities. To compare the experimental multiplicity to the predictions from the EHF and Fermi break-up calculations, the detector efficiency must also be taken into consideration. For a compound nucleus formation, the products are emitted isotropically in the center of mass, therefore, the probability to detect $i$ $\alpha$-particles in an event where $j$ $\alpha$-particles are emitted is given by the binomial probability:
\begin{eqnarray}
\epsilon_{i,j} = {{j}\choose{i}} p^{i} (1-p)^{j-i} \label{eq:effmatrix},
\end{eqnarray}
with the probability of detection $p$ taking the value of 0.28 which best reproduces the data and is also the solid angle coverage fraction. The single event $\mathbf{M}$ is then:
\begin{eqnarray}
\mathbf{M}=\mathbf{\epsilon} \mathbf{M^{\prime}},
\end{eqnarray}
with the elements of the efficiency matrix $\mathbf{\epsilon}$, being those given in Eq.~\ref{eq:effmatrix} and $\mathbf{M^{\prime}}$ being the true multiplicity. 

\begin{figure}[h]
\includegraphics[width=0.5\textwidth]{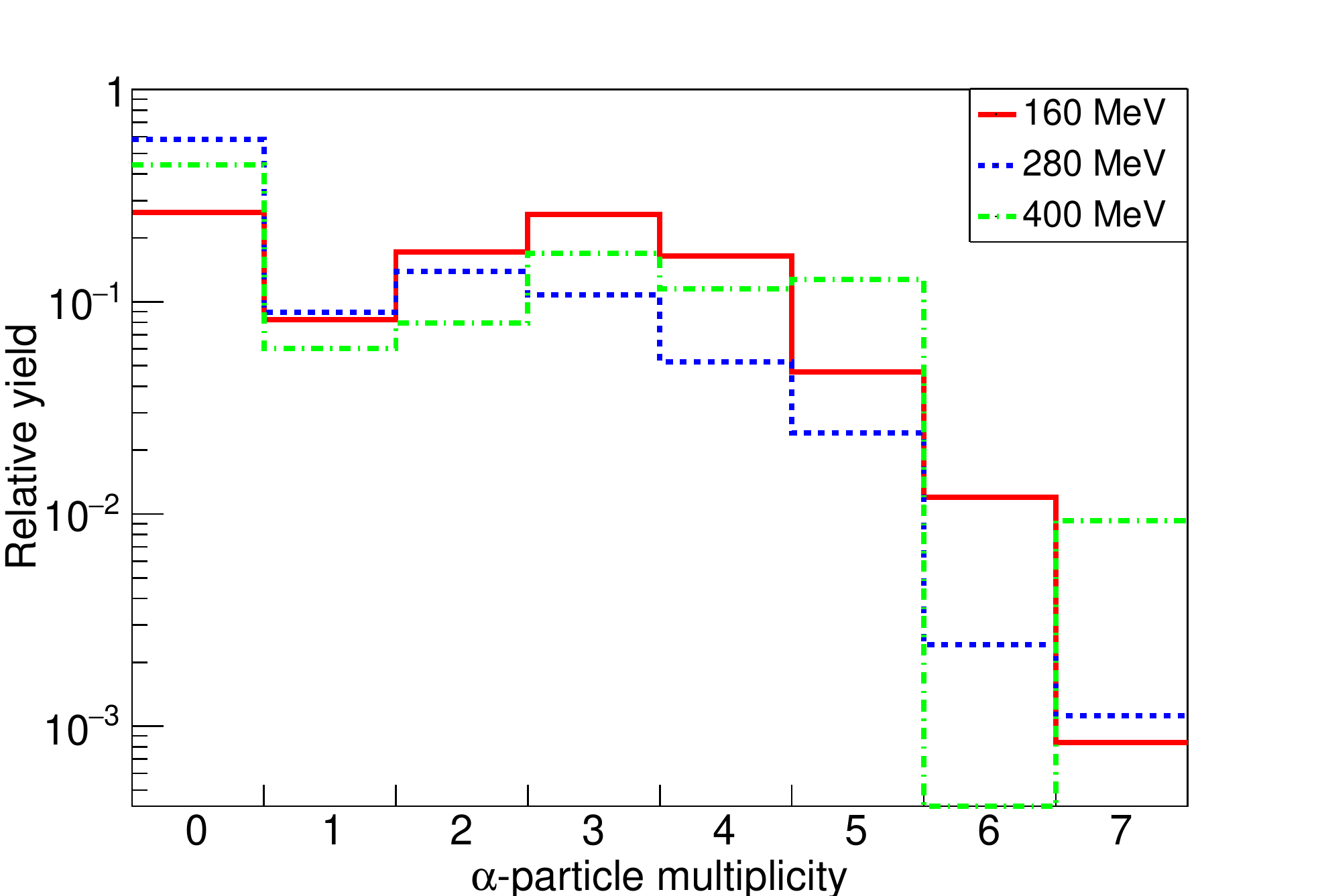}
\caption{True $\alpha$-particle multiplicity for the three beam energies of 160 (red), 280 (blue), and 400 (green) MeV. For each beam energy the total number of counts is normalized to 1. \label{fig:true_mult}}
\end{figure}

Inversion of the efficiency matrix then gives the true multiplicity as:
\begin{eqnarray}
\mathbf{M^{\prime}}=\mathbf{\epsilon}^{-1} \mathbf{M}.
\end{eqnarray}
The result of this true multiplicity inversion is given in Fig.~\ref{fig:true_mult}. This inversion is such that there is no restriction of the elements of $\mathbf{M^{\prime}}$ being positive. This only occurs for the 400 MeV data where the multiplicity for 6 $\alpha$-particles has a small negative value ($-2.6 \times 10^{-3}$).
For all beam energies, the multiplicity can be seen to well exceed that predicted by the sequential decay model (Fig.~\ref{fig:EHFmult}) with the multiplicities peaking around 2 or 3 $\alpha$-particles with a reasonable fraction of events generating a 7 $\alpha$-particle final state. The events with 0 $\alpha$-particles still constitute the largest fraction of events for all three energies. The most important result here is that the multiplicity for the three beam energies show a very similar behavior, far in contrast to the predictions from the EHF which demonstrated a strong dependence on the beam energy.
\section{\label{sec:reconstruction}Reconstruction}
In order to understand the reaction, the excited states populated during the decay process can be examined. As discussed in Section~\ref{sec:FBU}, a signature of an $\alpha$-gas state can be extracted from the 7-$\alpha$ break-up channels. Therefore, in events with sufficient multiplicity, the N-$\alpha$ excitation energies were calculated from reconstruction of the $N$ $\alpha$-particles.
\begin{equation}
E_{x} =\sum_i E_i - \frac{(\sum_i \vec{p_i})^2}{2 \sum_i m_i}  - Q.
\end{equation}
The results from these excitation spectra are separated by the nucleus studied.
\subsection{\label{sec:Be8}Beryllium-8}
\begin{figure}[h]
\includegraphics[width=0.5\textwidth]{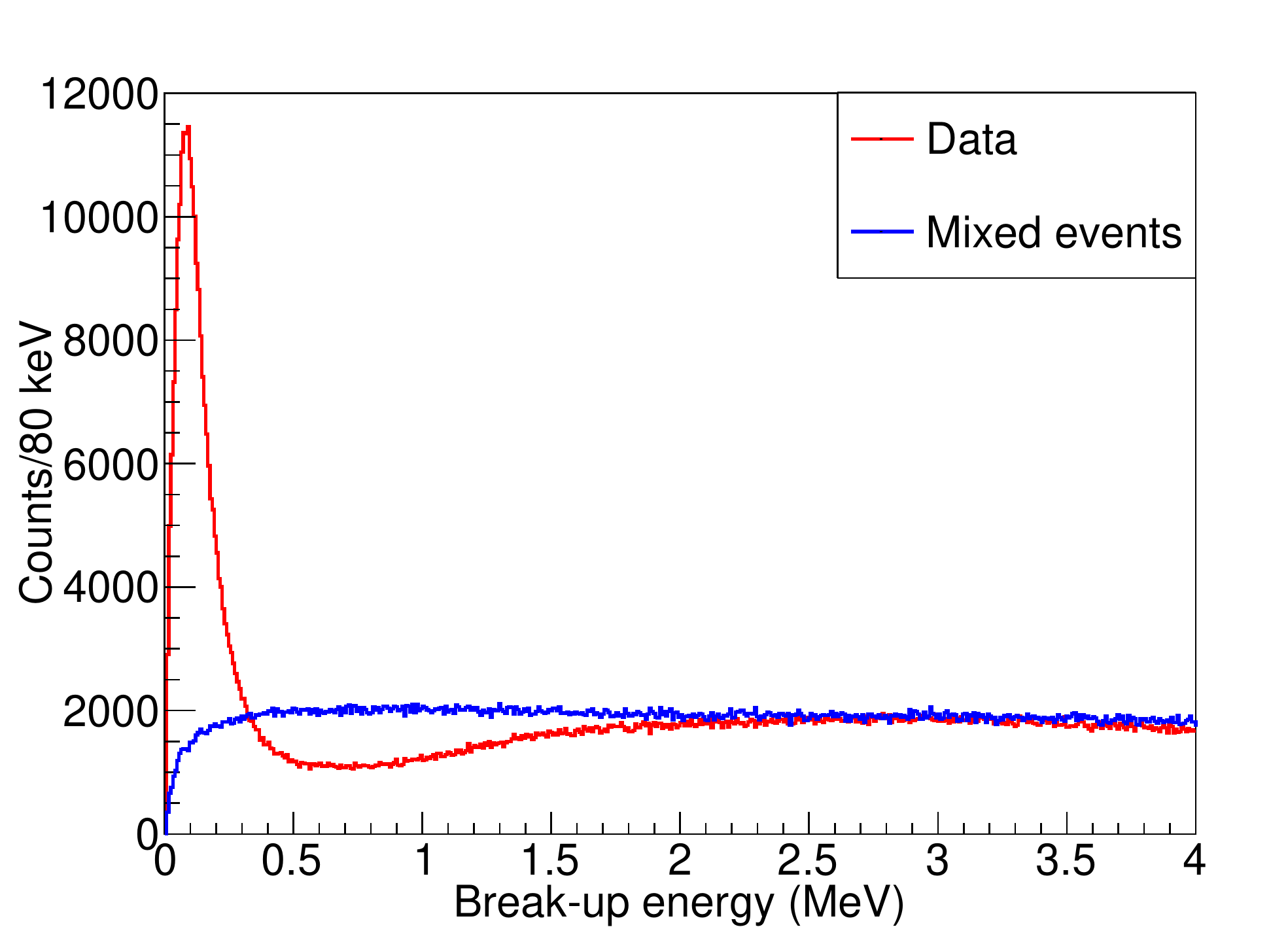}
\caption{Excitation function for $^{8}\mathrm{Be}$ (red) and the contribution from the event-mixing (blue) for a beam energy of 160 MeV.\label{fig:8Be}}
\end{figure}

The 2-$\alpha$ excitation function is shown in Fig.~\ref{fig:8Be} where the ground-state energy can be clearly seen at 92 keV (the Q-value of the break-up is omitted here in order to clearly see the ground-state). There is also a broad contribution at $\sim$ 3 MeV which may be ascribed to the first excited state, a $2_{1}^{+}$ resonance at 3.03 MeV with a large width of $\Gamma$ = 1.5 MeV. In order to understand whether this yield is correctly designated as arising from this broad state, event-mixing was used \cite{eventmixing} which sought to separate the contribution from correlated and uncorrelated events. Here, the scale of the event-mixed events was set such that the total integral of the spectra were the same. Event-mixing takes particles from different events and merges them into a new event where they pass through the same cuts and reconstruction as the non-mixed data.  By taking $\alpha$-particles from different events, one can therefore understand the importance of uncorrelated events (whether from statistical decay processes from the continuum or from mixing of $\alpha$-particles from different parent nuclei). 
This contribution above 2 MeV can be seen to be well reproduced by the event-mixing process demonstrating the contribution from the first excited state of $^{8}\mathrm{Be}$ is not the cause of this bump. The three beam energies of 160, 280, and 400 MeV all show this identical behavior with the ground-state being well populated and any higher energy bumps described by uncorrelated $\alpha$-particles. The dip between 0.5 MeV and 1.5 MeV can be understood as arising from a `Coulomb hole' where $\alpha$-particles with a small relative energy are suppressed by the presence of the Coulomb barrier (see \cite{eventmixing}) which cannot be accounted for by mixing.
\subsection{\label{sec:12C}Carbon-12}
\begin{figure}[h]
\includegraphics[width=0.5\textwidth]{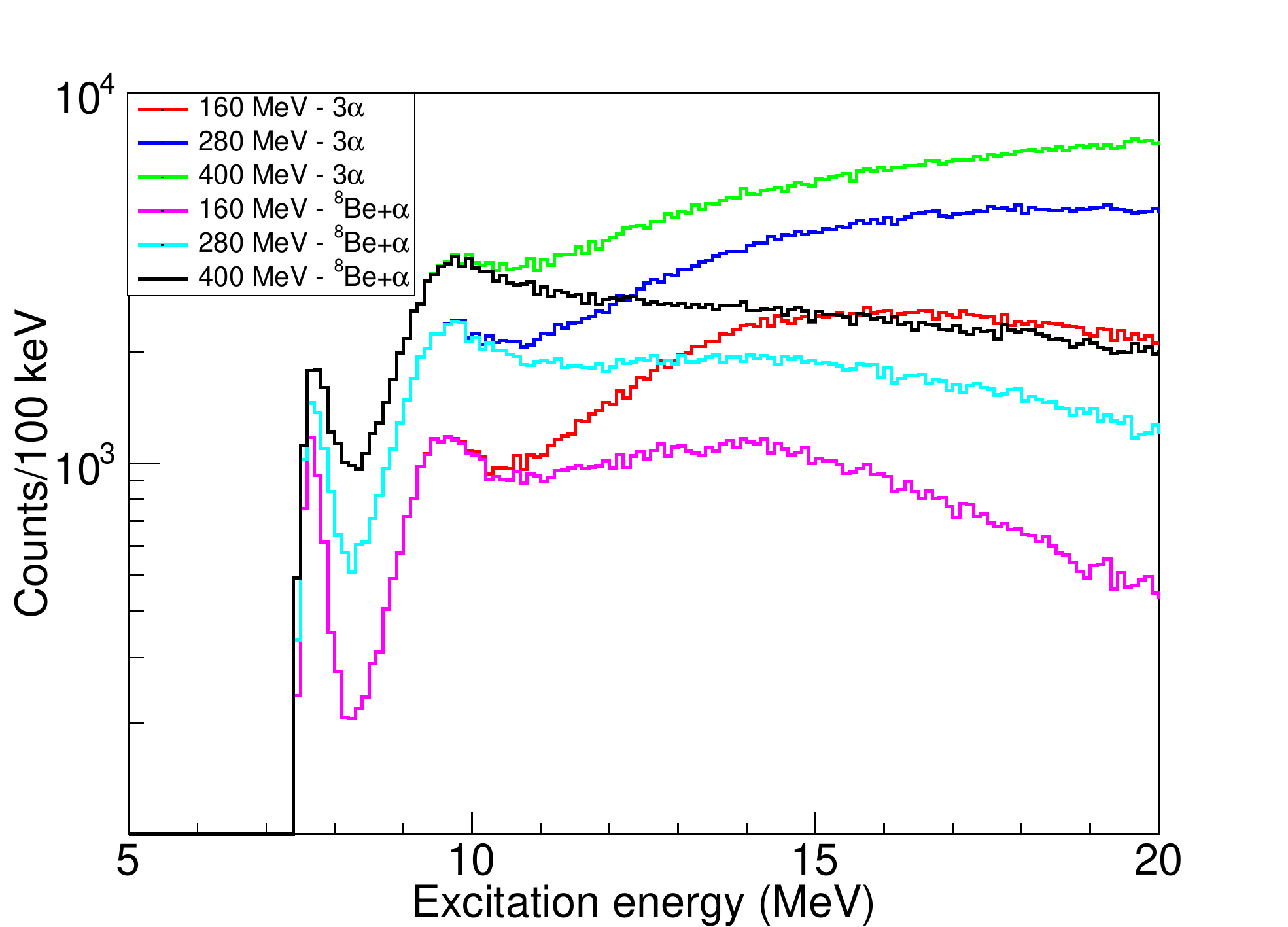}
\caption{Excitation function for $^{12}\mathrm{C}$ showing the results for the three beam energies of 160, 280, and 400 MeV as well as those events where the intermediate decay product corresponds to the $^{8}\mathrm{Be}$(g.s).\label{fig:12C}}
\end{figure}

Taking events where the $\alpha$-particle multiplicity is greater than or equal to 3, resonances in $^{12}\mathrm{C}$ can be reconstructed. To further reduce the background, events were taken where 2 of these 3 $\alpha$-particles can also be reconstructed to form $^{8}\mathrm{Be}$(g.s). The results from these two different possibilities are shown in Fig.~\ref{fig:12C} for the three beam energies.
Once again, the results for the three beam energies are very similar. There is a very strong population of two resonances above the triple-alpha threshold. The first of these is at 7.65 MeV which is the well known Hoyle state, a resonance that is of interest as discussed above. This is well populated in all three beam energies and is easily separable from the higher energy peak seen at 9.7 MeV. This peak is much broader than the Hoyle state suggesting this contribution is not solely due to the $3_{1}^{-}$ state at 9.64 MeV which has a measured width of 48 keV\cite{Tzany3minus}. Using the Monte Carlo simulations to study the expected resolution response as a function of excitation energy (and scaling the expected response to the experimental width of the Hoyle state), the width of this state averaged over the three beam energies is $\Gamma = 600(60)$ keV. This is a value which is incompatible with the $3_{1}^{-}$ and may therefore correspond to a contribution from the rotational excitation of the Hoyle state, $2_{2}^{+}$, which has been measured to have an excitation energy of 9.75(15) MeV with a width of 750(150) keV \cite{Freer2+,Freer2+2} from inelastic scattering experiments. Reanalysis of the photodissociation breakup reaction $^{12}\mathrm{C}(\gamma,3\alpha)$ \cite{Zimmerman} where the $2^{+}$ was unambiguously identified gave a much larger value of 1.6(1)~MeV \cite{FreerRMP}.\par
\begin{figure}[h]
\includegraphics[width=0.5\textwidth]{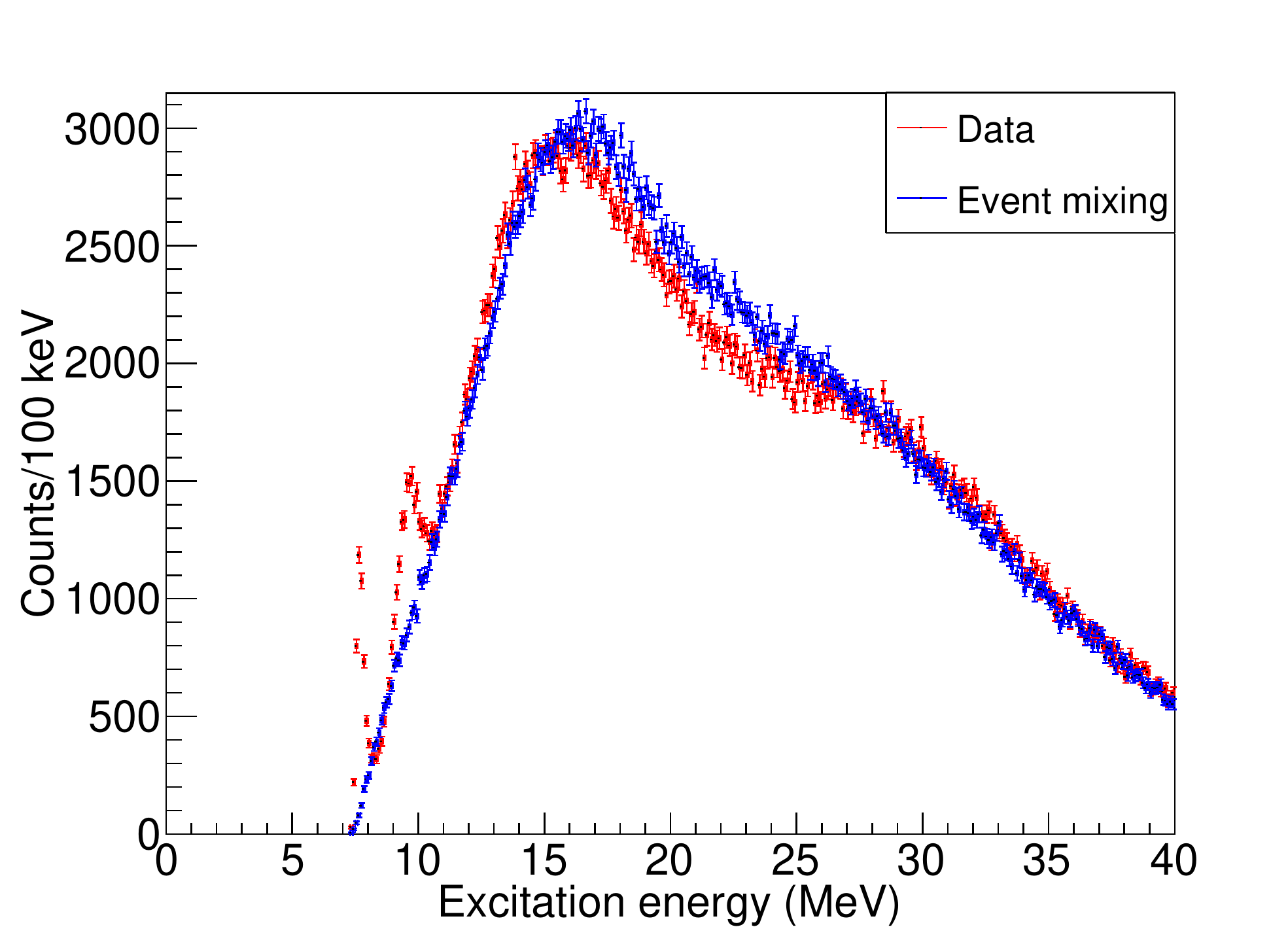}
\caption{Event mixing (blue) applied to the 3-$\alpha$ excitation function for a beam energy of 160 MeV scaled to give the same total area as the data. The smooth continuum in $^{12}\mathrm{C}$ seen in the data (red) can be described by uncorrelated $\alpha$-particles as demonstrated by event mixing.\label{fig:12C_mix}}
\end{figure}

As with $^{8}\mathrm{Be}$, event-mixing can be used to examine the smooth contribution above these two resonances which shows a shift in excitation energy with increasing beam energy. Figure~\ref{fig:12C_mix} shows the result of this process indicating the broad contribution can be described by event mixing. While the resonances are not reproduced as one would expect (showing they are due to correlated $\alpha$-particles), the double humped continuum is well matched by the event mixing contribution apart from a small excess around 25-30 MeV.
\subsection{\label{sec:O16}Oxygen-16}
\begin{figure}[h]
\includegraphics[width=0.5\textwidth]{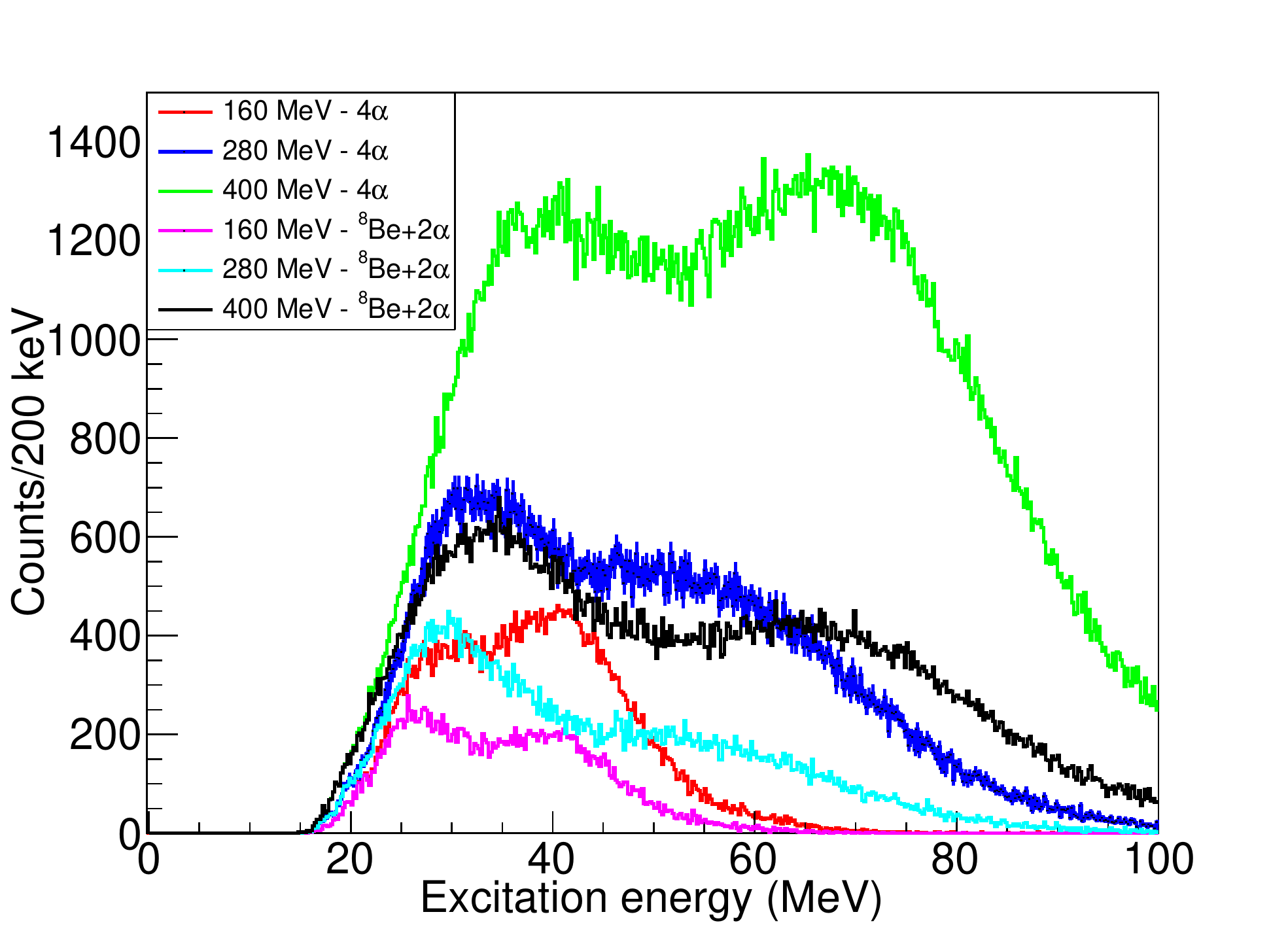}\\
\includegraphics[width=0.5\textwidth]{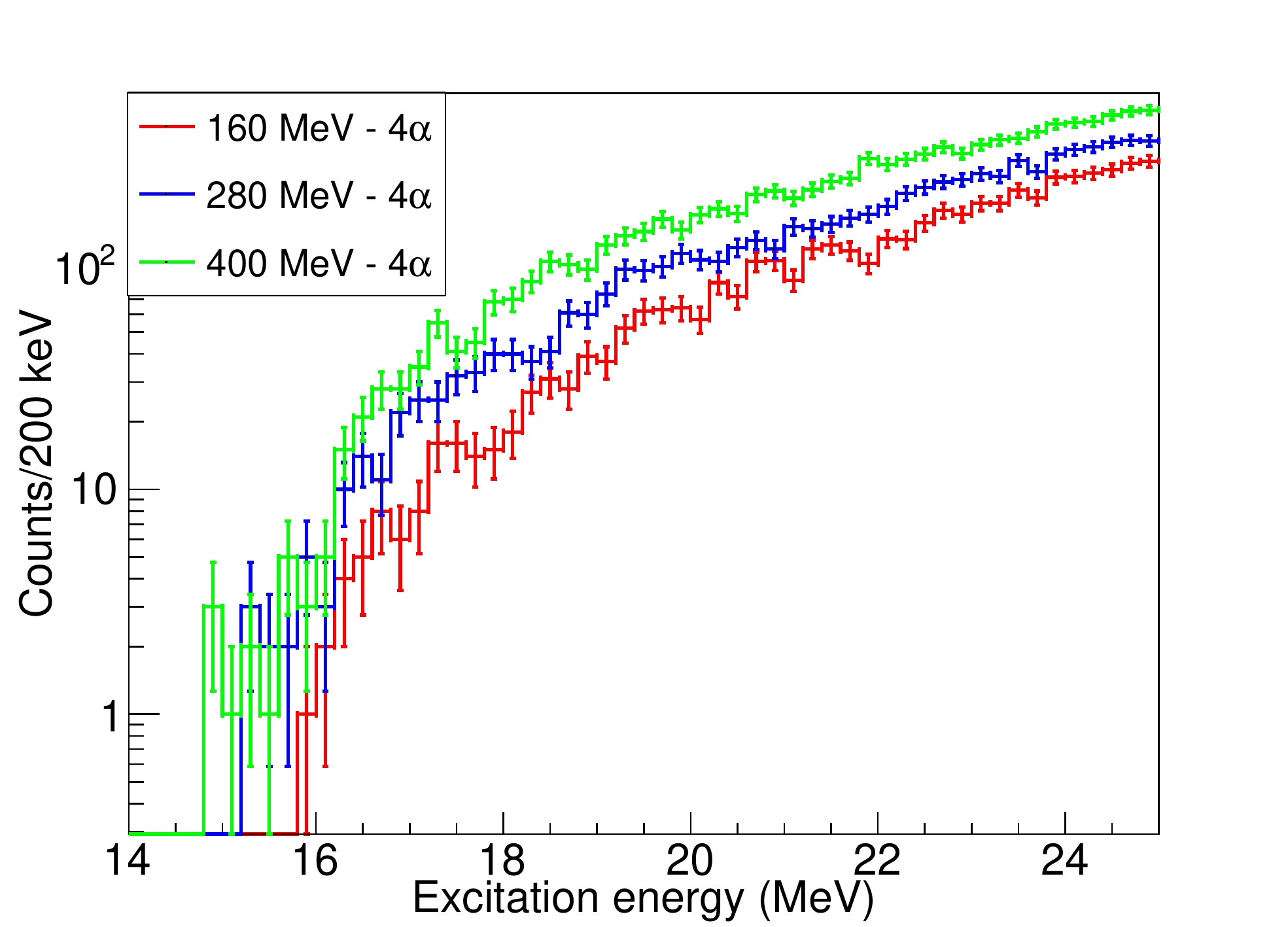}
\caption{Excitation for $^{16}\mathrm{O}$ for the three beam energies showing all decays to 4 $\alpha$-particles in lin-lin (top) for a wide excitation range and log-lin (bottom) for near the 4-$\alpha$ threshold. Also shown (top) is the contribution where a pair of the 4 $\alpha$-particles form $^{8}\mathrm{Be}$(g.s). \label{fig:16O_4a}}
\end{figure}

Following the formulations described for $^{8}\mathrm{Be}$ and $^{12}\mathrm{C}$, taking events with a multiplicity greater than or equal to 4, the excited states in $^{16}\mathrm{O}$ which decay via the 4$\alpha$, $^{8}\mathrm{Be}$+2$\alpha$, $^{12}\mathrm{C}(0_2^+) {+} \alpha$, and $^{8}\mathrm{Be}{+}^{8}\mathrm{Be}$ channels were examined. While the experimental resolution in this channel is expected to be $\sim$ few 100 keV (depending on the exact excitation energy and the decay path used to produce it), any strong resonances should stick out above the smooth background. The first two of these decay channels can be seen in Figure~\ref{fig:16O_4a}. A broad double humped structure can be seen for all the three beam energies (albeit at different excitation energies) in the 4-$\alpha$ channel however no narrow peaks are evident in the data. The bottom part of Fig.~\ref{fig:16O_4a} shows the excitation function near the 4-$\alpha$ barrier where the counts can be seen to very quickly approach zero around 15 MeV with no obvious enhancement in this region. One would not expect to see any strong `traditionally' $\alpha$-clustered states until over the Coulomb barrier at $\sim$ 18 MeV. To understand the yield around this important area, event mixing was employed to demonstrate the importance of uncorrelated $\alpha$-particles.\par
\begin{figure}[h]
\includegraphics[width=0.5\textwidth]{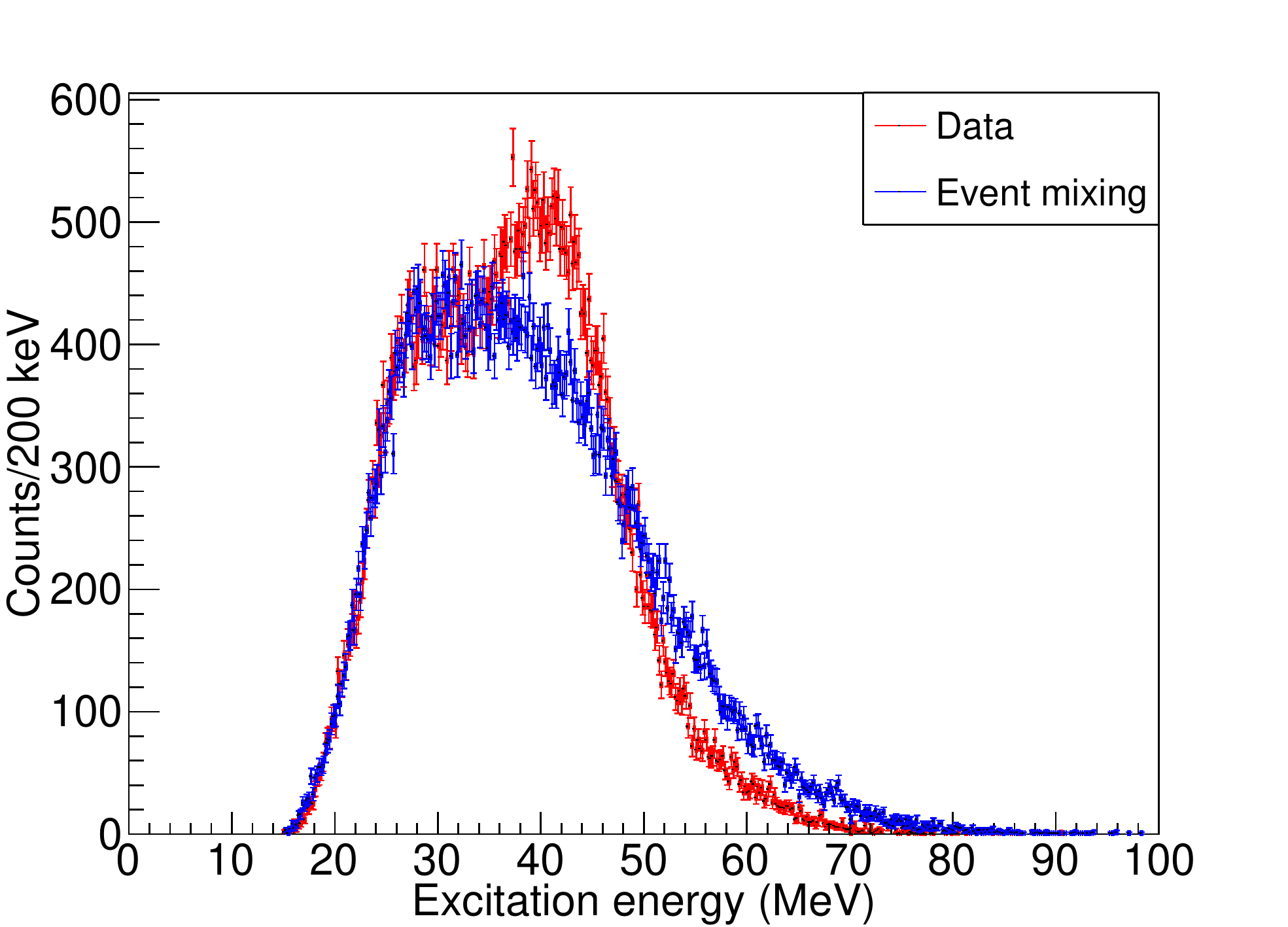}
\caption{Event-mixing (blue) for $^{16}\mathrm{O}$ for a beam energy of 160 MeV in comparison to the experimental data (red). The event-mixing describes the majority of the structures seen apart from the secondary peak in the data and the shape of the high energy tail.\label{fig:16O_mix}}
\end{figure}

Figure~\ref{fig:16O_mix} shows the event mixing for the 160 MeV beam energy. This can describe the seen distribution extremely well, particularly at lower excitation energies where the smooth decrease in counts towards the 4-$\alpha$ threshold is well reproduced. The second peak in the distribution is not reproduced by this mixing suggesting this corresponds to a correlated contribution to the total excitation spectrum. This is reproduced for all three beam energies suggesting the observed enhancement is not a particularly well populated state but a manifestation of the break-up mechanism as the excitation energy changes with the different beam energy.\par
\begin{figure}[h]
\includegraphics[width=0.5\textwidth]{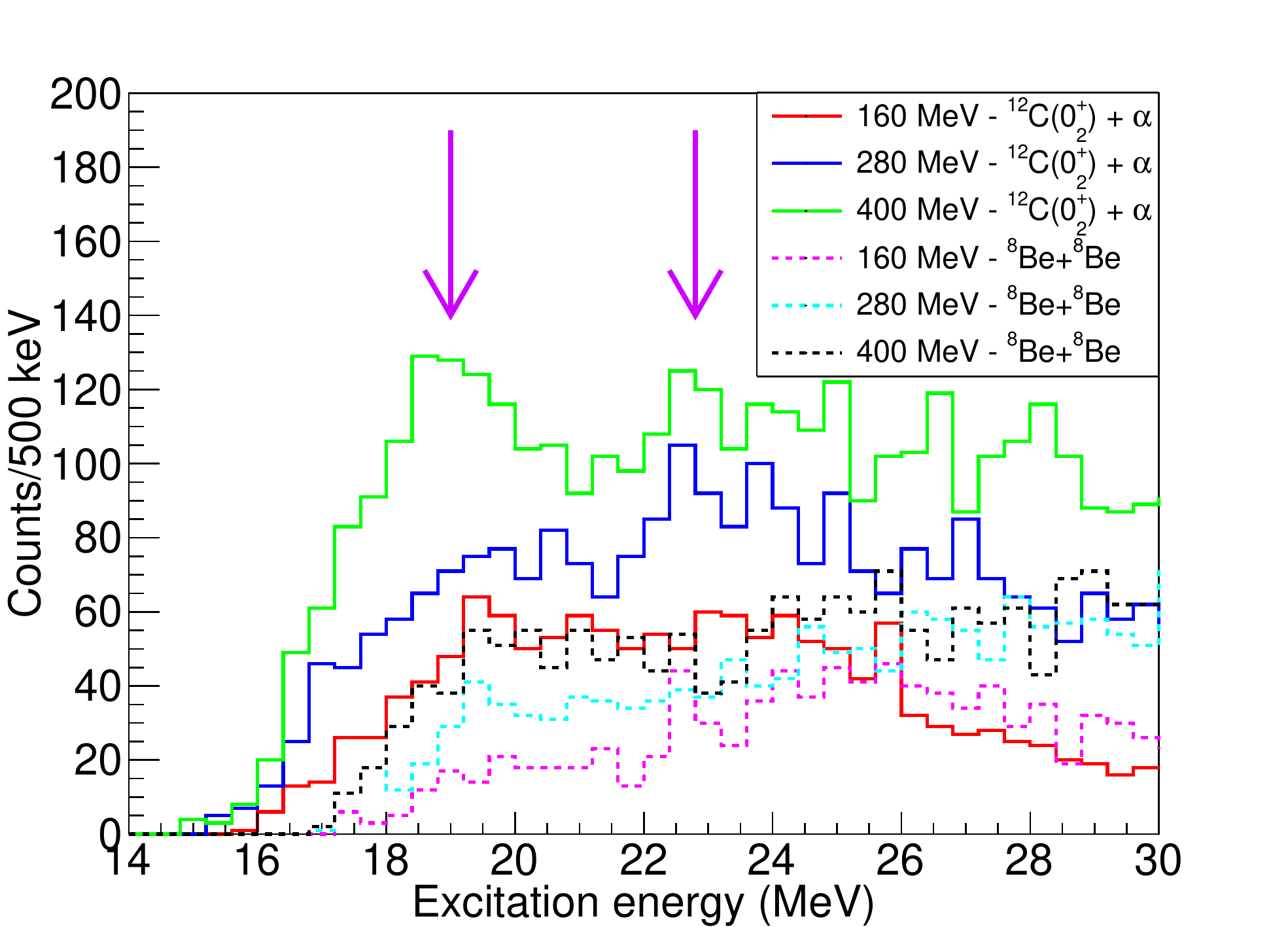}
\caption{Excitation function in $^{16}\mathrm{O}$ for the clustered $^{12}\mathrm{C}(0_2^+) {+} \alpha$ (solid lines) and $^{8}\mathrm{Be}{+}^{8}\mathrm{Be}$ decay paths (dashed) for the 160 (red/magenta), 280 (blue/cyan), and 400 (green/black) MeV data for the ($^{12}\mathrm{C}(0_2^+){+}\alpha$/$^{8}\mathrm{Be}{+}^{8}\mathrm{Be}$) channel. The violet arrows indicate a possible increased yield in the $^{12}\mathrm{C}(0_{2}^{+})+\alpha$ channel.\label{fig:16O_clustered}}
\end{figure}

To better understand the role of the continuum against genuine 4-$\alpha$ resonances, the $^{12}\mathrm{C}(0_2^+) {+} \alpha$ and $^{8}\mathrm{Be}{+}^{8}\mathrm{Be}$ paths were also investigated where the contribution from mismatching and uncorrelated $\alpha$-particles is expected to be greatly reduced. The results from this can be seen in Fig.~\ref{fig:16O_clustered}. In these decay channels, evidence of structure starts to manifest. For the data from beam energies of 160 and 400 MeV, there is evidence of a broad structure around 19 MeV in the $^{12}\mathrm{C}(0_2^+) {+} \alpha$ decay channel. This region corresponds to well known states which have been demonstrated to have a reasonable strength in several reactions employing different population methods \cite{Curtis13C}. The resonance, which has a possible $6^{+}$ spin-parity assignment \cite{Neil16O}, may therefore be preferentially populated by the $(2J+1)$ spin dependency of the cross section. Additionally, there is evidence of a broader component at 23-24 in the data from the 280 and 400 MeV beam energies. The statistics above the background here are bolstered by the occurrence of peaks at the same excitation energy with two different beam energies which provides an increased statistical significance. Therefore, there is evidence of population of resonances (or most probably a collection of resonances which cannot be experimentally resolved) in addition to the dominance of the non-correlated $\alpha$-particles which constitute the larger fraction of the total 4-$\alpha$ events.
\subsection{\label{sec:Ne20}Neon-20}
\begin{figure}[h]
\includegraphics[width=0.5\textwidth]{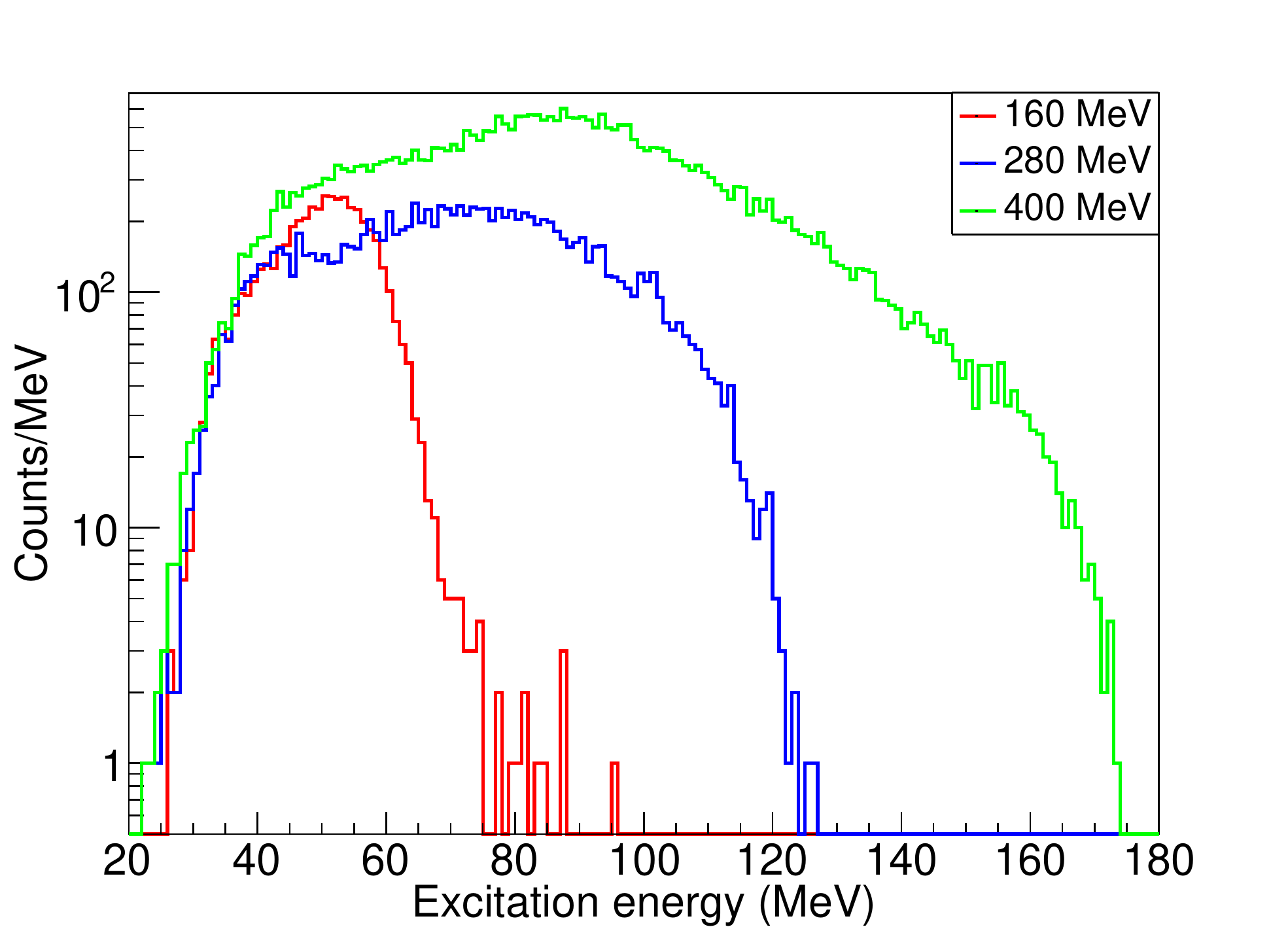}
\caption{Excitation function of $^{20}\mathrm{Ne}$ for states decaying to 5 $\alpha$-particles.\label{fig:20Ne_direct}}
\end{figure}

When increasing the size of the N-$\alpha$ system, the reduction in the significance of resonant states as well as the dominance of uncorrelated $\alpha$-particles and the continuum is noticeable. By examining the 5-$\alpha$ events corresponding to $^{20}\mathrm{Ne}$ in Fig.~\ref{fig:20Ne_direct}, it is apparent this trend continues. 
The low excitation energy region for $^{20}\mathrm{Ne}$ demonstrates a difference from the previously investigated nuclei however. The yield near the 5-$\alpha$ threshold at 19.2 MeV is not present.  This region is (as with the 15-18 MeV region in $^{16}\mathrm{O}$) below the Coulomb barrier, hence, the yield extends no lower than $\sim 24$ MeV, a result consistent with previous experiments of the 5-$\alpha$ channel where the lowest energy state seen was at 24.5 MeV in the $^{16}\mathrm{O}(\alpha,{^{12}\mathrm{C}(0_2^+)})^{8}\mathrm{Be}$ channel \cite{20NeTz}. Calculations from the EHF showed a large population was expected in this region suggesting the quenching of the 5-$\alpha$ decay channel for states in this region, instead preferring other decay paths such as $^{16}\mathrm{O}+\alpha$.\par
\begin{figure}[h]
\includegraphics[width=0.5\textwidth]{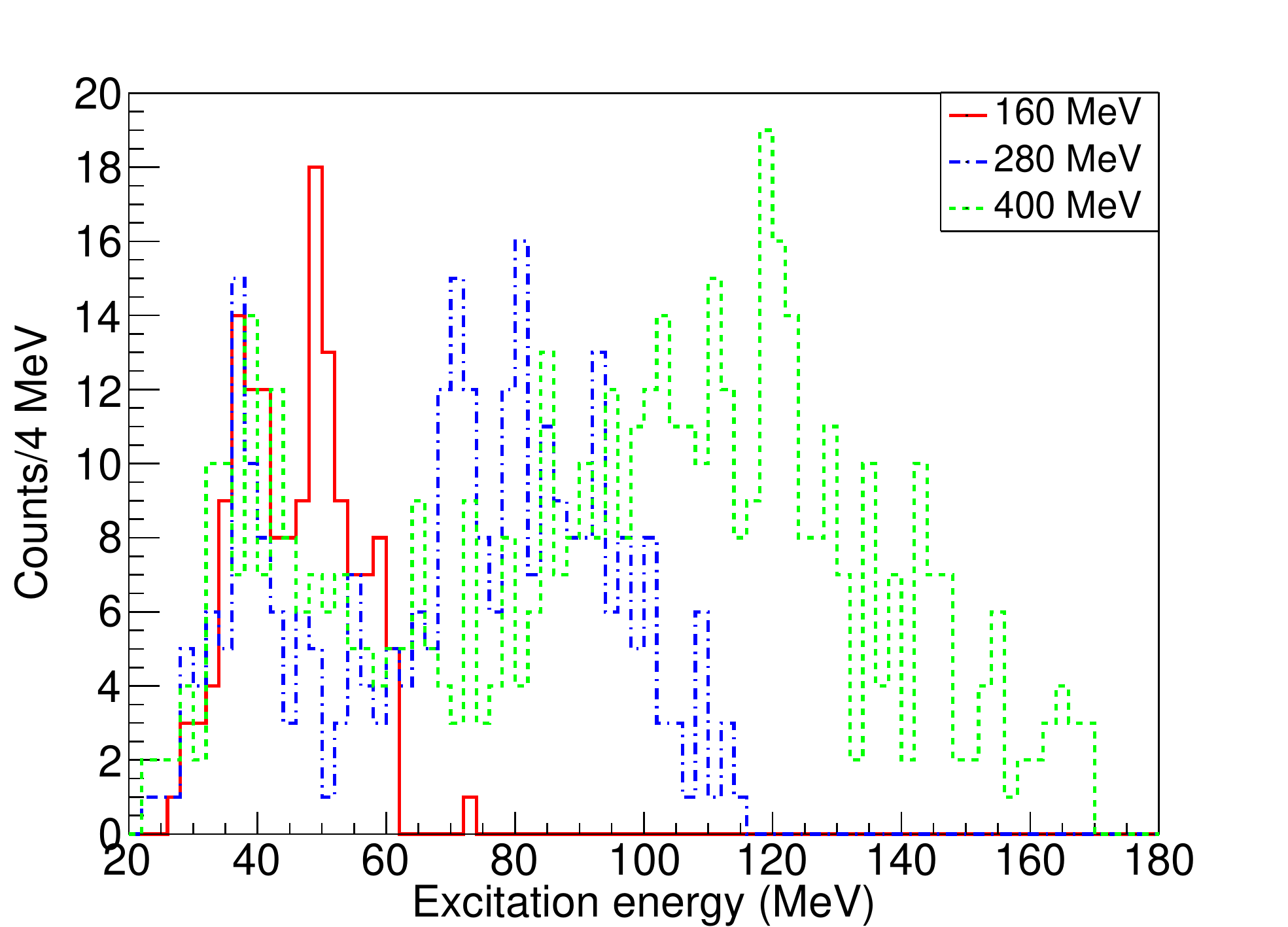}
\caption{Excitation function of $^{20}\mathrm{Ne}$ for states decaying to $^{12}\mathrm{C}(0_2^+){+}^{8}\mathrm{Be}$.\label{fig:20Ne_clustered}}
\end{figure}
The decays via highly clustered (potential $\alpha$-gas) decay channels, e.g. $^{12}\mathrm{C}(0_2^+){+}^{8}\mathrm{Be}$, also mirror this result with only a small fraction of the 5-$\alpha$ events belonging to this decay channel. The excitation energy for this decay path can be seen in Fig.~\ref{fig:20Ne_clustered}. The 160, 280, and 400 MeV data show a possible broad contribution from one or several states around 38 MeV matching a doublet of states seen previously \cite{FreerNe} although the statistics are limited here. An additional possible state is visible (in the 160 MeV data only) with an energy of around 50 MeV. There is also a much larger contribution from higher excitation energies outside of these peaks which is perhaps indicative that the $^{12}\mathrm{C}(0_2^+)$ and $^{8}\mathrm{Be}$ do not originate mainly from the decay of the $^{20}\mathrm{Ne}$ system but instead arise from separate sequential decay stages or from a multi-particle break-up as described by the Fermi break-up model. The three excitation functions for the different beam energies can be seen to essentially be a stretching of the 160 MeV structure to higher energies corresponding to a larger phase space.
\section{\label{sec:Origins}Origin of $\alpha$-clustered states}
In order to better understand the reaction mechanisms which generate the observed $\alpha$-clustered states, it is important to study where these resonances originate from. To do this, the role of the two-body break-up of the compound nucleus can be studied to understand the possible role of direct reactions. By reconstructing the missing momentum following the measurement of a given $\alpha$-conjugate state, the Q-value for the reaction can be examined. These will be briefly discussed for each nucleus.
\subsection{\label{sec:Origins8Be}Origins of beryllium-8}
\begin{figure}[h]
\includegraphics[width=0.5\textwidth]{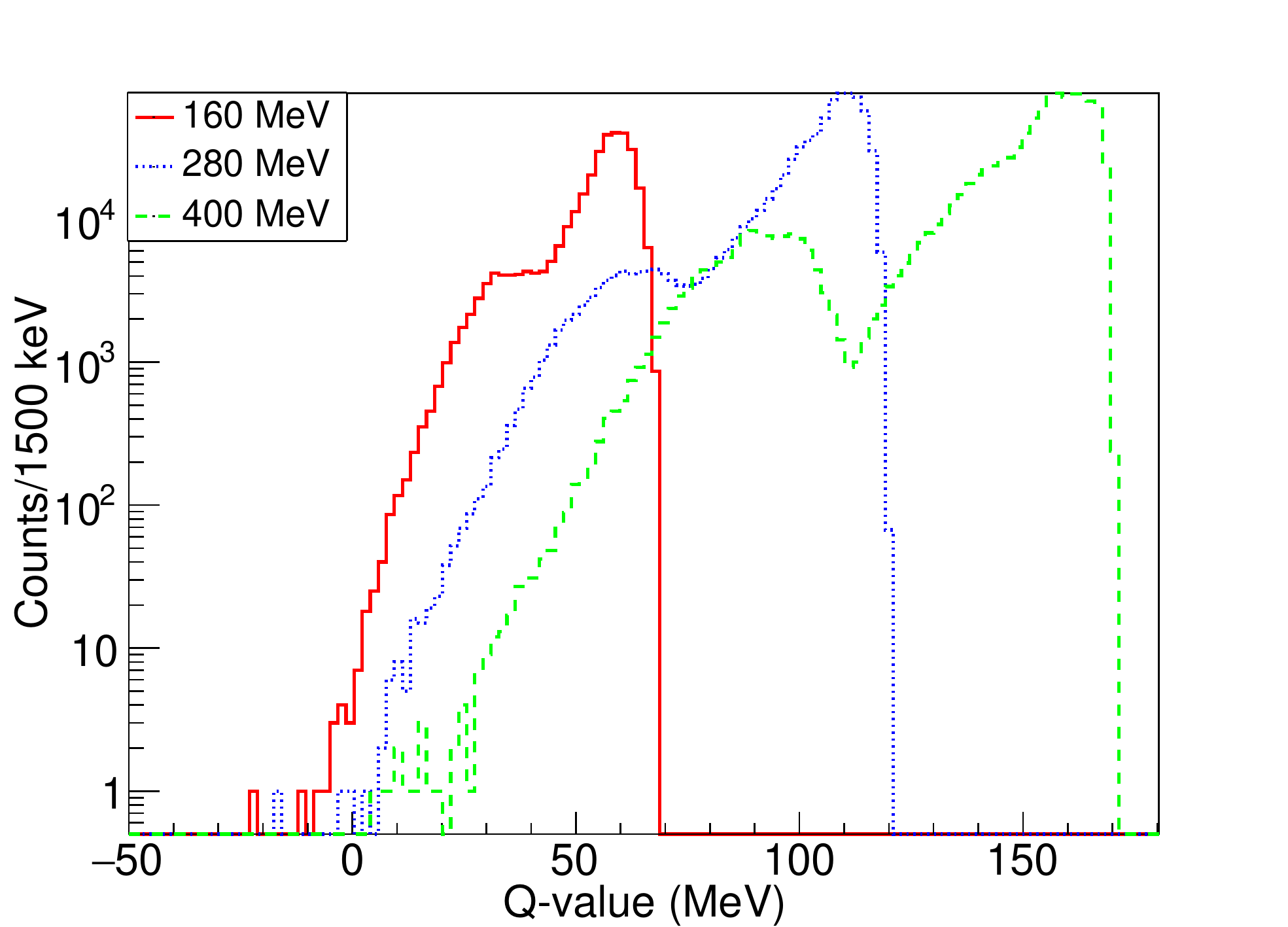}
\caption{(Negative) Q-value for the break-up of the compound nucleus into $^{8}\mathrm{Be}$(g.s) {+} $^{20}\mathrm{Ne}$ by reconstructing the $^{20}\mathrm{Ne}$ from the measurement of the $^{8}\mathrm{Be}$(g.s) from 2 $\alpha$-particles.\label{fig:Qvalue_8Be20Ne}}
\end{figure}

By selecting events where 2 $\alpha$-particles can be reconstructed to form $^{8}\mathrm{Be}$(g.s) and then using the missing momentum to form $^{20}\mathrm{Ne}$, the total Q-value for the reaction is calculated. For a simple binary decay from the compound nucleus, the difference between the expected Q-value and the observed Q-value is therefore due to the excitation energy of the $^{20}\mathrm{Ne}$ which can therefore be observed without measuring its decay.  For ease of visualization, the negative Q-value is plotted so excitation energy is congruent with the Q-value. The results for this are shown in Fig.~\ref{fig:Qvalue_8Be20Ne}. An obvious path that would be a strong signature of $\alpha$-condensation here is evidence of a peak at $\sim$ 20 MeV, corresponding to an $\alpha$-gas state in $^{20}\mathrm{Ne}$ (which are predicted to exist just above the N-$\alpha$ breakup thresholds). As discussed above, a good signature that a state is an $\alpha$-condensate states is the decay into two sub-units which are also $\alpha$-condensates. Such an enhancement is clearly absent for the three beam energies which show an extremely similar form for the different energies albeit translated by virtue of the modification to the initial beam momentum. Such results are highly indicative that there is no $^{8}\mathrm{Be}$(g.s.) {+} $^{20}\mathrm{Ne}^{\star}$ path with only $2 \times 10^{-3} \%$ of the measured $^{8}\mathrm{Be}$ reproducing the correct Q-value for the $^{12}\mathrm{C}(^{16}\mathrm{O},^{8}\mathrm{Be}$(g.s))$^{20}\mathrm{Ne}$(g.s.) break-up of $-$2.6 MeV for the 160 MeV beam energy (and no events for the 280 and 400 MeV beam energy data). The absence of the mutual ground state break-up mirrors well the results of the EHF calculations which show this decay path is expected to be very weak.

\subsection{\label{sec:Origins12C}Origins of carbon-12}
\begin{figure}[h]
\includegraphics[width=0.5\textwidth]{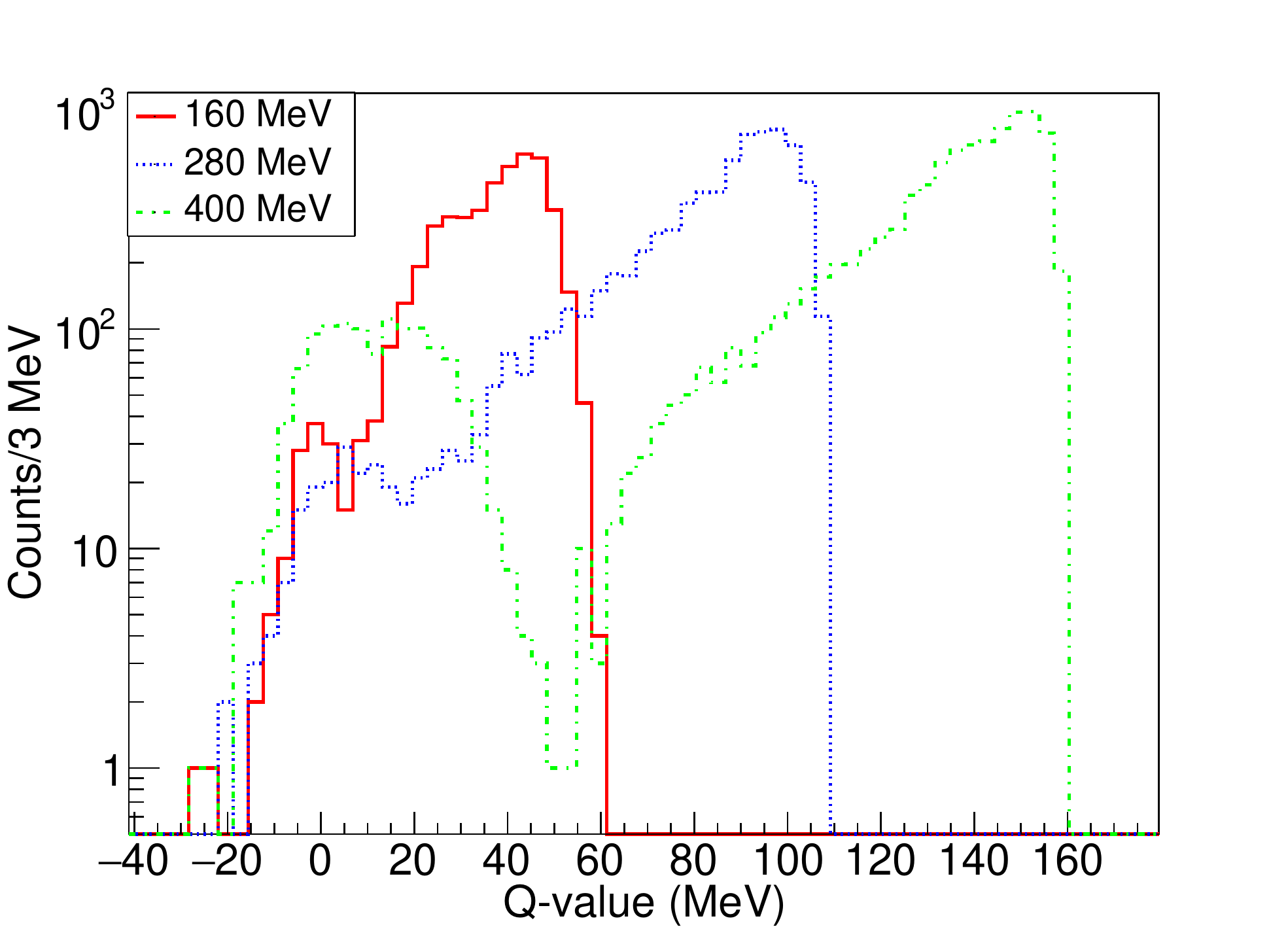}
\caption{(Negative) Q-value for the break-up of the compound nucleus into $^{12}\mathrm{C}(0_2^+)$ + $^{16}\mathrm{O}$ by reconstructing the $^{16}\mathrm{O}$ from the measurement of the $^{12}\mathrm{C}(0_2^+)$ from 3 $\alpha$-particles.\label{fig:Qvalue_Hoy}}
\end{figure}

It was demonstrated in Section~\ref{sec:12C} that a large number of the observed $^{8}\mathrm{Be}$(g.s) events corresponded to the decay of states in $^{12}\mathrm{C}$, therefore, examination of the Q-value to study the binary fission mode $^{12}\mathrm{C}^{\star}{+}^{16}\mathrm{O}^{\star}$ (which is also shown to be dominant in the EHF calculations) may indicate a large fraction of the $^{12}\mathrm{C}(0_2^+)$ and $^{12}\mathrm{C}(3_1^-)$ originates from this decay mode. One may also determine the amount of inelastic scattering (where the beam is left in its ground state) in this way. The results of the (negative) Q-value from reconstructing the missing $^{16}\mathrm{O}^{\star}$ from measuring the Hoyle state from 3 $\alpha$-particles are shown in Fig.~\ref{fig:Qvalue_Hoy}.\par

For the three beam energies, two components can be seen. The lowest component is a broad peak centered around 0 MeV with a width of 1.2 MeV for the 160 MeV beam energy data but increasing to 12 MeV for a beam energy of 400 MeV. For the 400 MeV data in particular, this lower energy component can be seen to correspond well with the clustered 4-$\alpha$ yields seen in Fig.~\ref{fig:16O_clustered} which terminate around 40 MeV. This is therefore highly indicative of a $^{12}\mathrm{C}(0_2^+) {+} ^{16}\mathrm{O}^{\star}$ component to the compound nucleus decay or alternatively a contribution from inelastic scattering. These events are clearly a small fraction of the total events in the plot for each beam energy therefore show that (even if all of these events are inelastic scattering rather than compound nucleus decay) the direct contribution is small in comparison to the other observed paths. The events that lie in the secondary peak, which has a smoothly increasing structure followed by an abrupt drop-off (at the largest value seen for each beam energy), correspond to paths which are not $^{12}\mathrm{C}(0_2^+) {+} ^{16}\mathrm{O}^{\star}$ and therefore cannot be inelastic scattering. These have a form well reproduced by Monte Carlo simulation of other reactions paths such as the sequential $\alpha$-decay via $^{24}\mathrm{Mg}, ^{20}\mathrm{Ne}, ^{16}\mathrm{O}, {^{12}\mathrm{C}(0_2^+)}$ or multi-particle break-up.\par
\begin{figure}[h]
\includegraphics[width=0.5\textwidth]{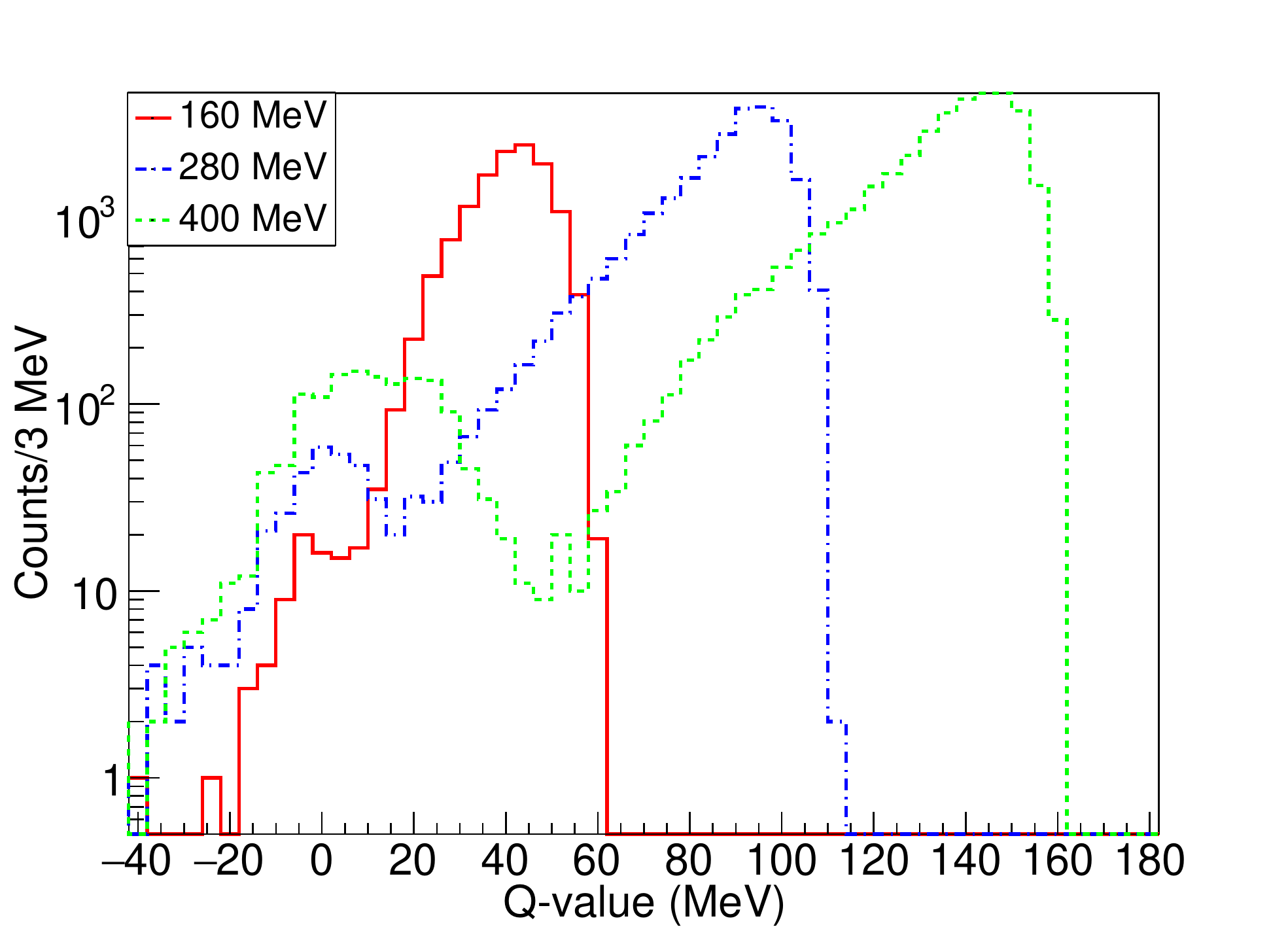}
\caption{(Negative) Q-value for the break-up of the compound nucleus into $^{12}\mathrm{C}(3_1^-)$ {+} $^{16}\mathrm{O}^{\star}$ by reconstructing the $^{16}\mathrm{O}$ from the measurement of the $^{12}\mathrm{C}(3_1^-)$ from 3 $\alpha$-particles.\label{fig:Qvalue_3m}}
\end{figure}

The same formulation can be used to study the mechanism of $^{12}\mathrm{C}(3_1^-)$ formation, as shown in Fig.~\ref{fig:Qvalue_3m}. The form is extremely similar to that seen for the Hoyle state albeit with a slightly smaller component corresponding to the direct decay and a larger component from more sequential paths. There is, despite being slightly weaker than the $^{12}\mathrm{C}(0_2^+) {+} ^{16}\mathrm{O}^{\star}$ component, a reasonable strength $^{12}\mathrm{C}(3_1^-) {+} ^{16}\mathrm{O}^{\star}$ path present in the data for all three beam energies.\par
These data therefore match the EHF calculations which predict a large $^{12}\mathrm{C}^{\star} {+} ^{16}\mathrm{O}^{\star}$ strength. There is no evidence of the predicted 15.1 MeV $\alpha$-gas state \cite{THSR16O} in the data as measured from the $^{12}\mathrm{C}(0_2^+)$ and a large portion of the seen $^{12}\mathrm{C}(0_2^+)$ and $^{12}\mathrm{C}(3_1^-)$ events do not originate from binary scission but instead arise from more sequential decays or multi-particle break-up.
\subsection{\label{sec:Origins16O}Origins of oxygen-16}
\begin{figure}[h]
\includegraphics[width=0.5\textwidth]{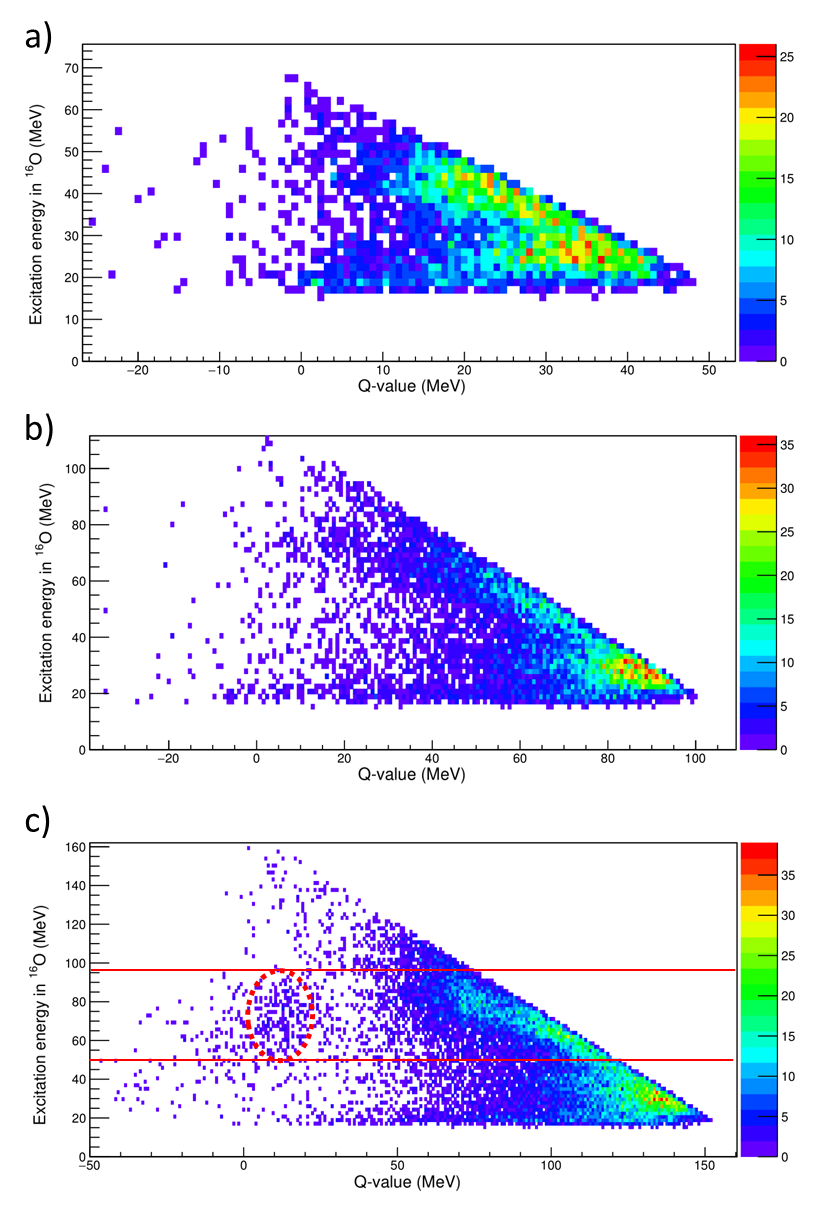}
\caption{(Negative) Q-value for $^{12}\mathrm{C}^{\star} {+} ^{16}\mathrm{O}^{\star}$ against the excitation energy in $^{16}\mathrm{O}$ from the 4-$\alpha$ channel. The results are shown for the three beam energies of 160 (a), 280 (b), and 400 (c) MeV. Red annotations denote the areas associated with excess yields in the data which are mentioned in the text.  \label{fig:Qvalue_16O12C}}
\end{figure}

\begin{figure}[h]
\includegraphics[width=0.5\textwidth]{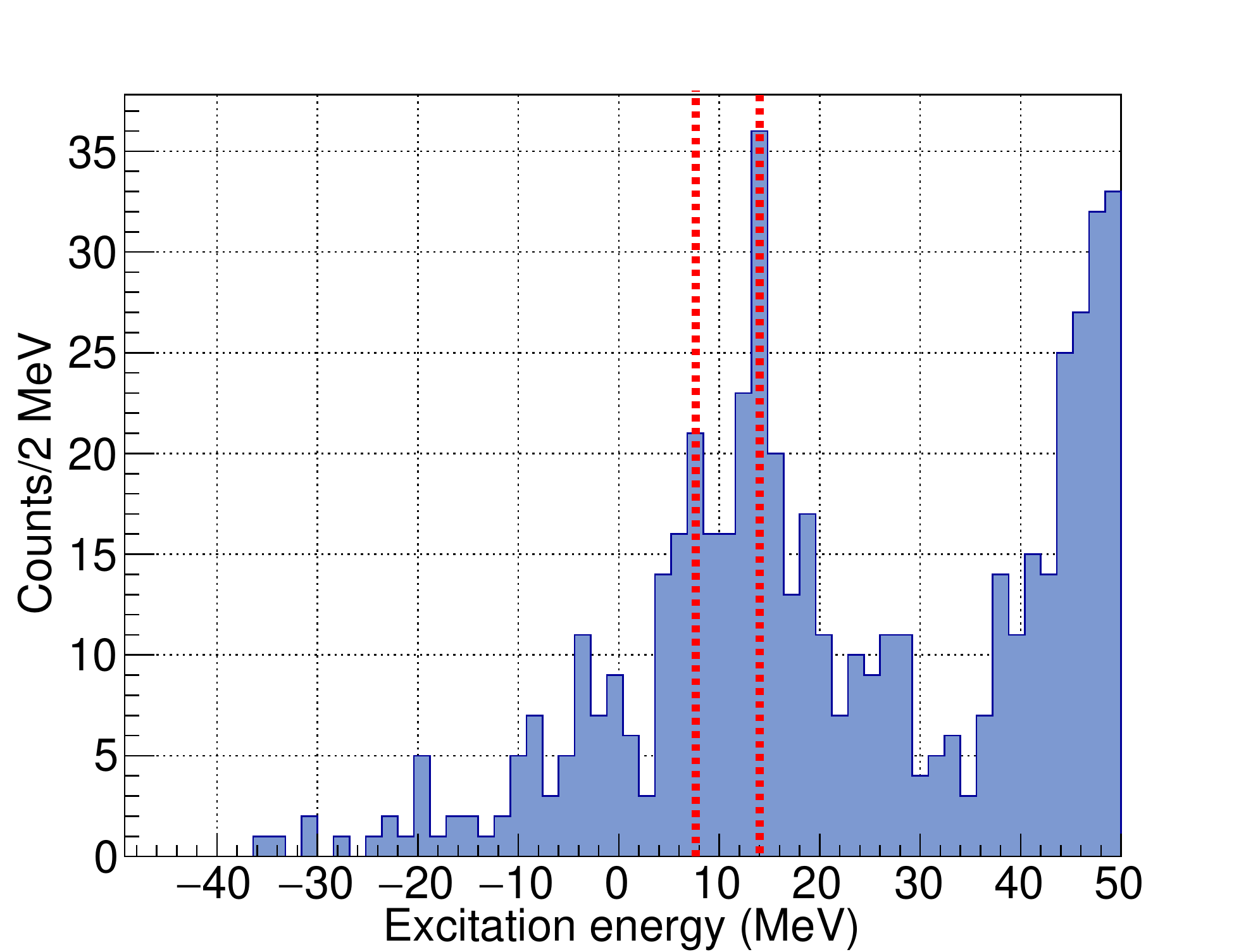}
\caption{Projection of the region between the red lines in Fig.~\ref{fig:Qvalue_16O12C}c showing the excitation energy in $^{12}\mathrm{C}$ as determined from the measurement of the 4-$\alpha$ breakup with excitation energies between 50 and 95 MeV in $^{16}\mathrm{O}$. The two red lines guide the eye to the locations of the enhancements at 7.65 MeV and 14 MeV which may be as a consequence of the highly-clustered states at those excitation energies. A small component corresponding to the $^{12}\mathrm{C}$(g.s) can also be seen at 0 MeV.\label{fig:projection}}
\end{figure}

While there are only a few resonances (although none of them near-threshold) in $^{16}\mathrm{O}$ observed in the 4-$\alpha$ breakup channel (and their contributions are weak), the same formulation can still be applied to the 4-$\alpha$ states observed to examine the two-body break-up contribution. Rather than selecting events within a certain excitation range in $^{16}\mathrm{O}$, the excitation energy in $^{16}\mathrm{O}$ in the 4$\alpha$ channel was plotted against the (negative) Q-value to give an indication of the excitation energy present in $^{12}\mathrm{C}$. This is shown in Fig.~\ref{fig:Qvalue_16O12C} for the three beam energies. For the 400 MeV data, an enhancement can be seen corresponding to the $^{12}\mathrm{C}(0_2^+) {+} ^{16}\mathrm{O}(50 \rightarrow 90 \, \mathrm{MeV})$ and $^{12}\mathrm{C}(14 \, \mathrm{MeV}) {+} ^{16}\mathrm{O}(50 \rightarrow 90 \, \mathrm{MeV})$ where the excitation energy projection between the red lines can be seen in Fig.~\ref{fig:projection}. While the resolution and statistics are limited so as not to allow for a more precise breakdown of the contribution of states here, there is some evidence of resonances populated in this region. This enhanced yield is also visible in the 280 MeV beam energy data although the statistics are limited here that a definite assertion cannot be made. Examination of the 160 MeV beam data shows no significant peak here, partially due to the obscuring of the signal by the strong diagonal seen at all three beam energies which can be ascribed to non $^{12}\mathrm{C}^{\star} {+} ^{16}\mathrm{O}^{\star}$ break-up. This agrees with the previous conclusion that the secondary peak in $^{16}\mathrm{O}$ was not described well by event-mixing demonstrating it had a correlated $\alpha$-particle component.\par
The majority of events for $^{16}\mathrm{O}$ are seen to lie on the diagonal loci in Fig.~\ref{fig:Qvalue_16O12C} where, as verified by the Monte Carlo simulation, contributions from non-binary decays of $^{28}\mathrm{Si}$ lie involving multi-particle breakup and sequential decay. This also suggests that the contribution of inelastic scattering is not dominant over compound nucleus formation in this experiment for all beam energies.
\subsection{\label{sec:Origins20Ne}Origins of neon-20}
\begin{figure}[h]
\includegraphics[width=0.5\textwidth]{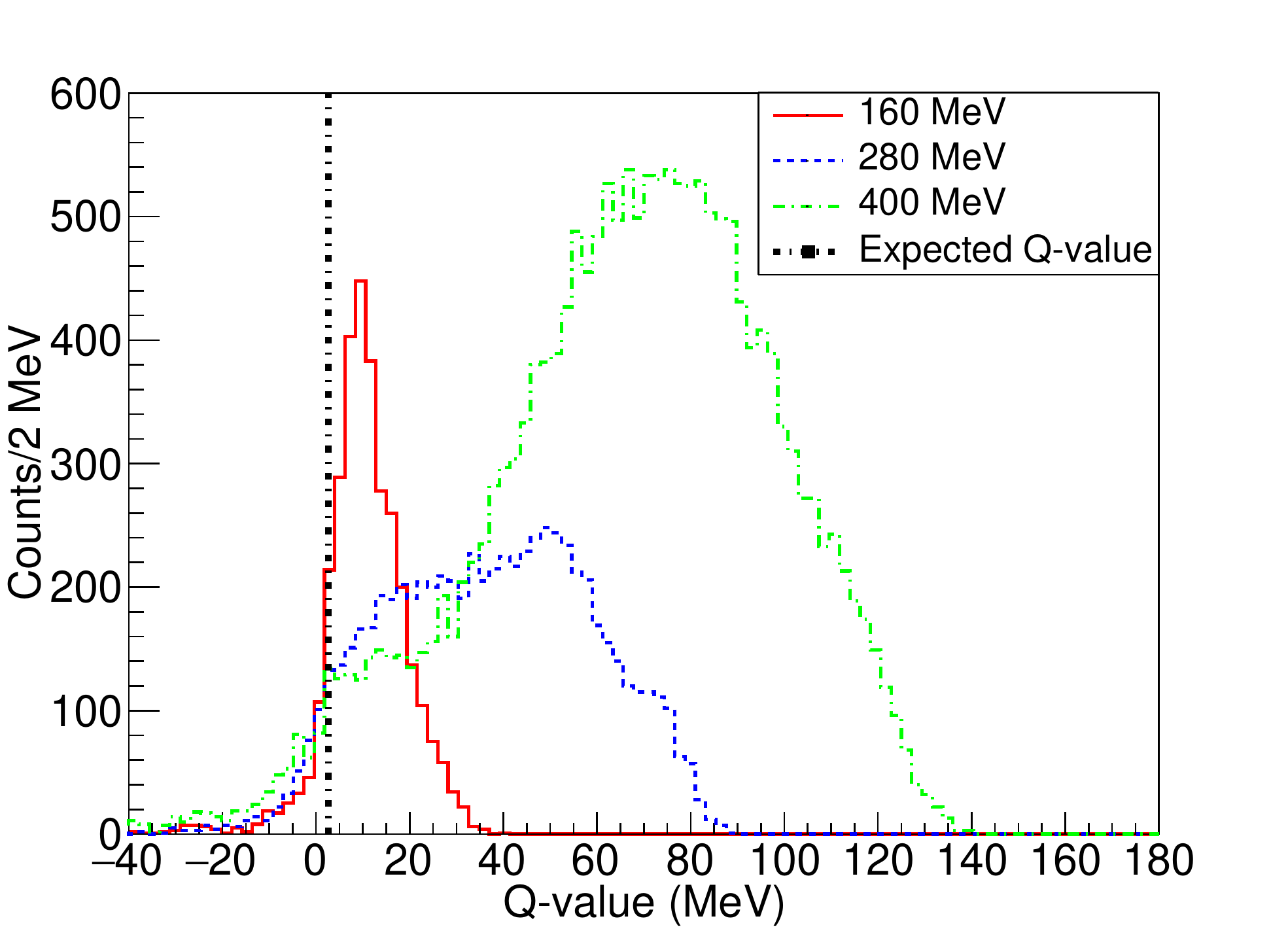}
\caption{Q-value for $^{20}\mathrm{Ne}^{\star} {+} ^{8}\mathrm{Be}^{\star}$ while reconstructing the missing $^{8}\mathrm{Be}$ from the measured $^{20}\mathrm{Ne}$ from 5 $\alpha$-particle events.\label{fig:hQvalue_20Ne8Be}}
\end{figure}

The EHF calculations also expect a large decay path via the intermediate $\alpha$-decay from $^{24}\mathrm{Mg}$ to $^{20}\mathrm{Ne}$. To verify this conclusion, the missing $^{8}\mathrm{Be}^{\star}$ can also be reconstructed from the measured $^{20}\mathrm{Ne}^{\star}$ from the decay to 5 $\alpha$-particles. Agreeing with the previous results, Fig.~\ref{fig:hQvalue_20Ne8Be} shows the results of this where the expected Q-value is not seen in the data for any of the three beam energies. Additionally, there is no component corresponding to higher excitations of $^{8}\mathrm{Be}$. This demonstrates these 5-$\alpha$ events either originate from sequential decay through $^{24}\mathrm{Mg}$ or from multi-particle break-up, both of which correspond to a Q-value not located along the expected Q-value labeled on Fig.~\ref{fig:hQvalue_20Ne8Be}.
\subsection{\label{sec:Origins24Mg}Origins of magnesium-24}
\begin{figure}[h]
\includegraphics[width=0.5\textwidth]{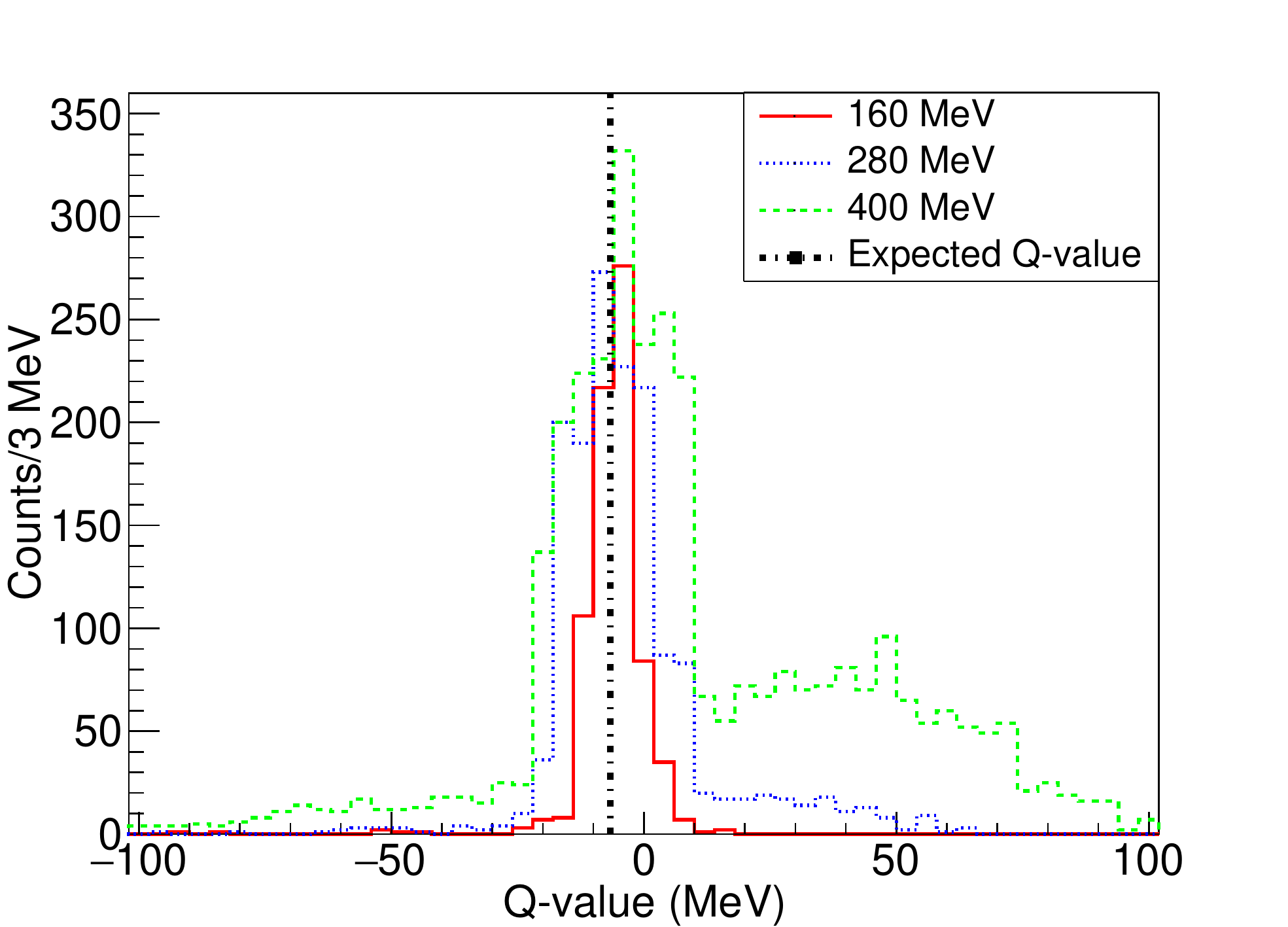}
\caption{(Negative) Q-value for $^{12}\mathrm{C}(^{16}\mathrm{O},{^{24}\mathrm{Mg}^{\star}}$) while reconstructing the missing $\alpha$ from $^{24}\mathrm{Mg}^{\star}$.\label{fig:hQvalue4}}
\end{figure}

The excitation energy spectrum for $^{24}\mathrm{Mg}$ shows no resonances in any determinable channel (6-$\alpha$, $^{12}\mathrm{C}(0_{2}^{+})+{^{12}\mathrm{C}(0_2^+)}$, $^{8}\mathrm{Be}+{^{8}\mathrm{Be}}+{^{8}\mathrm{Be}}$) with the primary decay mode being sequential $\alpha$-particle emission from $^{28}\mathrm{Si}$ to the continuum. Taking multiplicity 6 events and reconstructing for a missing $\alpha$-particle, the Q-value can be formulated, the results of which are seen in Fig.~\ref{fig:hQvalue4}. \par
To verify that the 6 $\alpha$-particles measured correspond to a complete 7-$\alpha$ decay, the majority of events lie around the expected Q-value showing the majority of these multiplicity 6 events are multiplicity 7 events with the final $\alpha$-particle being undetected. This matches well with the EHF calculations which demonstrated the multiplicity 6 yield is expected to be very small by virtue of the need to break apart an $\alpha$-particle (only becoming more common at a beam energy of 400 MeV, agreeing with the result seen here). This also demonstrates the multiplicity 6 events seen are genuine and not a result of pileup, matching well with the work in Section~\ref{sec:mult}.
\section{\label{sec:expsigres}Experimental signatures results}

The dominance of the uncorrelated $\alpha$-particles, particularly in the heavier N-$\alpha$ systems and the emergence of the near-threshold resonant states in $^{12}\mathrm{C}$ and $^{8}\mathrm{Be}$ where the known N-$\alpha$ structures appear, can be understood from the data. Aspects of the decay can be well described by sequential $\alpha$-decay predominantly through $^{24}\mathrm{Mg}$ or by multi-particle break-up as demonstrated by Fig.~\ref{fig:hQvalue4}. The EHF predictions for such a sequential $\alpha$-decay model however demonstrate that the predicted multiplicities are inconsistent with the experimental data and the Fermi break-up model better describes the higher multiplicities seen via multi-particle break-up.  An important signature of $\alpha$-condensation is the equivalency of the different decay paths into $\alpha$-gas substructures. Taking this, the phase spaces for the different partitions can therefore be ascribed an equal transition probability so the experimental yields can be directly compared to these phase spaces. To account for the experimental effects, most notably the efficiency, the different final states from the Fermi break-up model were efficiency-corrected using the Monte Carlo code to give a predicted experimental yield for the 7-$\alpha$ final state. The comparison between these predictions and the experimental data is shown in Table~\ref{tab:FBU}.\par
\begin{table*}[!ht]
\setlength\extrarowheight{2.0pt}
\caption{Experimental and theoretical branching ratios for the 7-$\alpha$ break-up modes for the different beam energies. The theoretical branching ratios for the 7-$\alpha$ break-up modes are calculated with the Fermi break-up model and efficiency-corrected using the Monte Carlo simulations. The error on the theoretical values is estimated at $\sim$ 1$\%$ of the total branching ratio from the uncertainty in the MC efficiency.\label{tab:FBU}}
\begin{ruledtabular}
\begin{tabular}{ccccccccc}
\multicolumn{2}{c}{Decay Path} & \multicolumn{3}{c}{Exp. branching ratio $\%$}  & \multicolumn{3}{c}{Theor. branching ratio $\%$}\\
\colrule
Label&Constituents&160 MeV&280 MeV&400 MeV&160 MeV&280 MeV&400 MeV\\ 
\colrule
I &$^{12}\mathrm{C}(0_2^+)$ + $^{12}\mathrm{C}(0_2^+)$ + $\alpha$& 0.0(0.0) & 0.0(0.0) & 0.2(0.2) & 7.1 & 0.3 & 0.0  \\ 

II &$^{12}\mathrm{C}(0_2^+)$ + $^{8}\mathrm{Be}$ + $^{8}\mathrm{Be}$& 0.0(0.0) & 0.0(0.0) & 0.0(0.0) & 1.0 & 0.3 & 0.1 \\ 

III &$^{12}\mathrm{C}(0_2^+)$ + $^{8}\mathrm{Be}$ + 2$\alpha$& 1.8(1.3) & 2.8(1.1) & 1.1(0.5) & 43.3 & 11.5 & 5.0 \\ 

IV &$^{12}\mathrm{C}(0_2^+)$ + 4$\alpha$& 4.5(2.0) & 2.8(1.1) & 1.5(0.6) & 11.7 & 16.7 & 10.8 \\ 

V &$^{8}\mathrm{Be}$ + $^{8}\mathrm{Be}$ + $^{8}\mathrm{Be}$ + $\alpha$& 3.6(1.8) & 0.4(0.4) & 0.4(0.3) & 20.7 & 5.2 & 2.3\\

VI &$^{8}\mathrm{Be}$ + $^{8}\mathrm{Be}$ + 3$\alpha$& 33.0(6.3) & 13.8(2.5) & 10.1(1.5) & 7.0 & 8.4 & 4.6 \\

VII &$^{8}\mathrm{Be}$ + 5$\alpha$& 45.5(7.7) & 37.8(4.5) & 32.8(3.0) & 9.2 & 57.3 & 76.7\\

VIII &7$\alpha$& 11.6(10.4) & 42.5(5.4) & 53.9(3.5) & 0.0 & 0.3 & 0.5 \\

\end{tabular}
\end{ruledtabular}
\end{table*}
\subsection{\label{sec:7alpha}Comparison of 7-$\alpha$ events with Fermi break-up predictions}

As the beam energy increases in the Fermi break-up calculation, the dominant decay modes change as the energy dependency becomes more important than the permutation factor for the $^{8}\mathrm{Be} {+} 5\alpha$ (VII) path. This decay mode is suppressed by a factor of $(5! = 120)$ in comparison to the $^{12}\mathrm{C}(0_2^+) {+} ^{8}\mathrm{Be} {+} 2\alpha$ (III) path which only has a suppression factor of 2. A similar situation can be seen for the 7$\alpha$-particle (VIII) decay which, while a 7-body decay, has a high power dependence on the energy, is suppressed by a factor of $(7! = 5040)$. The lower beam energies therefore correspondingly favor a small $n$-body decay.\par

It is apparent from Table~\ref{tab:FBU} that the decay modes seen experimentally for $E_b = 160$ MeV heavily favor the $^{8}\mathrm{Be}$ + $^{8}\mathrm{Be}$ (VI), single $^{8}\mathrm{Be}$ (VII) and 7-$\alpha$ (VIII) decay modes far in excess of those predicted theoretically. Part of this discrepancy can be attributed to modeling additional excited states in $^{12}\mathrm{C}$ above 7.65 MeV. While these states will have an increasingly reduced phase space as a consequence of the decrease in the kinematically available energy, they may also have a large spin component which partially compensates this effect as in Eq.~\ref{eq:FBU}. This angular momentum consideration and all aspects of penetrability are omitted in the Fermi break-up model. A large phase space to higher excitations in $^{12}\mathrm{C}$ would therefore manifest itself as replacing the $^{12}\mathrm{C}(0_2^+)$ strength with $^{8}\mathrm{Be}+\alpha$ and 3-$\alpha$ strength. Repositioning the branching ratio from the $^{12}\mathrm{C}(0_2^+) {+} ^{8}\mathrm{Be}$ and single-$^{12}\mathrm{C}(0_2^+)$ to these decay modes (i.e. III and IV move to VI and VII) then creates a better agreement with the experimentally observed data. It is not possible to include these additional contributions in the calculations as one cannot be certain of the density and spin of states once one enters the continuum. Additionally the transition matrix elements may start to deviate drastically from the constant value used under the assumption of an $\alpha$-condensate \cite{Tzany}.\par

An additional explanation for the discrepancy between the experimental and theoretical values is that the Hoyle state is not well described as an $\alpha$-condensate and as such, the transition matrix to break-up into the Hoyle state and $\alpha$-condensed systems is reduced or the mechanism is better described by a series of sequential decays rather than a direct break-up into constituent $\alpha$-particles. This latter explanation has been demonstrated to be unlikely as the experimental data suggest a reaction mechanism that differs from the EHF calculations due to the measured $\alpha$-particle multiplicities and the disagreement over the strength of binary scission modes which were shown to be extremely weak experimentally. The reaction mechanism has been demonstrated to be described well by aspects of both the Fermi break-up model and the statistical decay model showing the importance of both of these decay paths.\par
The 280 and 400 MeV data show a very similar pattern to those at 160 MeV albeit with a movement of strength in the $^{8}\mathrm{Be} {+} ^{8}\mathrm{Be}$ (VI) channel to the 7-$\alpha$ channel (VII). Even directly assigning the strength from the channels with excitations in $^{12}\mathrm{C}$ to the 7-$\alpha$ path, the strength is still in excess of that expected. This suggests that some of the yield seen in the single $^{8}\mathrm{Be}$ is therefore also sent to the 7-$\alpha$ channel. This could perhaps be by virtue of populating higher excitations in $^{8}\mathrm{Be}$ (i.e. the $\mathrm{^{8}\mathrm{Be}}(2_{1}^{+}$)) where limited statistics and resolution mean this contribution is difficult to observe although there is no evidence of such a strong contribution in the data. Again, this channel is difficult to calculate as the assumption of an identical transition matrix is no longer valid.
\par
Overall, the Fermi break-up description can be seen to show a reasonable agreement with the experimental data provided the points above regarding modeling additional higher excitations are taken into consideration. 

\subsection{\label{sec:conclusions}Decay mechanisms}
By modeling the decay processes via sequential decay using the EHF and modeling multi-particle break-up with the Fermi break-up model, the experimentally observed $\alpha$-particle multiplicities are seen to be in greater agreement with the Fermi break-up model. This can also be seen by the consistency of the population of the resonances in $^{12}\mathrm{C}$ (Fig.~\ref{fig:12C}) where as in Fig.~\ref{fig:EHFalphaKE}, as the beam energy increases, the EHF code predicts the dominance of sequential $\alpha$-particle emission rather than the population of discrete states. Additionally, the relative agreement between the 7-$\alpha$ channels discussed above in Section~\ref{sec:7alpha} with the discussed improvements to the model demonstrate the $\alpha$-particle decays are mainly driven by the multi-particle break-up decay mode. The sequential decay model does however have some success in describing the binary fission products extremely well and a lack of describing the importance of light cluster emissions in the form of $^{8}\mathrm{Be}$ and $^{12}\mathrm{C}(0_2^+)$ (which are shown to be very important in the Fermi break-up model) means this model fails to describe the high $\alpha$-particle multiplicity features seen experimentally. There is no evidence of an enhancement in the $\alpha$-gas multiplicity compared to those predicted by a non $\alpha$-condensate state. Previous investigations \cite{Akimune} which have stated such an enhancement over that predicted from the statistical decay processes require further examination and theoretical input into the exact decay mechanisms which incorporate many-particle break-up effects and light $\alpha$-conjugate cluster emission (i.e. $^{8}\mathrm{Be}$).

\subsection{\label{sec:thresholdstates}Measurement of near-threshold states}
Another key signature of $\alpha$-gas states is a highly clustered near-threshold state. While the previously-known highly clustered states were seen in $^{8}\mathrm{Be}$ and $^{12}\mathrm{C}$, no enhancement was seen in $^{16}\mathrm{O}$. This absence may be largely due to the decay mechanisms associated with such a state. The predicted energy of an $\alpha$-gas state in $^{16}\mathrm{O}$ is 15.1 MeV $(0_6^+)$ however the proximity of this state to the 4-$\alpha$ threshold at 14.44 MeV means that the preferential decay is via the $\alpha_0$ and $\alpha_1$ decay paths with the $\alpha_2$ ``signature-decay'' having a predicted branching ratio of $\sim 10^{-9}$ \cite{O16BR}. This calculation omits any modification to the Coulomb barrier by the mechanisms discussed in Section~\ref{sec:signatures}, but, however, demonstrates this state cannot be observed in the 4-$\alpha$ decay channel in the current experiment. Instead, this state was investigated by looking for the decay of the compound nucleus $^{28}\mathrm{Si}$ into $^{12}\mathrm{C}(0_2^+) {+} ^{16}\mathrm{O}(0_6^+)$ by measurement of the Hoyle state and reconstructing the $^{16}\mathrm{O}$ where the decay mode is no longer inhibitive. Figure~\ref{fig:Qvalue_Hoy} showed no such state populated in connection to the Hoyle state. Additionally, there was no evidence of resonant structure in the $^{12}\mathrm{C}$(g.s) + $\alpha$ excitation function demonstrating such a state is not well populated in this type of reaction despite the beryllium-8 ground state, the Hoyle state and its (possible) rotational excitation all being clearly visible above the continuum contribution.\par
In $^{20}\mathrm{Ne}$, a quenching of the 5-$\alpha$ strength was clearly evident in proximity to the 5-$\alpha$ barrier in agreement with previous experiments. This can be attributed to the dominance of the $^{16}\mathrm{O} + \alpha$ decay channels here mirroring the situation in $^{16}\mathrm{O}$ where such a near-threshold state cannot be detected in the 5-$\alpha$ decay channel until one is well over the Coulomb barrier ($\sim 24$ MeV). Indirect measurement of this state via measurement of $^{8}\mathrm{Be}$(g.s) and reconstruction of $^{20}\mathrm{Ne}$ also demonstrated no increased yield above the 5-$\alpha$ threshold where such a state would be expected to manifest and the limitations of measuring the characteristic 5-$\alpha$ decay are not required.
\section{\label{sec:summary}Summary}
In summary, the reaction mechanism of the $^{12}\mathrm{C}(^{16}\mathrm{O},^{28}\mathrm{Si}^{\star})$ channel was studied at three beam energies of 160, 280, and 400 MeV. This experiment was performed to search for $\alpha$-gas states in $^{12}\mathrm{C}$, $^{16}\mathrm{O}$, $^{20}\mathrm{Ne}$, $^{24}\mathrm{Mg}$ and $^{28}\mathrm{Si}$. No such states were found for systems heavier than $^{12}\mathrm{C}$. In the case of heavier isotopes, there were various reasons discussed in this work in detail which could have obscured the observation of those states. While the Hoyle state ($^{12}\mathrm{C}(0_2^+)$) is the best candidate for such an $\alpha$-gas state (from theoretical predictions), following other recent experiments (e.g. of the direct 3$\alpha$-decay of the Hoyle state \cite{Robin1,Danielle1}), we can conclude that although theoretically this state fits well with an $\alpha$-gas description, experimentally it appears that the underlying fermionic structure of the $\alpha$-particles is sufficient to prevent this $\alpha$-gas nature and none of the expected signatures for $\alpha$-condensation were seen in the current analysis.\par
This type of high-energy reaction requires input from multi-particle breakup models to describe the observed $\alpha$-particle multiplicities which are in excess of those predicted from Hauser Feshbach calculations. The enhancements are not sufficient however to support the observation of $\alpha$-condensate states.

\section{Acknowledgments}
The authors would like to thank the operators and technicians at Catania for their support and for providing a stable beam during the experiment. J.B. would also like to thank LNS for financial support during his attendance at the experiment.
\bibliography{apssamp}

\end{document}